\documentclass[11pt]{article}
\usepackage[margin=1in]{geometry}

\usepackage{amsfonts,amsmath,amssymb,mathtools} 
\usepackage{graphicx}
\usepackage{booktabs}

\usepackage[numbers]{natbib}
\usepackage[colorlinks=true,linkcolor=blue,citecolor=blue,urlcolor=blue]{hyperref}

\usepackage[nameinlink,noabbrev]{cleveref}

\usepackage{algorithm}
\usepackage{algpseudocode}
\usepackage[caption=false]{subfig}
\newtheorem{theorem}{Theorem}[section]
\newtheorem{lemma}{Lemma}[section]
\newtheorem{proposition}{Proposition}[section]
\newtheorem{corollary}{Corollary}[section]
\newtheorem{definition}{Definition}[section]
\newtheorem{example}{Example}[section]
\newtheorem{assump}{Assumption}[section]

\crefname{algorithm}{Algorithm}{Algorithm}
\crefname{figure}{Figure}{Figure}
\crefname{table}{Table}{Table}

\crefname{theorem}{Theorem}{Theorem}
\crefname{example}{Example}{Example}
\crefname{proposition}{Proposition}{Proposition}
\crefname{definition}{Definition}{Definition}
\crefname{corollary}{Corollary}{Corollary}
\crefname{lemma}{Lemma}{Lemma}
\crefname{assump}{Assumption}{Assumption} 
\crefrangemultiformat{assump}{Assumptions~#3#1#4--#5#2#6}{ and~#3#1#4--#5#2#6}{, #3#1#4--#5#2#6}{ and~#3#1#4--#5#2#6}
\crefname{equation}{}{}
\creflabelformat{equation}{#2(#1)#3}

\newcommand{\tA}{{\mathcal{A}}}
\newcommand{\ftA}{\boldsymbol{\mathcal{A}}}
\newcommand{\tB}{{\mathcal{B}}}
\newcommand{\ftB}{\boldsymbol{\mathcal{B}}}

\newcommand{\ftC}{\boldsymbol{\mathcal{C}}}
\newcommand{\tI}{{\mathcal{I}}}
\newcommand{\tT}{{\mathcal{T}}}
\newcommand{\ftT}{\boldsymbol{\mathcal{T}}}
\newcommand{\tE}{{\mathcal{E}}}
\newcommand{\ftE}{\boldsymbol{\mathcal{E}}}
\newcommand{\tU}{{\mathcal{U}}}
\newcommand{\ftU}{\boldsymbol{\mathcal{U}}}
\newcommand{\tV}{{\mathcal{V}}}
\newcommand{\ftV}{\boldsymbol{\mathcal{V}}}
\newcommand{\tS}{{\mathcal{S}}}
\newcommand{\ftS}{\boldsymbol{\mathcal{S}}}
\newcommand{\tP}{{\mathcal{P}}}
\newcommand{\tQ}{{\mathcal{Q}}}
\newcommand{\tX}{{\mathcal{X}}}
\newcommand{\ftX}{\boldsymbol{\mathcal{X}}}
\newcommand{\tM}{{\mathcal{M}}}
\newcommand{\tZ}{{\mathcal{Z}}}
\newcommand{\ftZ}{\boldsymbol{\mathcal{Z}}}
\newcommand{\tTheta}{{\Theta}}

\newcommand{\te}{{\vec{e}}}
\newcommand{\tte}{\stackrel{\circ}{e}}

\newcommand{\mA}{{A}}
\newcommand{\fmA}{\boldsymbol{A}}

\newcommand{\fmM}{\boldsymbol{M}}
\newcommand{\mB}{{B}}
\newcommand{\fmB}{\boldsymbol{B}}

\newcommand{\fmC}{\boldsymbol{C}}

\newcommand{\fmP}{\boldsymbol{P}}

\newcommand{\mT}{{T}}
\newcommand{\fmT}{\boldsymbol{T}}

\newcommand{\fmE}{\boldsymbol{E}}

\newcommand{\fmU}{\boldsymbol{U}}

\newcommand{\fmV}{\boldsymbol{V}}

\newcommand{\fmS}{\boldsymbol{S}}

\newcommand{\fmX}{\boldsymbol{X}}

\newcommand{\fmZ}{\boldsymbol{Z}}
\newcommand{\fmW}{\boldsymbol{W}}
\newcommand{\fmQ}{\boldsymbol{Q}}

\newcommand{\fmTheta}{\boldsymbol{\Theta}}
\newcommand{\Bernstein}{\mathrm{Bernstein}}
\newcommand{\Chernoff}{\mathrm{Chernoff}}
\newcommand{\BE}{\mathrm{BE}}
\newcommand{\diag}{\mathrm{diag}}
\newcommand{\bdiag}{\mathrm{bdiag}}
\newcommand{\bcirc}{\mathrm{bcirc}}
\newcommand{\unfold}{\mathrm{unfold}}
\newcommand{\fold}{\mathrm{fold}}
\newcommand{\overbar}[1]{\mkern 1.5mu\overline{\mkern-1.5mu#1\mkern-1.5mu}\mkern 1.5mu}
\newcommand{\toD}{\overset{\text{d}}{\longrightarrow}}

\usepackage{authblk}
\title{Uncertainty Quantification for Noisy Low-tubal-rank Tensor Completion
}
\author[1]{Jiuqian Shang \thanks{Email: \texttt{jiuqian@umich.edu}}}
\author[1]{Jingyang Li \thanks{Email: \texttt{jjyyli@umich.edu}}}
\author[1]{Yang Chen \thanks{Email: \texttt{ychenang@umich.edu}}}
\affil[1]{Department of Statistics, University of Michigan, Ann Arbor, MI.}

\begin{document}
\maketitle
\vspace{-0.3cm}
\begin{abstract}
High-dimensional tensor data often exhibit strong temporal correlations that appear as low-dimensional structures in the frequency domain. While the low-tubal-rank tensor model effectively captures these spectral features, making it potentially suitable for geophysical data, existing methods primarily focus on point estimation. Uncertainty quantification (UQ) of imputed values and rigorous statistical inference for these models remain largely unexplored. In this work, we propose a flexible inference framework for linear forms of high-dimensional tensors. Employing a double-sample debiasing technique followed by a low-rank projection, we construct asymptotically Gaussian estimators that give valid statistical inference under mild assumptions. More precisely, we can perform hypothesis testing and construct confidence intervals with this result. We validate the theoretical results through extensive simulations and demonstrate the method’s practical effectiveness in completing the global total electron content data. We demonstrate, using those numerical results, that our entrywise confidence intervals are robust and reliable, yielding informative uncertainty quantification that captures underlying variability.
\end{abstract}

\textbf{Key words.} Debiasing, spectral method, asymptotic normality, confidence interval.

\textbf{MSC codes.} 62F10, 62F12, 62F25, 62P35.

\section{Introduction}\label{sec:introduction}
\subsection{Background and motivation}
The recovery of missing entries in high-dimensional tensors, known as \emph{tensor completion}, is a fundamental challenge in the analysis of complex multi-way data \cite{bi2021tensors}. 
Without imposing structural constraints, this problem is infeasible, as the unobserved entries can take arbitrary values. To make the problem tractable, standard approaches assume that the underlying signal possesses a low-dimensional structure, most commonly formalized as low-rank. 
Extensive literature on tensor completion has leveraged various low-rank structures to recover missing data. These approaches include methods based on the canonical polyadic (CP) rank \cite{jain2014provable, barak2016noisy, cai2019nonconvex}, the Tucker rank \cite{huang2015provable,xia2019polynomial,tong2022scaling}, and tensor-train (TT) rank \cite{cai2022provable,cai2022tensor}. While these symmetric rank notions are powerful for general data representation, they treat all tensor modes equivalently. This isotropy assumption is often unrepresentative of \emph{spatiotemporal data} (such as time series of satellite imaging data), which are typically third-order tensors with two spatial modes and one temporal mode. The temporal dimension often exhibits distinct dynamics, such as periodicity or smoothness. These temporal patterns are not well captured by standard CP or Tucker decompositions.

To address the unique geometry of spatiotemporal data, the \emph{low-tubal-rank} tensor model has emerged as a robust alternative. Introduced within the t-product framework \cite{kilmer2013third} and further developed in subsequent completion algorithms \cite{zhang2016exact,liu2019low,jiang2019robust,song2023low}, this approach transforms the tensor along the temporal mode into the frequency domain, thereby naturally capturing periodic temporal dynamics. In this spectral representation, the tensor decouples into a sequence of frontal slices, and the tubal rank is defined as the maximum rank among these frequency-domain slices. This structure is intuitively compelling for physical processes: it implies that, at each temporal frequency, the spatial data lie in a low-dimensional subspace, thereby modeling the interplay between spatial correlations and temporal dependency structures such as periodicity.

From a computational standpoint, the recovery of low-tubal-rank tensors is well studied. Algorithms generally fall into two categories: convex approaches that minimize the tensor nuclear norm (TNN) \cite{zhang2016exact, lu2019low}, and nonconvex methods based on tensor factorization, such as Riemannian optimization \cite{song2023riemannian} or alternating minimization \cite{liu2019low}. While these techniques provide efficient and accurate point estimates, they do not inherently assess the reliability of the reconstruction.

A critical methodological gap remains: Uncertainty Quantification (UQ). While statistical inference for low-rank matrix completion is now well-established \cite{chen2019inference, xia2021statistical}, and recent works have extended these rigorous guarantees to low-CP-rank \cite{cai2022uncertainty} and low-Tucker-rank \cite{ma2024statistical} models, the statistical properties of the low-tubal-rank model remain unexplored. The complex spectral structure of the tubal rank introduces unique dependencies that cannot be addressed by simply applying existing tensor methods.

This gap is particularly acute in geophysical monitoring, specifically in the reconstruction of global \emph{Ionospheric Total Electron Content (TEC, \cite{sun2022matrix,sun2023complete})} maps. Although TEC signals naturally exhibit the spectral sparsity captured by our model, observations are often spatially sparse due to the limited coverage of the ground receiver \cite{mendillo2006storms}. Existing completion procedures focus solely on point estimation, yet domain scientists require rigorous uncertainty quantification (UQ) to distinguish between reliable data and reconstruction artifacts. Furthermore, general-purpose UQ alternatives, such as tensor conformal prediction \cite{sun2025conformalized}, are often computationally prohibitive or lack coverage validity in this high-dimensional, sparse setting.

\subsection{Our contributions}
In this work, we bridge this gap by developing a statistical inference framework for low-tubal-rank tensor completion. We propose a procedure for constructing confidence intervals and performing hypothesis tests for general linear functionals of the underlying tensor, including individual entries and regional averages.

From an algorithmic perspective, our method is independent of the specific solver used for initialization, provided that the solver is uniformly accurate at the scale of the observation noise. Starting from this consistent but biased initialization, we apply a double-sample debiasing step followed by a spectral projection. This refines the estimator into one whose linear functionals are asymptotically normal with consistently estimable variance. This result enables valid confidence intervals and hypothesis tests for any fixed linear functional of the tensor.

From a technical perspective, our analysis addresses the challenges posed by the complex spectral domain by deriving a novel perturbation bound for the tensor singular value decomposition (t-SVD). This result characterizes the stability of singular subspaces in the frequency domain and may be of independent interest for broader low-tubal-rank tensor problems. Furthermore, we establish non-asymptotic convergence rates for the estimated linear forms, providing finite-sample guarantees that complement the asymptotic normality results. We demonstrate the effectiveness of this framework on the TEC completion problem, providing the first spatially resolved uncertainty quantification for low-tubal-rank reconstructions of the videos of the global TEC maps.

\subsection{Paper organization}
The remainder of the paper is organized as follows. \Cref{subsec:tubal-rank} defines the tubal rank and t-product algebra, while \cref{sec:model-set-up} details the observation model.
Our main methodological and theoretical results are presented in \cref {sec:method_property}, where we describe the proposed debiasing procedure (\cref{sec:estimating_linear_forms}), establish asymptotic normality (\cref{subsec:Asymptotic Normality}), and derive uncertainty quantification (\cref{subsec:Inferences about Linear Forms}). We validate the methodology through simulations in \cref{sub:simu} and apply it to TEC data in \cref{subsec:TEC}.

\section{Preliminaries}

This section sets up the notation for third-order tensors and reviews the t-product framework used throughout the rest of the paper. We also introduce the basics of the {noisy tensor completion} model: the observation operator, sampling mechanism, and noise assumptions. 

\subsection{Tensor tubal rank and notations}
\label{subsec:tubal-rank}
In this subsection, we formalize the algebraic structure of the tubal rank, which relies on the tensor-tensor product (t-product) and the associated tensor singular value decomposition. We follow the framework established by \cite{kilmer2011factorization,kilmer2013third}.
We begin by defining the basic operators in \cref{def:ops} that facilitate the algebraic manipulation of third-order tensors.
\begin{definition}[basic operators]\label{def:ops}
For a tensor {$\tT\in\mathbb{C}^{d_1\times d_2\times d_3}$} (where $\mathbb{C}$ denotes the complex domain) with its frontal slices $\mT^{(t)}=\tT(:,:,t)\in\mathbb{C}^{d_1\times d_2}$ for $t=1,\dots,d_3$, we define the unfolding and folding operators as
\begin{displaymath}
    \unfold(\tT)=\begin{bmatrix}\mT^{(1)}\\ \mT^{(2)}\\ \vdots\\ \mT^{(d_3)}\end{bmatrix}\in\mathbb{C}^{(d_1d_3)\times d_2},\qquad
\fold(\unfold(\tT))=\tT.
\end{displaymath}
Additionally, we define the block-diagonal
\begin{displaymath}
\bdiag(\tT)=
\begin{bmatrix}
\mT^{(1)} & & & \\
& \mT^{(2)} & & \\
& & \ddots & \\
& & & \mT^{(d_3)}
\end{bmatrix},
\end{displaymath}
and block-circulant liftings
\begin{displaymath}
    \bcirc(\tT)=
\begin{bmatrix}
\mT^{(1)} & \mT^{(d_3)} & \cdots & \mT^{(2)}\\
\mT^{(2)} & \mT^{(1)} & \cdots & \mT^{(3)}\\
\vdots & \vdots & \ddots & \vdots\\
\mT^{(d_3)} & \mT^{(d_3-1)} & \cdots & \mT^{(1)}
\end{bmatrix}
\in\mathbb{C}^{(d_1d_3)\times(d_2d_3)}.
\end{displaymath}
\end{definition}
The t-product generalizes matrix multiplication to tensors via the circular convolution of tubes, as given by \cref{def:tprod}. 
\begin{definition}[t-product]\label{def:tprod}
For tensors $\tA\in\mathbb{R}^{d_1\times d\times d_3}$ and $\tB\in\mathbb{R}^{d\times d_2\times d_3}$, the t-product $\tA\tB$ is defined as
\begin{displaymath}
    \tA\tB=\fold\!\left(\bcirc(\tA)\,\textsf{unfold}(\tB)\right)\in\mathbb{R}^{d_1\times d_2\times d_3}.
\end{displaymath}
\end{definition}
The computational efficiency and theoretical properties of the t-product stem from its relationship with the Discrete Fourier Transform (DFT) along the third mode.
Let $F_{d_3}\in\mathbb{C}^{d_3\times d_3}$ be the unnormalized DFT matrix with$
    (F_{d_3})_{\ell s}=\omega^{(\ell-1)(s-1)}$,where $\omega=e^{-2\pi \sqrt{-1}/d_3}.$
\begin{definition}[Mode-3 DFT]\label{def:fft3}
The mode-3 DFT is the linear map $\mathcal{F}_3:\mathbb{R}^{d_1\times d_2\times d_3}\to\mathbb{C}^{d_1\times d_2\times d_3}$ defined by $\ftT=\mathcal{F}_3(\tT)
$ such that \begin{displaymath}
{\ftT(i,j,:)=F_{d_3}}\tT(i,j,:), \quad\forall (i,j)\in[d_1]
\times[d_2],
\end{displaymath}
where $[d] = \{1,\ldots, d\}$ for $d=d_1, d_2$. Then the inverse transform $\mathcal{F}_3^{-1}$, is given by $\tT=\mathcal{F}_3^{-1}(\ftT)$ such that 
\begin{displaymath}
    \tT(i,j,:)=F_{d_3}^{-1}\ftT(i,j,:),\ \ \quad \forall (i,j)\in[d_1]
\times[d_2],
\end{displaymath}
where $F_{d_3}^{-1} = \frac{1}{d_3}F_{d_3}^H$, and the superscript $H$ denotes the matrix conjugate transpose.
\end{definition}
Throughout this paper, we denote the tensor in the frequency domain by bold calligraphic letters (e.g., $\ftA, \ftB, \ftT$) and its frontal slices (matrices) by bold symbols (e.g., $\fmA^{(t)},\fmB^{(t)},\fmT^{(t)}$). In this slice-wise representation, the t-product decouples into matrix multiplication, as described in the following proposition.
\begin{proposition}\label{prop:tprod}
Let $\mathcal{C}=\mathcal{A}\mathcal{B}$. Then, for each frequency slice $t=1,\ldots,d_3$,
$\fmC^{(t)}=\fmA^{(t)}\,\fmB^{(t)}.$
This relationship implies $\bdiag(\ftC) = \bdiag(\ftA)\bdiag(\ftB)$.
\end{proposition}
The proof follows directly from the properties of block circulant matrices \cite{kilmer2011factorization,kilmer2013third}. In alignment with this literature, \cref{def:structure} extends standard matrix concepts, such as transpose, orthogonality, and factorizations to the tensor setting.
\begin{definition}[structural definitions]\label{def:structure}
\begin{enumerate}
    \item \textit{Conjugate transpose}: Let $\tT\in\mathbb{C}^{d_1\times d_2\times d_3}$; the conjugate transpose $\tT^{\dagger}$ is obtained by taking the conjugate transpose of each frontal slice of $\mathcal{T}$ and reversing the order of the slices $2,\cdots,d_3$. 
    \item \textit{Identity Tensor}: The identity tensor $\tI_d\in\mathbb{R}^{d\times d\times d_3}$ has $\tI_d(:,:,1)=I_d$ and $\tI_d(:,:,t)=0$ for $t>1$. It satisfies $\tA\tI_d=\tI_d\tA=\tA$ for any $\tA\in\mathbb{R}^{d\times d\times d_3}$.
    \item \textit{Orthogonal Tensor}: A tensor $\tQ\in\mathbb{R}^{d\times d\times d_3}$ is orthogonal if $\tQ^{\dagger}\tQ=\tQ\tQ^{\dagger}=\tI_d$. Equivalently, each frequency slice $\fmQ^{(t)}$ is unitary, $t=1,\cdots,d_3$.
    \item \textit{f-diagonal Tensor}: A tensor $\tS\in\mathbb{C}^{d_1\times d_2\times d_3}$ is f-diagonal if each frontal slice $\tS(:,:,k)$ is diagonal.
\end{enumerate} 
\end{definition}
This algebraic structure enables the tensor Singular Value Decomposition (t-SVD, see \cref{prop:tSVD}), which serves as the basis for the tubal rank given in \cref{def:tubrank}.
\begin{proposition}[t-SVD, Theorem~2.2 in \cite{lu2019tensor}]\label{prop:tSVD}
Any tensor $\tT\in\mathbb{R}^{d_1\times d_2\times d_3}$ admits a factorization $\tT = \tU\tS\tV^{\dagger}$,
where $\tU\in\mathbb{R}^{d_1\times d_1\times d_3}$ and
$\tV\in\mathbb{R}^{d_2\times d_2\times d_3}$ are orthogonal and
$\tS\in\mathbb{R}^{d_1\times d_2\times d_3}$ is f-diagonal. 
\end{proposition}
This decomposition is computed by performing standard matrix SVDs on each frontal slice in the frequency domain, followed by the inverse mode-3 DFT. Specifically, let $\fmT^{(t)}$  be the $t$-th frontal slice of the mode-3 DFT transformed tensor $\ftT$. We compute its matrix SVD as:
\begin{equation*}
    \fmT^{(t)}=\fmU^{(t)}\fmS^{(t)}\big(\fmV^{(t)}\big)^H, \quad t = 1,\cdots,d_3.
\end{equation*}
Next, we stack the resulting matrices $\{\fmU^{(t)}\}_{t=1}^{d_3}$, $\{\fmS^{(t)}\}_{t=1}^{d_3}$ and $\{\fmV^{(t)}\}_{t=1}^{d_3}$ along the third dimension to form the tensors $\ftU$, $\ftS$ and $\ftV$, respectively. Finally, applying the inverse mode-3 DFT to these tensors yields the final components $\tU$, $\tS$, and $\tV$. Efficient computation of t-SVD appears in \cite[Algorithm~2]{lu2019tensor}. 

Counting the number of non-zero diagonal tubes in $\tS$ provides an intrinsic notion of rank within this algebraic framework. This leads us to the formal concept of tubal rank, which is detailed in \cref{def:tubrank}.
\begin{definition}[tubal rank and skinny t-SVD]\label{def:tubrank}
The tubal rank of $\tT$, denoted $\operatorname{rank}(\tT)$, is the number of non-zero singular tubes in $\mathcal{S}$ in the original domain, i.e., $\#\{i: \mathcal{S}(i,i,:)\neq \boldsymbol{0}\}$, or $\#\{i: \boldsymbol{\mathcal{S}}(i,i,:)\neq \boldsymbol{0}\}$ in the frequency domain. Equivalently, it is the maximum rank of the spectral slices:
$$\operatorname{rank}(\mathcal{T}) = \max_{t=1,\ldots, d_3} \left\{\operatorname{rank}(\widehat{\mathbf{T}}^{(t)})\right\}.$$
For tensors with tubal rank $r$, we employ the skinny t-SVD $\mathcal{T} = \mathcal{U}\mathcal{S}\mathcal{V}^{\dagger}$, where $\mathcal{U} \in \mathbb{R}^{d_1\times r\times d_3}$, $\mathcal{V} \in \mathbb{R}^{d_2\times r\times d_3}$ are orthogonal, and $\mathcal{S} \in \mathbb{R}^{r\times r\times d_3}$ is f-diagonal.
\end{definition}
We will use the skinny t-SVD $\tT = \tU\tS\tV^{\dagger}$ throughout the paper unless otherwise stated. 

Finally, we define the inner product $\langle \tA,\tB\rangle := \sum_{k=1}^{d_3} \operatorname{Tr} \ ((\mA^{(k)})^{H}\mB^{(k)})$, and the Frobenius norm $\|\tA\|_{\mathrm{F}}=\sqrt{\langle\tA,\tA\rangle}$. 
Recall that $\mA^{(k)}$ denotes the $k$-th frontal slice of $\tA$ in the original domain, and thus this definition is equivalent to the standard Frobenius norm. Due to the scaling of the DFT, we have the relationship $\|\tA\|_{\mathrm{F}}=\|\bdiag(\ftA)\|_{\mathrm{F}}/\sqrt{d_3}$.
We also define the tensor spectral norm $\|\tA\|$ as the maximum spectral norm of its frequency slices:
\begin{align*}
    \|\tA\|:= \|\bdiag(\ftA)\| = \max_t \|\fmA^{(t)}\|,
\end{align*}
where $\ftA$ is the DFT of $\tA$ along the third mode, and $\fmA^{(t)}$ are its frontal slices.
Here, the spectral norm of a complex matrix $\fmA^{(t)}$ is defined as its largest singular value. 

\subsection{Model Set-up}
\label{sec:model-set-up}
We formulate the noisy tensor completion problem for a tensor $\tT \in \mathbb{R}^{d_1 \times d_2 \times d_3}$ with tubal rank $r$. Suppose we observe $n$ independent pairs $\{(\tX_i, Y_i)\}_{i=1}^n$ satisfying
\begin{equation}
    \label{eq:model}  Y_i=\langle\tT,\tX_i\rangle+\xi_i,
\end{equation}  
where $\tX_i$ is the canonical basis tensor that selects the $(j_i, k_i, l_i)$ entry of $\mathcal{T}$,  i.e., {$\tX_i(j_i,k_i,l_i) =1$ and all other entries in $\tX_i$ are equal to zero, yielding} $Y_i = \mathcal{T}{(j_i, k_i ,l_i)} + \xi_i$. Thus, $Y_i$ is a scalar equal to the true entry $\mathcal{T}(j_i, k_i ,l_i)$ corrupted by stochastic noise $\xi_i$. In a typical tensor completion problem, $n< d_1d_2d_3$. In this paper, we assume the following missingness mechanism.
\begin{assump}[sampling mechanism]\label{assump:sampling}
The indices $(j_i,k_i,l_i)$ corresponding to {the non-zero entry in} $\tX_i$ are sampled independently and uniformly with replacement from $[d_1] \times [d_2] \times [d_3]$, where $[d]=\{1,\ldots, d\}$ for $d=d_1,d_2,d_3$.
\end{assump} \Cref{assump:sampling} imposes an independent and uniform sampling design. This is the canonical model in the completion literature \cite{candes2012exact, cai2019nonconvex, xia2021statistically} and ensures the sampling operator is isotropic in expectation. While non-uniform or adaptive sampling designs \cite{klopp2014noisy, krishnamurthy2013adaptive} are relevant for specific applications, the uniform design serves as a theoretical benchmark, enabling us to establish fundamental statistical guarantees that serve as a baseline for more complex scenarios.

Complementing the sampling design, we now specify the distributional properties of the measurement noise.
\begin{assump}[noise model]\label{assump:noise}
    The noise variables $\{\xi_i\}_{i=1}^n$ are independent and identically distributed, and are independent of $\{\tX_i\}$. Each $\xi_i$ has mean zero and variance $\sigma_\xi^2$, and is sub-Gaussian with parameter $\sigma_\xi^2$, i.e., satisfying $\mathbb{E}e^{s\xi} \leq e^{s^2\sigma_\xi^2/2}$ for all $s \in \mathbb{R}$.
\end{assump}

\section{Methodology and theoretical properties}
\label{sec:method_property}

Although the point estimation problem of recovering a partially observed tensor $\tT$ is well-studied \cite{zhang2016exact,jiang2019robust,liu2019low,song2023riemannian}, a critical objective in scientific applications is to quantify the uncertainty associated with these estimates. Point estimates alone are often insufficient for reliable decision-making, particularly in geophysical monitoring, where it is essential to distinguish true physical signals from artificial errors introduced by the reconstruction process~\cite{sun2023complete}. 
By choosing appropriate test tensors $\tM$, this flexible formulation enables the construction of valid confidence intervals for individual entries, fiber averages, or other regional quantities of interest, thereby providing a comprehensive statistical characterization of the reconstructed tensor.

\subsection{Estimating linear forms}
\label{sec:estimating_linear_forms}
Wen ow construct an estimator for the linear form $\langle \tT,\tM\rangle$, starting from an initial tensor completion estimator that satisfies the consistency requirement to be stated formally in \cref{assump:init_entry}.
Our methodology combines a one-step debiasing correction with a low-rank retraction, implemented in \Cref{alg:debias}. It is designed to improve estimation accuracy for the linear form and to facilitate its subsequent distributional analysis.
\begin{algorithm}[t]
\caption{Sample–splitting debiasing and retraction}
\label{alg:debias}
\begin{algorithmic}[1]
\Statex \textbf{Input:} Observations $\{(\tX_i,Y_i)\}_{i=1}^n$, dimensions $(d_1,d_2,d_3)$, rank $r$, test tensor $\tM$, an initial tensor completion algorithm.
\Statex \textbf{Output:} The estimated linear form $\langle \widehat{\tT},\tM\rangle$.
\State \textit{Initialization}
\Statex Randomly split the sample into $\mathfrak{D}_1 = \left\{(\tX_i, Y_i)\right\}_{i=1}^{\lceil n/2 \rceil}$ and $\mathfrak{D}_2 = \left\{(\tX_i, Y_i)\right\}_{i=\lceil n/2 \rceil+1}^n.$

\Statex Obtain initial estimators $\widehat{\tT}_{\mathrm{init},1}$ and $\widehat{\tT}_{\mathrm{init},2}$ by applying the initial estimation algorithm to $\mathfrak{D}_1$ and $\mathfrak{D}_2$, respectively.

\State \textit{Debiasing}
\begin{equation}\label{alg-debias}
    \left\{
\begin{aligned}
\widehat{\tT}_{\mathrm{unbs},1} &\gets \widehat{\tT}_{\mathrm{init},1}+\frac{d_1 d_2 d_3}{|\mathfrak{D}_2|}
\sum_{i=\lceil n/2 \rceil+1}^{n}\bigl(Y_i-\langle \widehat{\tT}_{\mathrm{init},1},\tX_i\rangle\bigr)\tX_i; \\
\widehat{\tT}_{\mathrm{unbs},2} &\gets \widehat{\tT}_{\mathrm{init},2}+\frac{d_1 d_2 d_3}{|\mathfrak{D}_1|}
\sum_{i=1}^{\lceil n/2 \rceil}\bigl(Y_i-\langle\widehat{\tT}_{\mathrm{init},2},\tX_i\rangle\bigr)\tX_i.
\end{aligned}
\right.
\end{equation}
\State \textit{Retraction (Projection)}
\Statex \textbf{for} {$a=1,2$} \textbf{do}
  \Statex Perform a t-SVD on $\widehat{\tT}_{\mathrm{unbs},a}$, keeping only the leading $r$ components to obtain $\big(\widehat{\tU}_a, \widehat{\tS}_a, \widehat{\tV}_a\big)$.
  \Statex Form the projected estimator: $\widehat{\tT}_{\mathrm{proj},a} \gets \widehat{\tU}_a \widehat{\tS}_a \widehat{\tV}_a^{\dagger}$.
\Statex \textbf{end for}

\State \textit{Averaging}
\Statex Form the final estimator $\widehat{\tT} \gets \big(\widehat{\tT}_{\mathrm{proj},1}+\widehat{\tT}_{\mathrm{proj},2}\big)/{2}$.
\Statex Compute the estimated linear form $ \langle \widehat{\tT},\tM\rangle$.
\end{algorithmic}
\end{algorithm}
\paragraph{Debiasing and implication for inference}
Let $d^*:=d_1d_2d_3$, and let $\widehat{\tT}_{\mathrm{init}}$ be an initial consistent estimator of $\tT$ (e.g., obtained via Riemannian optimization~\cite{song2023riemannian}), which is assumed to satisfy \cref{assump:init_entry}. In general, the exact distribution of $\widehat{\tT}_{\mathrm{init}}$ is intractable due to the complexity of nonconvex solvers~\cite{song2023riemannian} and the structure of nuclear norm regularization~\cite{zhang2016exact,lu2019low}. We adopt a one-step debiasing scheme and define
\begin{equation}\label{eq:errorDe}
\widehat{\tT}_{\mathrm{unbs}}
:= \widehat{\tT}_{\mathrm{init}}
+ \frac{d^*}{n}\sum_{i=1}^n
\bigl(Y_i - \langle \widehat{\tT}_{\mathrm{init}},\tX_i\rangle\bigr)\tX_i.
\end{equation}
One-step corrections of this form have been successfully used in low-rank matrix and tensor settings \cite{xia2021statistical, ma2024statistical, duan2025online}.
Plugging the sampling model $Y_i=\langle\tT,\tX_i\rangle+\xi_i$ into \cref{eq:errorDe} yields
\begin{align}
    \widehat{\tT}_{\mathrm{unbs}}= \tT + 
    \underbrace{\widehat{\tT}_{\mathrm{init}}- \tT- \frac{d^*}{n}\sum_{i=1}^n\langle\widehat{\tT}_{\mathrm{init}}-\tT,\tX_i\rangle \tX_i}_{\tE_{\mathrm{init}}} + 
   \underbrace{\frac{d^*}{n}\sum_{i=1}^n \xi_i \tX_i}_{\tE_{\mathrm{rn}}}.\label{eq:error-decom}
\end{align}
The term $\tE_{\mathrm{init}}$ collects the remainder arising from the initialization error, whereas $\tE_{\mathrm{rn}}$ captures the observational noise and can be expressed as a sum of independent mean-zero random tensors. Hence, for any fixed $\tM$, the estimation error of using $\langle\widehat{\tT}_{\mathrm{unbs}},\tM\rangle$ for $\langle\tT,\tM\rangle$ decomposes into two parts,
\begin{equation*}
    \langle \widehat{\tT}_{\mathrm{unbs}}-\tT,\tM\rangle=
\langle \tE_{\mathrm{rn}},\tM\rangle+\langle \tE_{\mathrm{init}},\tM\rangle .
\end{equation*}
This decomposition facilitates a precise distributional analysis of the estimator $\langle \widehat{\tT}_{\mathrm{unbs}},\tM\rangle$. In particular, the term $\langle \tE_{\mathrm{rn}},\tM\rangle$ is responsible for the Gaussian approximation, and $\langle \tE_{\mathrm{init}},\tM\rangle$ is controlled separately and enters the final result as a remainder term; as detailed in \cref{thm:norm} and its proof.

This observation motivates our theoretical analysis: we establish high-probability control of the remainder term and then derive the distributional guarantee of the debiased linear functional in \cref{thm:norm}.

\paragraph{Sample–splitting debiasing and retraction}
To facilitate theoretical control of $\tE_{\mathrm{init}}$ and to characterize the distribution of the leading noise term, we implement \cref{eq:errorDe} using sample splitting and cross-fitting; see \Cref{alg:debias}. Concretely, we split the sample into two disjoint subsets $\mathfrak{D}_1$ and $\mathfrak{D}_2$, compute $\widehat{\tT}_{\mathrm{init},1}$ on $\mathfrak{D}_1$ and $\widehat{\tT}_{\mathrm{init},2}$ on $\mathfrak{D}_2$, and form the debiased estimators by applying the one-step correction on the held-out half as shown in \cref{alg-debias}.
{By construction, conditional on $\widehat{\tT}_{\mathrm{init},a}$ ($a=1,2$), the corresponding correction term formed from the other subsample is a sum of independent random tensors. This independence is convenient for applying concentration inequalities to control the remainder terms.}
Regarding the implementation details of the sample-splitting in \Cref{alg:debias}, while we employ an even two-fold split for notation simplicity, general $K$-fold cross-fitting
strategies~\cite{chernozhukov2018double} are also applicable, provided the subsamples are sufficiently large to maintain the required independence
structure and to ensure that the initialization bound (see \Cref{assump:init_entry}) holds on each fold.

The one-step correction in \cref{alg-debias} adds an empirical score term, which is a sum of rank-one design tensors supported on the observed entries, so $\widehat{\tT}_{\mathrm{unbs},a}$ $(a=1,2)$ is generally no longer rank-$r$ even when $\widehat{\tT}_{\mathrm{init},a}$ is rank-$r$. As a result, $\widehat{\tT}_{\mathrm{unbs},a}$ can contain high-rank noise, which may inflate the variance of downstream plug-in estimates $\langle \widehat{\tT},\tM\rangle$ if used directly. We thus retract each debiased tensor to be of tubal-rank-$r$ via truncated t-SVD, producing $\widehat{\tT}_{\mathrm{proj},a}$. In \cref{sub:simu}, simulation results confirm that the proposed retraction reduces the variance of the estimator for $\langle \widehat{\tT},\tM\rangle$ relative to using $\widehat{\tT}_{\mathrm{unbs},a}$ directly.
Finally, we average the two retracted estimators to obtain the final tensor estimator $\widehat{\tT}$ and report the plug-in estimate $\langle \widehat{\tT},\tM\rangle$.

We note that the steps after initialization, including debiasing, projection, and averaging, are non-iterative and computationally efficient
compared to the initial solver (which typically relies on iterative algorithms such as Riemannian optimization~\cite{song2023riemannian}); thus, the proposed inference framework adds negligible computational overhead to the original estimation procedure. 

\subsection{Asymptotic normality of the estimator}
\label{subsec:Asymptotic Normality}
In this section, we show that the estimator $\langle\widehat{\mathcal{T}},\mathcal{M}\rangle$ obtained from \cref{alg:debias} admits a Gaussian approximation and therefore can be used to derive confidence intervals for the true parameter $\langle{\mathcal{T}},\mathcal{M}\rangle$.  

We now state the following regularity conditions required to establish the Gaussian approximation. First, in \cref{assump:SNR}, we impose a lower bound on the signal magnitude to ensure it is distinguishable from the noise. 
To this end, let $\lambda_{\min}$ and $\lambda_{\max}$ denote the smallest and largest non-zero singular values of $\operatorname{bdiag}(\ftT)$, respectively, where $\ftT$ is the DFT of $\tT$.
\begin{assump}[signal-to-noise ratio]\label{assump:SNR}
There exists an absolute constant $C_{\mathrm{SNR}}>0$ such that
\begin{equation*}
    \frac{\lambda_{\min}}{\sigma_\xi}\ \ge\ C_{\mathrm{SNR}}\, \frac{d_3(d_1\vee d_2)^{3/2}}{\sqrt{n}}\,
    \sqrt{\log\!\big(d_1d_3+d_2d_3\big)}.
\end{equation*}
\end{assump}
\Cref{assump:SNR} specifies a lower bound on the signal-to-noise ratio, defined via SNR:$=\lambda_{\min}/\sigma_{\xi}$. 
To interpret this scaling, consider the full observation 
where $n =O( d_1 d_2 d_3)$. In this scenario, the noise tensor has a spectral norm of order $\sigma_\xi \sqrt{d_3(d_1 \vee d_2)}$ with high probability \cite{vershynin2018hdp,wainwright2019hds}. The right-hand side of the inequality simplifies to this order up to a logarithmic factor, indicating that the requirement of \cref{assump:SNR} is achievable. Intuitively, the condition requires that the intrinsic signal strength must dominate the spectral norm of the noise to ensure accurate recovery and valid inferences.

Next, we impose an entrywise accuracy condition on the initial estimators. This condition ensures that the initialization error term $\tE_{\mathrm{init}}$ in the debiasing decomposition \cref{eq:error-decom} admits an explicit bound and can therefore be incorporated into the subsequent analysis as a remainder term.
We define tensor max norm as $\|\tT\|_{\max}:= \max_{jkl}|\tT(j,k,l)|$.
\begin{assump}[{initialization}]\label{assump:init_entry}
There exists a sequence $\gamma_{n,d_1,d_2,d_3}\in (0,1]$ such that, with probability at least {$1-(d_1d_3+d_2d_3)^{-1}$},
\begin{equation*}
\|\tT_{\mathrm{init},1}-\tT\|_{\max}+\| \tT_{\mathrm{init},2}-\tT\|_{\max}\leq C_{\mathrm{init}}\gamma_{n,d_1,d_2,d_3}\cdot \sigma_{\xi}
\end{equation*}
for an absolute constant $C_{\mathrm{init}}>0$.
\end{assump}

For brevity, write $\gamma_n=\gamma_{n,d_1,d_2,d_3}$ hereafter. \Cref{assump:init_entry} requires the initial estimators $\tT_{\mathrm{init},1}$ and $\tT_{\mathrm{init},2}$ used in \cref{alg:debias} to achieve entrywise accuracy of order $O(\sigma_\xi\gamma_n)$ with high probability.
 Under this condition, the contribution of initialization to the error, represented by $\tE_{\mathrm{init}}$ in \cref{eq:error-decom}, can be quantified separately and enters the distributional bound in \cref{thm:norm} through an explicitly controlled remainder term. This separation isolates the noise term $\tE_{\mathrm{rn}}$, whose distributional behavior yields the Gaussian approximation.
 Related initialization assumptions have also been used in the matrix setting; see \cite{xia2021statistical}.

 While explicit entrywise bounds are standard in matrix and CP/Tucker completion literature \cite{chen2019inference, cai2022uncertainty, li2023online, wang2023implicit}, analogous theoretical results for the low-tubal-rank model are not yet fully established. However, this is primarily a gap in analysis rather than a fundamental limitation; such bounds are achievable in principle by adapting techniques like leave-one-out analysis \cite{chen2019inference, cai2022uncertainty, wang2023implicit} 
or online algorithms \cite{li2023online} to the t-product framework. Thus, we regard this as a mild requirement for the initial solver. 

To prevent the tensor from being highly concentrated in a few entries, which would make it unrecoverable via random sampling \cite{liu2019low,chen2019inference}, we impose the standard incoherence condition in \cref{assump:incoh}. Define $\te_{j_1}$ as the $d_1 \times 1 \times d_3$ column basis tensor whose only nonzero entry is $\te(j_1, 1,1)=1$, and define $\te_{j_2}$ as the $d_2 \times 1 \times d_3$ column basis tensor whose only nonzero entry is $\te(j_2, 1,1)=1$. Also recall the spectral structure: let $\mathcal{T} = \mathcal{U} \mathcal{S} \mathcal{V}^{\dagger}$ be the skinny t-SVD (\cref{def:tubrank}), where $\mathcal{U}, \mathcal{V}$ are orthogonal tensors (\cref{def:structure}). 
\begin{assump}[incoherence]\label{assump:incoh} 
There exists $\mu_{\max}>0$ so that,
\begin{equation*}
\max_{j_1\in[d_1]} \|\te_{j_1}^{\dagger} \, \tU \|  \leq \mu_{\max}\sqrt{\frac{r}{d_1}}, \quad \max_{{j_2}\in[d_2]} \|\te_{j_2}^{\dagger} \, \tV \|  \leq \mu_{\max}\sqrt{\frac{r}{d_2}},
\end{equation*}
\end{assump}
The incoherence parameter $\mu_{\max}$ ranges from a minimum of $1$, representing a perfectly incoherent state with uniformly distributed mass, up to a maximum of $\sqrt{\max\{d_1,d_2\}/r}$, representing a perfectly coherent state where mass is concentrated on a single fiber (proof provided in \cref{sec:ap_pre}). Consequently, for the derived statistical rates in our distributional guarantee (see, \cref{thm:norm}) to be sharp and for efficient recovery from limited samples, $\mu_{\max}$ is typically expected to be a small constant independent of the tensor dimensions. 

Finally, we require that the test tensor $\mathcal{M}$ has sufficient alignment with the singular subspaces of $\tT$. 
If the projection of $\mathcal{M}$ onto the singular subspaces of $\mathcal{T}$ is close to zero, the resulting estimator would have a degenerate variance, rendering inference impossible.
\begin{assump}[alignment]\label{assump:M}
    There exists $\alpha_{\tM}>0$ such that \begin{equation*}
         \|\tU^{\dagger}\tM\|_{\mathrm{F}}\geq \alpha_{\tM} \|\tM\|_{\mathrm{F}} \cdot \sqrt{\frac{r}{d_1}}, \quad
    \|\tV^{\dagger}\tM\|_{\mathrm{F}}\geq \alpha_{\tM} \|\tM\|_{\mathrm{F}}\cdot  \sqrt{\frac{r}{d_2}}.
    \end{equation*}
\end{assump}
This assumption is the tensor counterpart of the alignment condition used in matrix completion and inference \cite{chen2019inference,xia2021statistical}.
In \cref{sec:example}, we choose a specific $\tM$ based on entrywise inference to show that \cref{assump:M} is especially natural in common inference scenarios.

The following theorem provides a quantitative normal approximation for $\langle\widehat{\tT},\tM\rangle$ under \cref{assump:sampling,assump:noise,assump:SNR,assump:init_entry,assump:incoh,assump:M}. 
Recall that $\gamma_n$ is defined in \cref{assump:init_entry}, $\mu_{\max}$ is the incoherence parameter in \cref{assump:incoh}, and $\alpha_{\tM}$ is the alignment parameter in \cref{assump:M}.

\begin{theorem}[rate of convergence to normality]\label{thm:norm}
Suppose \cref{assump:sampling,assump:noise,assump:SNR,assump:init_entry,assump:incoh,assump:M} hold. There exist absolute constants \(C_1,\dots,C_5>0\) such that, when
\begin{equation*}
n>6\mu_{\max}^2rd_3(d_1\vee d_2)\log^2(d_1d_3 +d_2d_3),
\end{equation*}
we have
\begin{equation*}
\sup_{x\in\mathbb R}\Bigg|\,
\mathbb{P}\!\left(\frac{\langle\widehat{\tT},\tM\rangle-\langle\tT,\tM\rangle}
{\sigma_\xi (\|\tM\tV\|_{\mathrm{F}}^2+\|\tU^{\dagger}\tM\|_{\mathrm{F}}^2)\;\sqrt{d_1d_2d_3/n}}\le x\right) - \Phi(x)
\Bigg|
\;\le\; R.
\end{equation*}
Here, $\Phi$ denotes the cumulative distribution function of a standard Gaussian distribution. The total approximation error bound, $R$, decomposes into five terms (\(R=\sum_{j=1}^5 R_{j}\)) defined as follows:
\begin{align*}
&\begin{aligned}
R_1 &= \frac{8}{(d_1d_3+d_2d_3)^{2}}, & R_2 &= C_2\,\mu_{\max}\,\sqrt{\frac{r\,d_3\,(d_1\vee d_2)}{n}}, \\
R_3 &=  C_3\,\gamma_n\,\sqrt{\log(d_1d_3+d_2d_3)}, \qquad & R_4 &=C_4\,\frac{\mu_{\max}^4\,\|\tM\|^2_{\ell_1}}{\alpha_{\tM}^2\,\|\tM\|_{\mathrm{F}}^2}\,r\,\sqrt{\frac{\log(d_1d_3+d_2d_3)}{d_1\wedge d_2}},
\end{aligned} \\
&R_5 = C_5\,\frac{\mu_{\max}^2\,\|\tM\|_{\ell_1}}{\alpha_{\tM}\,\|\tM\|_{\mathrm{F}}}\,{\kappa_0}\,\frac{\sigma_\xi}{\lambda_{\min}}\,\frac{d_3^{3/2}(d_1\vee d_2)^{3/2}}{\sqrt{n}}\,\log(d_1d_3+d_2d_3),
\end{align*}where $\|\tM\|_{\ell_1}$ denotes the tensor $\ell_1$-norm, defined as the sum of the absolute values of all its elements; and $\kappa_0:=: \lambda_{\max}/\lambda_{\min}$ denotes the condition number of $\tT$.
\end{theorem}
The proof is given in \cref{proof:norm}.
\Cref{thm:norm} gives a non-asymptotic Gaussian approximation for the estimator $\langle\widehat{\tT}, \tM\rangle$. The decomposition $R=\sum_{j=1}^5 R_j$ separates the total approximation error into various sources.
In particular, $R_1$ represents the aggregate failure probability arising from our concentration inequalities used to control quantities such as the noise tensor.
The term $R_2$ is derived explicitly from the Berry-Esseen bound \cite{stein1972bound} for the projected random noise error term $\tE_{\mathrm{rn}}$, which is a {sum of independent random variables as shown in \cref{eq:error-decom}}. It scales with the canonical $n^{-1/2}$ rate and corresponds to the unavoidable stochastic error. 
The term $R_3$ quantifies the algorithmic bias of the initialization, confirming that the validity of the debiasing procedure remains coupled to the convergence quality ($\gamma_n$) of the initial estimator. 

Beyond these baseline terms, $R_4$ and $R_5$ arise from higher-order spectral perturbations, which measure the error incurred when projecting the data onto estimated singular subspaces. Notably, $R_5$ depends explicitly on the inverse SNR ($\sigma_\xi / \lambda_{\min}$), highlighting that subspace stability is governed by signal strength; namely, weaker signals amplify these curvature effects, making the singular subspaces more sensitive to noise.
In the high-dimensional regime, the total inference error is dominated by the interplay between the statistical fluctuation ($R_2$) and the SNR-dependent curvature cost ($R_5$), while the remaining terms vanish asymptotically.

The next corollary records the sufficient high-dimensional scaling conditions required for $R \to 0$, which allow the standardized linear functional to converge in distribution to $\mathcal{N}(0,1)$.
\begin{corollary}[asymptotic normality]\label{cor}
Assume the conditions of \cref{thm:norm}. Suppose that $\mu_{\max}, \alpha^{-1}_{\tM}, \kappa_0$ are uniformly bounded by absolute constants and that
\begin{equation}\label{eq:rate1}
    \frac{\|\tM\|^2_{\ell_1}}{\|\tM\|_{\mathrm{F}}^2}
    \cdot 
    \max \left\{\frac{r\sqrt{\log(d_1d_3+d_2d_3)}}{d_1\wedge d_2},\;\frac{\sigma_{\xi}}{\lambda_{\min}}\,\frac{d_3^{3/2}(d_1\wedge d_2)^{3/2}}{\sqrt{n}}\,\log(d_1d_3+d_2d_3)\right\}\;\longrightarrow\; 0,
\end{equation}
and
\begin{equation}\label{eq:rate2}
    \gamma_n \sqrt{\log (d_1d_3+d_2d_3)} \;\longrightarrow\; 0,
\end{equation}
as $d_1, d_2, n \to \infty$. Then the standardized linear form converges in distribution to the standard Gaussian, i.e.,\begin{displaymath}  \frac{\langle\widehat{\tT},\tM\rangle-\langle{\tT},\tM\rangle}
    {\sigma_{\xi}\left(\|\tM \tV\|_{\mathrm{F}}^2+\|\tU^{\dagger} \tM\|_{\mathrm{F}}^2\right)^{1 / 2} \cdot \sqrt{d_1 d_2 d_3 / n}}
    \;\toD\; \mathcal{N}(0,1),
\end{displaymath}
as $d_1$, $d_2$, and $n\to \infty$.
\end{corollary}

With $d_1,d_2\to\infty$ and the consistent initialization assumption in \cref{assump:init_entry}, all terms in \cref{eq:rate1,eq:rate2} vanish automatically except for the signal-to-noise (SNR) factor\begin{equation*} \frac{\sigma_{\xi}}{\lambda_{\min}}\,
    \frac{d_3^{3/2}(d_1\wedge d_2)^{3/2}}{\sqrt{n}}\,
    \log(d_1d_3+d_2d_3)
    \;\longrightarrow\; 0.
\end{equation*}
This condition strengthens the SNR requirement from \cref{assump:SNR}, making it more restrictive by an additional factor of order $\sqrt{d_3\log(d_1d_3+d_2d_3)}$. This increased signal separation is necessary to achieve asymptotically normal inference.

In the special case $d_3 = 1$, the low-tubal-rank tensor model reduces to the classical low-rank matrix model, and \cref{thm:norm}, together with \cref{cor}, recovers the main results of \cite{xia2021statistical}.

A simple and important specialization of \cref{cor} is entrywise inference, where the linear form extracts a single entry of the tensor. The following example shows how the alignment condition and the asymptotic normality statement simplify in this setting.

\begin{example}[entrywise inference]\label{sec:example}
    In entrywise inference, we set $\tM$ to be the coordinate mask supported on a single entry $(i_0,j_0,k_0)$, denoted by $\tM=\te_{i_0}\tte_{k_0}\te_{j_0}^\dagger$ with the notations defined in \cref{subsec:tubal-rank}. Then \cref{assump:M} reduces to\begin{equation}\label{example1-al}
    \|\tU^{\dagger}\te_{i_0}\|_{\mathrm{F}}\geq \alpha_{\tM}\sqrt{\frac{r}{d_1}}\quad\text{and}\quad \|\tV^{\dagger}\te_{j_0}\|_{\mathrm{F}}\geq \alpha_{\tM}\sqrt{\frac{r}{d_2}}.
\end{equation}That is, the singular vectors of $\tT$ must place non-negligible mass on the entry of interest. \Cref{example1-al} also aligns with the incoherence condition, \cref{assump:incoh}, which requires that the mass of the singular vectors of $\tT$ is reasonably spread out over row indices. Moreover,
\begin{equation*}
    \frac{\widehat{\tT}(i_0,j_0,k_0)-\tT(i_0,j_0,k_0)}{\sigma_{\xi}\left(\|\te^{\dagger}_{i_0}\tU\|_{\mathrm{F}}^2+\|\te^{\dagger}_{j_0}\tV\|_{\mathrm{F}}^2\right)^{1 / 2} \cdot \sqrt{d_1 d_2 d_3 / n}} \toD \mathcal{N}(0,1),
\end{equation*} as $d_1,d_2,n\to \infty$, under the conditions \cref{eq:rate1,eq:rate2} with $\|\tM\|_{\ell_1}/\|\tM\|_{\mathrm{F}}=1$.
\end{example}

\subsection{Inferences about linear forms}
\label{subsec:Inferences about Linear Forms}
\Cref{thm:norm,cor} establishes asymptotic normality of the linear form $\langle \widehat{\tT}, \tM\rangle$. To compute uncertainty intervals, we still need a consistent estimator of its asymptotic variance. Recall that\begin{equation*}
    \operatorname{Var}\big(\langle \widehat{\tT}, \tM\rangle\big)\asymp\sigma_{\xi}^{2}\, s_{\tM}^{2}\cdot \frac{d_1 d_2 d_3}{n},\qquad
s_{\tM}^{2}:=\|\tM\tV\|_{\mathrm F}^{2}+\|\tU^{\dagger}\tM\|_{\mathrm F}^{2},
\end{equation*}
where $(\tU,\tV)$ are the singular factors of $\tT$ from the (skinny) t-SVD.

We estimate the noise variance $\sigma_{\xi}^2$ using the residual sum of squares computed from the observed entries,
\begin{equation}
    \label{eq:plugin_sigma}\widehat{\sigma}_{\xi}^2=\frac{1}{n} \sum_{i=n_0+1}^n\left(Y_i-\left\langle\widehat{\tT}_{\text {init }, 1}, \tX_i\right\rangle\right)^2+\frac{1}{n} \sum_{i=1}^{n_0}\left(Y_i-\left\langle\widehat{\tT}_{\text {init }, 2}, \tX_i\right\rangle\right)^2,
\end{equation}
and the scalar factor $s_{\tM}=\|\tM \tV\|_{\mathrm{F}}^2+\|\tU^{\dagger} \tM\|_{\mathrm{F}}^2$ by a plugged-in estimator
\begin{equation}
\label{eq:plugin_s}    \widehat{s}_{\tM}^2=\frac12\left(
\|\tM\widehat{\tV}_1\|_{\mathrm{F}}^2+\|\widehat{\tU}_1^{\dagger} \tM\|_{\mathrm{F}}^2+
\|\tM\widehat{\tV}_2\|_{\mathrm{F}}^2+\|\widehat{\tU}_2^{\dagger} \tM\|_{\mathrm{F}}^2\right).
\end{equation}
\Cref{thm:infer} shows that the asymptotic normality remains valid if we replace the variance of $\langle\widehat{\tT},\tM\rangle$ with the estimates given in \cref{eq:plugin_sigma,eq:plugin_s}.

\begin{theorem}[asymptotic normality with plug-in estimates]\label{thm:infer}
     Under \cref{assump:sampling,assump:noise,assump:SNR,assump:init_entry,assump:incoh,assump:M}, suppose that $\mu_{\max}, \alpha_{\tM}, \kappa_0$ are uniformly bounded, and that the asymptotic conditions \cref{eq:rate1} and \cref{eq:rate2} hold. Then, as $d_1, d_2$ and $n \to \infty$, we have\begin{displaymath}
         \frac{\langle\widehat{\tT},\tM\rangle-\langle\tT,\tM\rangle}{\widehat{\sigma}_{\xi} \widehat{s}_{\tM} \cdot \sqrt{d_1 d_2 d_3/ n}} \xrightarrow{\mathrm{~d}} \mathcal{N}(0,1).
     \end{displaymath}
\end{theorem}
\Cref{thm:infer} yields the confidence interval for $\langle\tT,\tM\rangle$, stated in \cref{cor:CI}.
\begin{corollary}[confidence interval]\label{cor:CI}
Assume that the conditions of \cref{thm:infer} hold. For any $\alpha\in(0,1)$, define\begin{equation}\label{CI1}
    \widehat{\mathrm{CI}}_{\alpha,\tM}
:=\Big[
\langle\widehat{\tT},\tM\rangle
- z_{1-\alpha/2}\,\widehat{\sigma}_{\xi}\widehat{s}_{\tM} \sqrt{d_1 d_2 d_3/n},\;
\langle\widehat{\tT},\tM\rangle
+ z_{1-\alpha/2}\,\widehat{\sigma}_{\xi}\widehat{s}_{\tM} \sqrt{d_1 d_2 d_3/n}
\Big],
\end{equation}
where $z_{1-\alpha/2}=\Phi^{-1}(1-\alpha/2)$ and $\Phi$ are the standard normal cdf. Then\begin{displaymath}
    \lim_{d_1,d_2,d_3,n\to\infty}
\Pr\!\left(\langle\tT,\tM\rangle\in \widehat{\mathrm{CI}}_{\alpha,\tM}\right)
=1-\alpha .
\end{displaymath}
\end{corollary}
In addition, if one is interested in a confidence interval for the noisy observation,\begin{equation*}
    Y_{\tM} = \langle \tT,\tM\rangle+\xi,
\end{equation*}where the noise $\xi$ is sub-Gaussian, as stated in \cref{assump:noise}, a conservative $(1-\alpha)$-level interval can be obtained by inflating the asymptotic standard error by the noise level $\sigma_{\xi}$:\begin{equation}
\label{CI2}\widehat{\mathrm{CI}}_{\text{ob},\alpha,\tM}:=
\Big[\langle\widehat{\tT},\tM\rangle
- z_{1-\frac{\alpha}{2}}\big(\widehat{\sigma}_{\xi}\widehat{s}_{\tM} \sqrt{\frac{d_1d_2d_3}{n}}+\widehat{\sigma}_{\xi}\big)\,, \,\langle\widehat{\tT},\tM\rangle
+z_{1-\frac{\alpha}{2}}\big(\widehat{\sigma}_{\xi}\widehat{s}_{\tM} \sqrt{\frac{d_1d_2d_3}{n}}+\widehat{\sigma}_{\xi}\big)
\Big].
\end{equation}
Analogous hypothesis tests for the linear form follow directly from \cref{thm:infer} and are omitted for the sake of simplicity in the statements.

\section{Numerical experiments}

In this section, we first evaluate the proposed sample splitting debiasing and retraction procedure in \cref{alg:debias} via Monte Carlo simulations. We then show that the resulting estimator $\widehat{\tT}$ exhibits negligible bias and low variance through numerical results. Finally, the simulations validate the asymptotic normality of the estimated linear form $\langle \widehat{\tT}, \tM \rangle$ proven by \cref{thm:infer}, and the associated confidence intervals achieve near-nominal coverage. We further illustrate the method on the TEC dataset.

\subsection{Simulations}
\label{sub:simu}
\begin{figure}[]
  \centering
  \subfloat[Ground truth $\tT$]{\includegraphics[width=.22\linewidth]{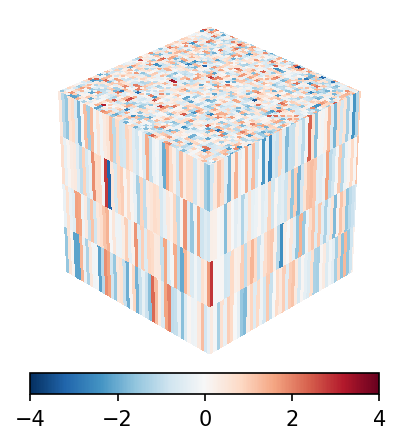}}
  \subfloat[Noise corrupted]{\includegraphics[width=.22\linewidth]{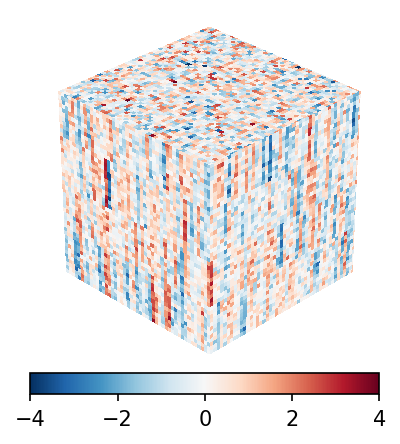}}
  \subfloat[Observed (40\%)]{\includegraphics[width=.22\linewidth]{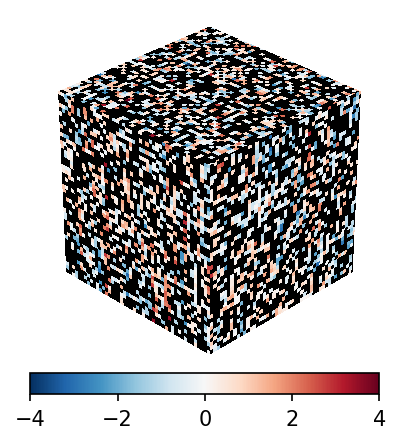}}
  \subfloat[Debiased estimation]{\includegraphics[width=.22\linewidth]{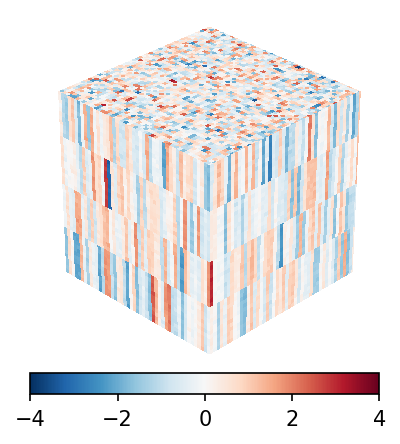}}
  \caption{\small Visualizations of the example tensor used in simulation. Panels show (a) the ground truth low-tubal-rank tensor $\tT\in\mathbb{R}^{500\times500\times500}$ with rank $r=4$; (b) $\tT$ corrupted by i.i.d. sub-Gaussian noise with standard deviation $\sigma_{\xi}=0.6$; (c) the observed tensor with 40\% of entries observed and missing entries marked in black; (d) the tensor imputed by the proposed procedure in \cref{alg:debias}.}
  \label{fig:v_tensors}
\end{figure}

We generate a fixed ground-truth tensor $\tT\in\mathbb{R}^{500\times500\times500}$ of tubal rank $r=4$ and use it throughout all simulation runs. {Specifically, $\tT$ is constructed in the Fourier domain by setting each frequency slice to $\fmT^{(t)} = \fmU \fmV_t^H$, where $\fmU\in\mathbb{R}^{500\times4}$ is a deterministic indicator matrix with four equally sized row groups, and $\fmV_t\in\mathbb{C}^{500\times4}$ is a frequency-specific loading matrix generated from Gaussian draws with mild phase and amplitude variation across $t$. After imposing conjugate symmetry across frequencies, an inverse FFT along the third mode yields a real-valued tensor that retains tubal rank $4$.} Because $\fmU$ contains only four distinct row patterns, each frontal slice of $\tT$ has four unique rows repeated across indices, making the low-rank structure visually transparent; panel (a) of \cref{fig:v_tensors} illustrates this construction. After an overall normalization, the entries of $\tT$ have a mean $9.31\times10^{-4}$ and a standard deviation $0.997$. The smallest nonzero singular value of $\tT$ is $\lambda_{\min}(\tT)=1325.16$ and the largest is $\lambda_{\min}(\tT)=6348.03$. This stylized design is adopted solely to facilitate visualization and comparison; it is not intended to represent the full class of low-tubal-rank tensors satisfying our assumptions, and the proposed methodology is not tailored to this specific pattern.

In each Monte Carlo replicate, we add i.i.d. Gaussian noise with standard deviation $\sigma_\xi=0.6$ to every entry and then mask $60\%$ of the entries uniformly at random (i.e., $40\%$ entries are observed). The signal-to-noise ratio in this setting is $\mathrm{SNR}=\lambda_{\min}/\sigma_\xi = 1325.16/0.6 \approx 2.21\times 10^3$.
In \cref{fig:v_tensors}, panels (b) and (c) display a noisy tensor and the corresponding observed tensor. We then apply the proposed debiasing and retraction procedure in \cref{alg:debias} to obtain $\widehat{\tT}$; an example appears in panel (d) in \cref{fig:v_tensors}. The initial estimator in the debiasing pipeline is computed via Riemannian Conjugate Gradient Descent (RCGD) \cite[Algorithm 2]{song2023riemannian}. 

\begin{figure}[tbhp]
    \centering
    \includegraphics[width=.9\linewidth]{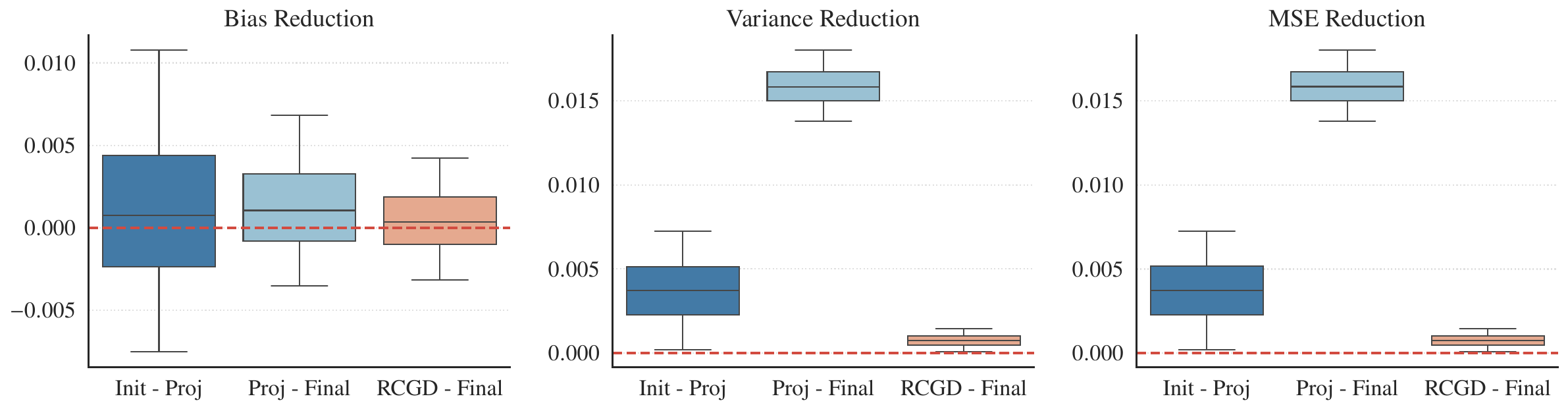}
    \caption{\small Pixel-wise performance gains of estimators across estimation stages: \texttt{Init} ($\tT_{\mathrm{init},a},\,a=1,2$), \texttt{Proj} ($\tT_{\mathrm{proj},a},\,a=1,2$), and \texttt{Final} ($\widehat\tT$), and comparison to {RCGD} \cite[Algorithm 2]{song2023riemannian}. Grouped boxplots display the pixel-wise reduction in empirical absolute Bias, Variance, and MSE evaluated at 100,000 randomly sampled tensor locations (1,000 simulations). Values above the red dashed line indicate an improvement. Boxplots of \texttt{Init}$-$\texttt{Proj} combine the projection steps from both initializations ($\tT_{\mathrm{init},a} - \tT_{\mathrm{proj},a},\,a=1,2$); \texttt{Proj}$-$\texttt{Final} shows the averaging to the final estimator; and \texttt{RCGD}$-$\texttt{Final} displays the overall gain over the baseline. Whiskers denote the 5th and 95th percentiles. Outliers are omitted.}
    \label{fig:boxplot}
\end{figure}

We evaluate the sample–splitting debiasing and retraction procedure in \cref{alg:debias} by comparing six estimators: the initial RCGD estimators $\tT_{\mathrm{init},1}$ (fitted on $\mathfrak D_1$) and $\tT_{\mathrm{init},2}$ (fitted on $\mathfrak D_2$), the corresponding debiased-and-projected estimators $\tT_{\mathrm{proj},1}$ and $\tT_{\mathrm{proj},2}$, the final averaged estimator $\widehat{\tT}$, and the baseline RCGD estimator fitted on the full data without debiasing. \Cref{fig:boxplot} reports, at 100,000 randomly selected tensor locations, the stage-wise reductions in empirical absolute bias, variance, and mean squared error (MSE). Positive values indicate that the second estimator in the comparison has a smaller error than the first.
The \texttt{Init}$-$\texttt{Proj} boxplots aggregate the entrywise gains from the two comparisons $(\tT_{\mathrm{init},a},\tT_{\mathrm{proj},a})$ $a\in\{1,2\}$, and therefore reflect the combined effect of cross-fitted debiasing and rank-$r$ retraction.
The \texttt{Proj}$-$\texttt{Final} boxplots aggregate the gains from $(\tT_{\mathrm{proj},1},\widehat\tT)$ and $(\tT_{\mathrm{proj},2},\widehat\tT)$, thereby capturing the contribution of the final averaging step. The \texttt{RCGD}$-$\texttt{Final} boxplots compare the final estimator from our procedure and that from the RCGD algorithm \cite[Algorithm 2]{song2023riemannian}. 

For bias reduction (left panel in \cref{fig:boxplot}), the medians of all three comparison groups lie above zero, indicating that, for more than half of the sampled pixels, the procedure reduces bias and that the final estimator improves upon the baseline RCGD.
The variance and MSE reductions are more pronounced, as given in the middle and right panels in \cref{fig:boxplot}: for all three comparisons, the medians are positive, and the interquartile ranges lie above zero, with the largest gains occurring in the \texttt{Proj}$-$\texttt{Final} step, where we average the two cross-fitted projected estimators. The close similarity between the variance and MSE panels further suggests that the overall MSE improvement is primarily driven by variance reduction rather than by large, uniform reductions in entrywise bias. Overall, \cref{fig:boxplot} shows that the transition from the initial estimators to the debiased-and-projected estimators, together with the final averaging step, yields a clear gain in estimation accuracy in the proposed procedure.

\begin{table}[]
\footnotesize
\centering
\setlength{\tabcolsep}{3pt}
\begin{tabular}{lccccccccccccc}
\toprule
& $\tT$
& \multicolumn{3}{c}{$\langle\tT,\tM^{(1)}\rangle$}
& \multicolumn{3}{c}{$\langle\tT,\tM^{(2)}\rangle$}
& \multicolumn{3}{c}{$\langle\tT,\tM^{(3)}\rangle$}
& \multicolumn{3}{c}{$\langle\tT,\tM^{(4)}\rangle$}
\\
\cmidrule(lr){2-2}\cmidrule(lr){3-5}\cmidrule(lr){6-8}\cmidrule(lr){9-11}\cmidrule(lr){12-14}
Method
& RMSE
& Bias & SD & MSE
& Bias & SD & MSE
& Bias & SD & MSE
& Bias & SD & MSE
\\
\midrule
Init-1   & {.1883} & .0149 & .1807 & .0329 & -.0132 & .1875 & .0353 & .0175 & .2580 & .0668 & .0100 & .3188 & .1017 \\
Init-2   & {.1883} & .0130 & .1923 & .0371  & -.0096 & .1813 & .0330  & .0095 & .2626 &  .0690 & .0157 & .3287 &  .1083 \\
Proj-1 & {.1781} & .0102 & .1775 & .0316  & \textbf{-.0066} & .1728 &  .0299 & \textbf{.0048} & .2499 & .0625  & .0118 & .3024 &  .0916 \\
Proj-2 & {.1781} & \textbf{.0030} & .1726 & .0298  & -.0070 & .1775 & .0315  & .0063 & .2476 & .0613  & \textbf{-.0028} & .3010 &  .0906 \\
Final    & \textbf{.1259}& .0066 & \textbf{.1219} & \textbf{.0149 }  & {-.0068} & \textbf{.1285} &  \textbf{.0165} & .0056 & \textbf{.1769} & \textbf{.0313} & .0045 & \textbf{.2158} &  \textbf{.0466} \\
RCGD & .1289& .0099& .1244& .0156 & -.0101 & .1318 & .0175 & .0095 & .1814 & .0330 & .0090 & .2224 & .0496
\\
\bottomrule
\end{tabular}
\caption{\small Comparison of tensor estimators. The rows correspond to the estimators under consideration: \texttt{Init-1} denotes $\tT_{\mathrm{init},1}$; \texttt{Init-2} denotes $\tT_{\mathrm{init},2}$; \texttt{Proj-1} and \texttt{Proj-2} denote $\tT_{\mathrm{proj},1}$ and $\tT_{\mathrm{proj},2}$, respectively; \texttt{Final} denotes $\widehat{\tT}$ produced by \cref{alg:debias}; and \texttt{RCGD} denotes the estimator fitted using the full data set without debiasing. The first column reports the root mean squared error, defined as $\mathrm{RMSE}=\sqrt{\mathbb E\|\widehat \tT-\tT\|_{\mathrm{F}}^2/(d_1d_2d_3)}$. For each linear form $\langle \tT,\tM^{(k)}\rangle$, where $k=1,\ldots,4$, the table reports the bias, defined as $\mathbb E\langle \widehat \tT-\tT,\tM^{(k)}\rangle$; the standard deviation, defined as $\sqrt{\mathrm{Var}\langle \widehat \tT,\tM^{(k)}\rangle}$; and the mean squared error, defined as the sum of the squared bias and the variance. Boldface indicates the smallest absolute value within each column. All entries are averages computed from 1000 independent Monte Carlo trials. The simulation parameters are $d_1=d_2=d_3=500$, $r=4$, $\sigma_\xi=0.6$, and $n=0.4\,d_1d_2d_3$.}
\label{tab:compare}
\end{table}

For the following results, we consider four illustrative linear forms $\langle \tT,\tM^{(k)}\rangle$, $k=1,2,3,4$. The masks are defined as follows.
$\tM^{(1)}$ has a single nonzero entry $\tM^{(1)}(1,1,1)=1$;
$\tM^{(2)}$ has a single nonzero entry $\tM^{(2)}(201,201,201)=1$;
$\tM^{(3)}$ has two ones at $(1,1,1)$ and $(201,201,1)$; and
$\tM^{(4)}$ has three ones at $(1,1,1)$, $(1,1,2)$, and $(1,1,3)$.
All other entries of $\tM^{(k)}$ are zero, so in each case $\langle\tT,\tM^{(k)}\rangle$ equals the sum of the indicated entries of $\tT$, respectively. These choices cover single-entry inference at two distant locations and slices, a two-entry cross-location sum within one slice, and a short within-tube sum across three slices. If the tensor is interpreted as spatial-temporal, they correspond to: one location at time 1, one location at time 201, two locations at a common time, and one location aggregated over three consecutive times.

\begin{figure}[htbp]
    \centering    \includegraphics[width=.7\linewidth]{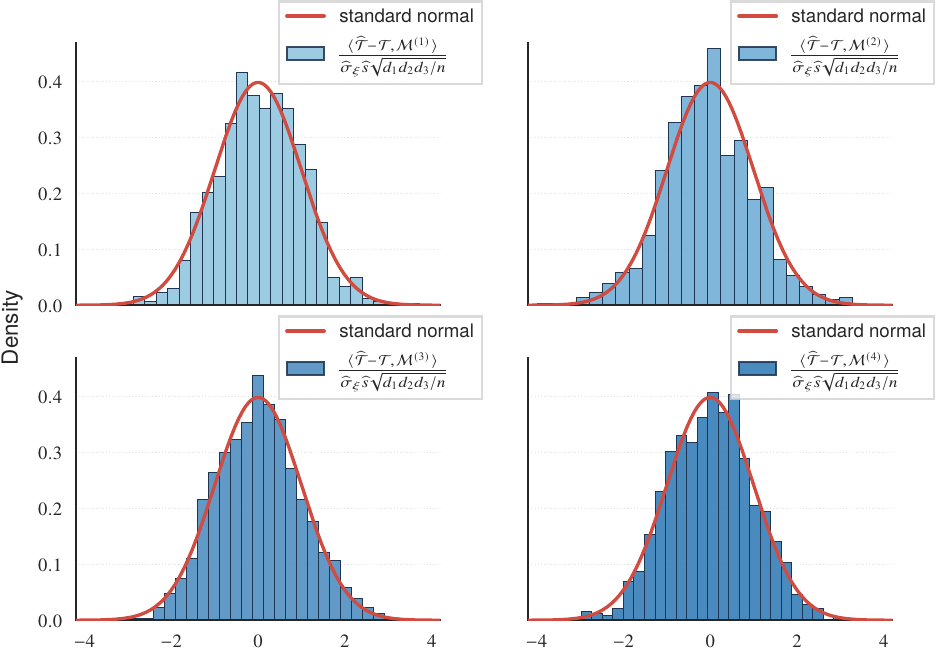}
    \caption{\small Normal approximation of $\frac{\langle\widehat{\tT},\tM\rangle-\langle{\tT},\tM\rangle}{\widehat{\sigma}_{\xi}\widehat{s}\sqrt{d_1d_2d_3/n}}$ with $\tM=\tM^{(1)}$ (upper left), $\tM^{(2)}$ (upper right), $\tM^{(3)}$ (bottom left), $\tM^{(4)}$ (bottom right). Each panel shows a density histogram of the standardized estimated statistics from 1000 independent simulations, and the red curve represents the p.d.f. of standard normal distributions. The simulation parameters are $d_1=d_2=d_3=500$, $r=4$, $\sigma_\xi=0.6$, and $n=0.4\,d_1d_2d_3$.}
    \label{fig:hist}
\end{figure}

We use these functionals to assess the sample-splitting debiasing and retraction procedure in \cref{alg:debias}. \Cref{tab:compare} reports the empirical mean RMSE and, for four linear forms $\langle\tT,\tM^{(k)}\rangle$ ($k=1,2,3,4$), the bias, standard deviation (SD), and mean-squared error (MSE) averaged over 1000 independent Monte Carlo trials. The debiased estimators reduce bias relative to their initializations. The subsequent projection step further stabilizes the estimates, leading to the final estimator with the smallest tensor RMSE and the lowest or near-lowest MSE across all four linear forms, outperforming the vanilla RCGD estimator.

We next assess the normal approximation in \cref{thm:infer} for the standardized linear form \begin{equation}\label{eq:hist_stats}
    \frac{\langle\widehat{\tT},\tM\rangle-\langle{\tT},\tM\rangle}{\widehat{\sigma}_{\xi}\widehat{s}_{\tM}\sqrt{d_1d_2d_3/n}}.
\end{equation}We evaluate this statistic for the same four masks $\{\tM^{(k)}\}_{i=1}^4$ introduced above.
We run $1000$ Monte Carlo replicates, plot the histogram of the statistic in \cref{eq:hist_stats} for each $\tM^{(k)}$, and compare it with the standard normal density. The results are shown in \cref{fig:hist}, where panel order matches the index: top left $k=1$; top right $k=2$; bottom left $k=3$; and bottom right $k=4$.
The histograms are centered near zero with variance close to one and align closely with the standard normal curve, providing empirical support for \cref{thm:infer}.

\begin{table}[tbhp]
\centering
\small
\begin{tabular}{lccccc}
\toprule
Estimator & CI width($\langle\tT,\tM\rangle$) & Coverage($\langle\tT,\tM\rangle$) & CI width($Y_{\tM}$) & Coverage($Y_{\tM}$) \\
\midrule
$\langle \widehat{\tT},\tM^{(1)}\rangle$ &0.4993$\pm$ 0.0007  &0.960&2.5169$\pm$ 0.0003 & 0.965 \\
$\langle \widehat{\tT},\tM^{(2)}\rangle$ &0.4857$\pm$0.0007 & 0.935& 2.5142$\pm$ 0.0003& 0.964 \\
$\langle \widehat{\tT},\tM^{(3)}\rangle$ & 0.6965$\pm$ 0.0007 & 0.956&3.5575$\pm$ 0.0004&  0.955\\
$\langle \widehat{\tT},\tM^{(4)}\rangle$& 0.8717$\pm$ 0.0018& 0.962&4.3608$\pm$ 0.0006 & 0.969 \\
\bottomrule
\end{tabular}
\caption{\small Empirical 95\% confidence interval (CI) performance over 1000 Monte Carlo replications. The columns “CI width($\langle\tT,\tM\rangle$)” and “Coverage($\langle\tT,\tM\rangle$)” report, respectively, the average interval width and empirical coverage of the confidence intervals in~\cref{CI1} for the target $\langle\tT,\tM^{(k)}\rangle$. The columns “CI width($Y_{\tM}$)” and “Coverage($Y_{\tM}$)” report the corresponding quantities for the noisy linear functional $Y_{\tM^{(k)}}=\langle\tT,\tM^{(k)}\rangle+\xi$ based on the intervals in~\cref{CI2}. The simulation parameters are $d_1=d_2=d_3=500$, $r=4$, $\sigma_\xi=0.6$, and $n=0.4\,d_1d_2d_3$.
}
\label{tab:CI}
\end{table}

\Cref{tab:CI} summarizes the finite–sample performance of the 95\% confidence intervals in \cref{CI1} and \cref{CI2} for four linear forms $\langle\tT,\tM^{(k)}\rangle$, $k=1,2,3,4$. For each mask $\tM^{(k)}$, the columns “CI width($\langle\tT,\tM\rangle$)” and “CI width($Y_{\tM}$)” report the Monte Carlo mean $\pm$ and the Monte Carlo standard error of the interval length over 1000 replications, while the “Coverage” columns report the corresponding empirical coverage probabilities.

For all four masks, the intervals for $\langle\tT,\tM^{(k)}\rangle$ have empirical coverage between $0.935$ and $0.962$. Thus, the procedures based on \cref{CI1} achieve coverage close to the nominal level $0.95$. 
The intervals for the noisy linear functionals $Y_{\tM^{(k)}}$ constructed from \cref{CI2} are substantially wider, as expected from the additional noise component. Nevertheless, the empirical coverage probabilities remain close to $0.95$ (between $0.955$ and $0.969$) for all four masks. Overall, these findings indicate that the proposed confidence intervals exhibit near-nominal coverage with reasonable lengths for both the latent signal $\langle\tT,\tM\rangle$ and the noise–corrupted quantity $Y_{\tM}$.

\subsection{Application on Total Electronic Content(TEC) data}
\label{subsec:TEC}

We apply our method to the task of reconstructing global TEC maps introduced in \cref{sec:introduction}. The data are taken from the open-access VISTA TEC database \cite{sun2023complete,sunData}, which is fully imputed using the VISTA algorithm \cite{sun2022matrix}. The spatial–temporal resolution is $1^{\circ}\text{ (latitude)}\times 1^{\circ}\text{ (longitude)}\times 5\text{ minutes}$. In the experiment, we focus on the 24-hour storm of September 8, 2017. The TEC video over this window forms a third-order tensor of dimension $181\times 361\times 288$. For each map, the second-dimension coordinates are converted from longitudes to local time (LT), and the columns are circularly shifted so that noon (12 LT) is centered. The resulting VISTA TEC tensor serves as the ground truth in our study.

To evaluate our method, we randomly mask $60\%$ of the entries in the ground-truth tensor, independently removing each pixel with a probability of $0.6$. This missing rate is higher than that typically observed in real TEC maps \cite{sun2022matrix}. Throughout this experiment, we impose a tubal-rank constraint of $5$.

\begin{figure}[]
    \centering    \includegraphics[width=0.8\linewidth]{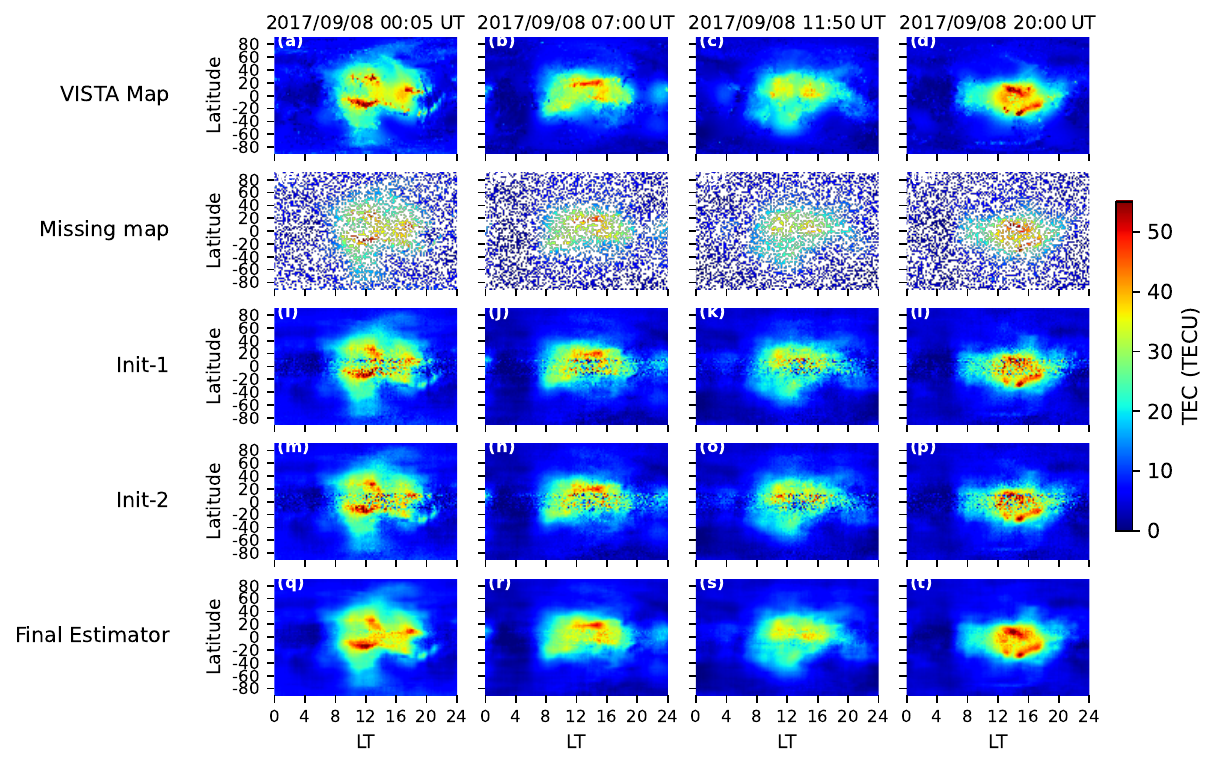}
    \caption{\small TEC reconstruction for the 24-hour storm of 2 September 2017. Panels (a)–(d) show the VISTA TEC maps (ground truth) at four representative times (00:05, 07:00, 11:50, and 20:00 UT), displayed as latitude versus local time (LT) with 12 LT centered. Panels (e)–(h) show the corresponding masked data, where each pixel is removed independently with probability 0.6. Panels (i)–(l) and (m)–(p) display two initial low-tubal-rank reconstructions (Init-1 and Init-2) based on the masked tensor. Panels (q)–(t) show the final reconstruction obtained by our proposed method (\cref{alg:debias}). Color represents TEC in TEC units (TECU).}
    \label{fig:TEC1}
\end{figure}

\begin{figure}
    \centering
    \includegraphics[width=0.8\linewidth]{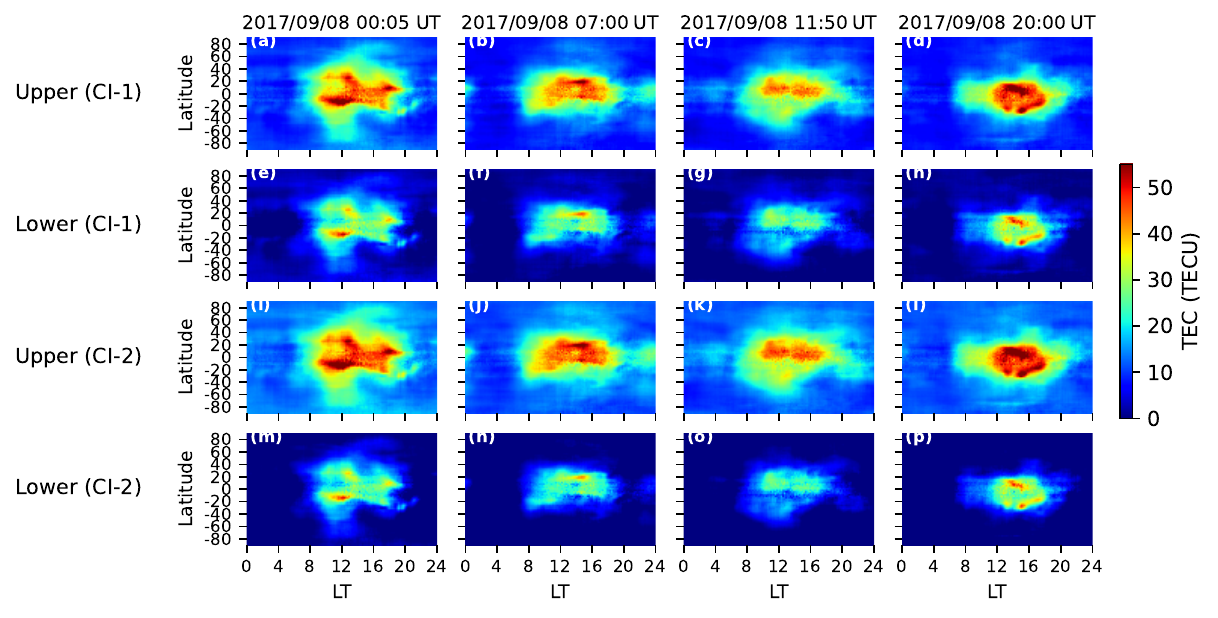}
\caption{\small Pointwise upper and lower confidence bounds (with significance level 95\%) for storm-time TEC on 8 September 2017 at four representative times (00:05, 07:00, 11:50, and 20:00 UT). Panels (a)–(d) and (e)–(h) show, respectively, the upper and lower bounds of the pointwise $(1-\alpha)$–confidence intervals targeting the latent TEC map, constructed from \cref{cor:CI}. Panels (i)–(l) and (m)–(p) display the corresponding upper and lower bounds for confidence intervals targeting the noisy observations, obtained by inflating the variance with the additional noise component. Latitude is plotted against local time (LT), with 12 LT centered, and color indicates the TEC value in TEC units (TECU).}
    \label{fig:TEC2}
\end{figure}

\Cref{fig:TEC1} illustrates the reconstruction performance of our method for the September 8, 2017 geomagnetic storm at four representative time points. The first row displays the VISTA TEC maps (ground truth), revealing the underlying large-scale spatial structure and its evolution over time. The second row shows the corresponding masked maps with 60\% missingness. 
The third and fourth rows present two initializations (Init-1 and Init-2) obtained by applying RCGD \cite{song2023riemannian} to two subsets of the `observed' pixels in the masked tensor under the tubal-rank-5 constraint. Both initializations recover roughly the large-scale patterns but contain visible artifacts, likely arising from enforcing an exact low-rank structure. 
In contrast, the bottom row shows the reconstruction from our proposed method (\cref{alg:debias}). Visually, it removes most artifacts from the initializations and aligns much more closely with the ground truth. It preserves the overall morphology of the storm-time TEC distribution and restores the intensity of high-TEC regions. These comparisons suggest that, under an aggressive 60\% missing-at-random scenario, the combination of a low-tubal-rank model with our debiasing procedure yields a substantial improvement over naive low-rank fits and provides an accurate representation of the underlying TEC map.

\Cref{fig:TEC2} summarizes the spatial uncertainty quantification provided by our procedure. For each of the four representative time points, the first pair of panels shows the pointwise 95\% confidence bounds for the latent TEC map, constructed from \cref{CI1}. The corresponding pair of panels displays the bounds of the confidence intervals targeting the noisy observations defined in~\cref{CI2}, where the variance has been inflated by the measurement–noise component. Notably, the overall lengths of the confidence intervals are not excessively large, indicating that our procedure yields informative uncertainty quantification without being overly conservative. In all cases, the upper and lower bounds inherit the large-scale spatial structure of the storm-time TEC distribution, indicating that the inferred uncertainty is concentrated in regions with high underlying TEC or stronger spatial variation. 

\section{Conclusion and Discussion}
\label{sec:conclusion}

This article develops a framework for statistical inference in low-tubal-rank tensor completion. Although t-product and tubal rank algebra have been widely used for estimation, statistical inference for this class of models has remained unexplored. The proposed sample–splitting debiasing and retraction procedure (see \cref{alg:debias}) converts any entrywise consistent initialization into an asymptotically normal estimator for general linear functionals of the tensor.

The core theoretical contribution is the establishment of asymptotic normality for linear forms of the estimated tensor (see \cref{thm:norm,cor}), which enables the construction of valid confidence intervals (see \cref{thm:infer,cor:CI}). Simulation studies demonstrate that the resulting confidence intervals achieve nominal coverage rates, thereby validating the theoretical derivations.
From a practical perspective, the proposed framework offers advantages for analyzing spatiotemporal data.
The application of Total Electron Content (TEC) data demonstrates the method's utility in geophysical contexts.
Future research should extend this framework beyond the current uniform sampling assumption to accommodate the non-uniform mechanisms frequently encountered in real-world data.

\appendix
\section{Appendix Preliminaries}
\label{sec:ap_pre}
Before presenting the proofs in the subsequent appendix sections, we collect the notation and basic facts that will be used repeatedly.

\paragraph{Fourier notation and frontal slices}
For a real-valued third-order tensor $\tA \in \mathbb{R}^{d_1 \times d_2 \times d_3}$, let the bold letter $\ftA = \mathcal{F}_3(\tA) \in \mathbb{C}^{d_1 \times d_2 \times d_3}$ denote its discrete Fourier transform (DFT) along the third mode. The $t$-th frontal slice of $\ftA$ is denoted by $\fmA^{(t)} \in \mathbb{C}^{d_1 \times d_2}$ for $t=1,\dots,d_3$.

\paragraph{Canonical basis tensors and tube extraction}
et $\te_j \in \mathbb{R}^{d \times 1 \times d_3}$ denotes the tensor whose $(j,1,1)$ entry is $1$ and whose remaining entries are $0$. Throughout the appendix, we use $\te_j$ with a slight abuse of notation: the first dimension $d$ may vary to match the dimensions of the tensor under consideration.

For $\tA \in \mathbb{R}^{d_1 \times d_2 \times d_3}$, the product
\begin{equation*}
    \te_{j_1}^{\dagger} \tA \te_{j_2} \in \mathbb{R}^{1 \times 1 \times d_3}
\end{equation*}
extracts the $(j_1,j_2)$ tube of $\tA$. Here, again by slight abuse of notation, $\te_{j_1} \in \mathbb{R}^{d_1 \times 1 \times d_3}$ and $\te_{j_2} \in \mathbb{R}^{d_2 \times 1 \times d_3}$.

\paragraph{A basic bound for an extracted tube}
We will repeatedly use the elementary bound
\begin{equation*}
    \|\te_{j_1}^{\dagger} \tA \te_{j_2}\|_{\max}
    \le
    \|\te_{j_1}^{\dagger} \tA \te_{j_2}\|,
\end{equation*}
for any $\tA \in \mathbb{R}^{d_1 \times d_2 \times d_3}$ and $j_1 \in [d_1]$, $j_2 \in [d_2]$. Indeed, write the entries of $\te_{j_1}^{\dagger} \tA \te_{j_2}$ as $(a_1,\dots,a_{d_3})$, and the entries of $\mathcal{F}_3(\te_{j_1}^{\dagger} \tA \te_{j_2})$ as $(\hat a_1,\dots,\hat a_{d_3})$. By the inverse DFT, for any $j \in [d_3]$,
\begin{equation*}
    |a_j|
    =
    \left|
    \frac{1}{d_3}\sum_{k=1}^{d_3} \hat a_k \omega^{-(j-1)(k-1)}
    \right|
    \le
    \frac{1}{d_3}\sum_{k=1}^{d_3} |\hat a_k|
    \le
    \max_{1 \le k \le d_3} |\hat a_k|,
    \qquad
    \omega = e^{-2\pi\sqrt{-1}/d_3}.
\end{equation*}
By the definition of tSVD in \cref{subsec:tubal-rank}, $\max_{1 \le k \le d_3} |\hat a_k|$ is the largest singular value of the $1 \times 1 \times d_3$ tube $\te_{j_1}^{\dagger} \tA \te_{j_2}$. Therefore,
\begin{equation*}
    \|\te_{j_1}^{\dagger} \tA \te_{j_2}\|_{\max}
    =
    \max_{1 \le j \le d_3} |a_j|
    \le
    \max_{1 \le k \le d_3} |\hat a_k|
    =
    \|\te_{j_1}^{\dagger} \tA \te_{j_2}\|.
\end{equation*}

\paragraph{Block-diagonal notation}
For any matrix $A \in \mathbb{R}^{d_1 \times d_2}$, we write
\begin{equation*}
    \diag_{d_3}(A) = \diag(A,\dots,A) \in \mathbb{R}^{d_1 d_3 \times d_2 d_3}
\end{equation*}
for the block-diagonal matrix obtained by repeating $A$ along the diagonal $d_3$ times.

\paragraph{Admissible range of the incoherence parameter}
We also record proof of the admissible range of the incoherence parameter in \cref{assump:incoh}, which helps clarify the mildness of \cref{assump:incoh}.

The statement is that if $\mu_{\max}$ is chosen as the smallest constant for which \cref{assump:incoh} holds, then
\begin{equation*}
    1 \le \mu_{\max} \le \sqrt{\frac{\max\{d_1,d_2\}}{r}}.
\end{equation*}
The lower endpoint corresponds to the perfectly incoherent case, while the upper endpoint corresponds to the perfectly coherent case.

The proof is as follows.
Let $\ftU=\mathcal{F}_3(\tU)$ and $\ftV=\mathcal{F}_3(\tV)$, and write
$\fmU^{(t)}$ and $\fmV^{(t)}$ for their $t$-th frontal slices.
Since $\tU$ and $\tV$ are the singular tensors in the tSVD, they are orthogonal, so for every $t\in[d_3]$,
\begin{equation*}
    (\fmU^{(t)})^H \fmU^{(t)} = I_r,
    \qquad
    (\fmV^{(t)})^H \fmV^{(t)} = I_r.
\end{equation*}
Moreover, by the definition of the tensor spectral norm,
\begin{equation*}
    \|\te_{j_1}^{\dagger}\tU\|
    =
    \max_{t\in[d_3]}
    \|\te_{j_1}^{\top}\fmU^{(t)}\|_{\mathrm{F}},
    \qquad
    \|\te_{j_2}^{\dagger}\tV\|
    =
    \max_{t\in[d_3]}
    \|\te_{j_2}^{\top}\fmV^{(t)}\|_{\mathrm{F}},
\end{equation*}
where we used the fact that a row vector has a single singular value, equal to its Euclidean norm and hence to its Frobenius norm.
Fix $t\in[d_3]$. Since $\fmU^{(t)}$ has orthonormal columns,
\begin{equation*}
    \sum_{j_1=1}^{d_1}
    \|\te_{j_1}^{\top}\fmU^{(t)}\|_{\mathrm{F}}^2
    =
    \|\fmU^{(t)}\|_{\mathrm{F}}^2
    =r.
\end{equation*}
Hence, the average squared row norm is $r/d_1$, and therefore
\begin{equation*}
    \max_{j_1\in[d_1]}
    \|\te_{j_1}^{\top}\fmU^{(t)}\|_{\mathrm{F}}
    \ge
    \sqrt{\frac{r}{d_1}}.
\end{equation*}
On the other hand, for every $j_1\in[d_1]$,
\begin{equation*}
    \|\te_{j_1}^{\top}\fmU^{(t)}\|_{\mathrm{F}}^2
    =
    \te_{j_1}^{\top}
    \fmU^{(t)}(\fmU^{(t)})^H
    \te_{j_1}
    \le 1,
\end{equation*}
because $\fmU^{(t)}(\fmU^{(t)})^H$ is an orthogonal projector. Thus
\begin{equation*}
    \sqrt{\frac{r}{d_1}}
    \le
    \max_{j_1\in[d_1]}
    \|\te_{j_1}^{\top}\fmU^{(t)}\|_{\mathrm{F}}
    \le
    1.
\end{equation*}
Taking the maximum over $t\in[d_3]$ yields
\begin{equation*}
    \sqrt{\frac{r}{d_1}}
    \le
    \max_{j_1\in[d_1]}
    \|\te_{j_1}^{\dagger}\tU\|
    \le
    1.
\end{equation*}
By the same argument,
\begin{equation*}
    \sqrt{\frac{r}{d_2}}
    \le
    \max_{j_2\in[d_2]}
    \|\te_{j_2}^{\dagger}\tV\|
    \le
    1.
\end{equation*}
Therefore,
\begin{equation*}
    1
    \le
    \mu_{\max}
    \le
    \max\left\{
    \sqrt{\frac{d_1}{r}},
    \sqrt{\frac{d_2}{r}}
    \right\}
    =
    \sqrt{\frac{\max\{d_1,d_2\}}{r}}.
\end{equation*}

Finally, both endpoints are attainable. The lower endpoint $\mu_{\max}=1$ is attained exactly when every row of every $\fmU^{(t)}$ has norm $\sqrt{r/d_1}$, and every row of every $\fmV^{(t)}$ has norm $\sqrt{r/d_2}$. For example, this happens when each $\fmU^{(t)}$ and $\fmV^{(t)}$ is formed by
$r$ columns of a unitary Fourier matrix, since then all row norms are equal. At the opposite extreme, the upper endpoint $\mu_{\max}=\sqrt{\max\{d_1,d_2\}/r}$ is attained in the coordinate-aligned case; for example, when the factor corresponding to the larger ambient dimension is formed by $r$ columns of the identity matrix, some row norm equals $1$. In the rank-one case, this means that all mass is concentrated on a single fiber.

\section{SVD Representation}
\label{Asec:SVDrepresentation}
We adopt a spectral analysis, thus the key objects are the singular subspaces of the low-rank signal tensor, $\tT$, and those of the debiased estimators, $\widehat{\tT}_{\mathrm{unbs},1}$ and $\widehat{\tT}_{\mathrm{unbs},2}$, which we defined in our sample–splitting debiasing and retraction procedure (\cref{alg:debias}).  
To obtain distributional bounds for the estimator, we need a precise description of how these singular subspaces change under perturbations.  
The SVD representation used in \cite{xia2021normal} provides such a description by expressing the difference between empirical and population spectral projectors as a convergent series in the perturbation.  
This section records the corresponding representation in the tensor setting, using the t-SVDs and the corresponding Fourier-domain slice-wise SVDs.

Define perturbation tensors $\tE_i$ as\begin{equation*}
    \tE_a = \widehat{\tT}_{\mathrm{init},a} - \tT, \quad a=1,2,
\end{equation*}
where \(\tT\) has tubal rank \(r\) and t-SVD $\tT = \tU\tS\tV^{\dagger}$,  and $\widehat{\tT}_{\mathrm{unbs},a}$ has left and right singular subspaces $\widehat{\tU}_a$ and $\widehat{\tV}_a$, respectively.  
For simplicity of notation, in this section, we omit the subscript $a$ since the two initial estimators share the same structural properties. Thus we write \begin{equation*}
    \tE = \tE_a,\quad \widehat{\tU} = \widehat{\tU}_a,\quad \widehat{\tV} = \widehat{\tV}_a,
\end{equation*}
with the understanding that $a\in\{1,2\}$ is arbitrary.

We work in the Fourier domain along the third mode. Recall the notation that $\ftE \in \mathbb{C}^{d_1 \times d_2 \times d_3}$ is the discrete Fourier transform (DFT) of $\tE$ along the third dimension, and denote by $\fmE^{(t)}$ the $t$-th frontal slice of $\ftE$ for $t=1,\dots,d_3$.  
Similarly, we write $\fmT^{(t)}, \fmU^{(t)}, \fmV^{(t)}, \fmS^{(t)}, \widehat{\fmU}^{(t)}$ and $\widehat{\fmV}^{(t)}$ for the corresponding frontal slices in the Fourier domain.

By the definition of the tensor t-SVD, for each frequency $t$, $\fmT^{(t)}$ has matrix rank $r_t\leq r$, so we have
\begin{equation*}
\fmU^{(t)} \fmS^{(t)} \fmV^{(t)H} = \operatorname{SVD}(\fmT^{(t)})
\quad\text{and}\quad
\widehat{\fmU}^{(t)} \widehat{\fmS}^{(t)} \widehat{\fmV}^{(t)H}
= \operatorname{SVD}_{r_t}(\fmT^{(t)} + \fmE^{(t)}),
\end{equation*}
where $\operatorname{SVD}_{r_t}(\cdot)$ denotes the rank-$r_t$ truncated singular value decomposition.  
According to \cite[Theorem~2.2]{lu2019tensor}, the $r_t$ diagonal entries of $\fmS^{(t)}$ are real, positive singular values of the complex matrix $\fmT^{(t)}$. Hence, the matrix perturbation tools of \cite{xia2021normal} can be applied to each slice $\fmT^{(t)}$ in the Fourier domain, provided an appropriate spectral norm regularity condition holds.

To this end, for each $(t=1,\dots,d_3)$ we introduce the Hermitian dilation
\begin{equation*}
\fmA^{(t)} =
\begin{pmatrix}
0        & \fmT^{(t)} \\
\fmT^{(t)H} & 0
\end{pmatrix},
\qquad
\fmP^{(t)} =
\begin{pmatrix}
0        & \fmE^{(t)} \\
\fmE^{(t)H} & 0
\end{pmatrix},
\end{equation*}
so that $\fmA^{(t)}$ is a Hermitian matrix of rank $2r$ and $\fmP^{(t)}$ is its Hermitian perturbation.  
We also define the block-diagonal matrices of singular vectors\begin{equation*}
\fmTheta^{(t)} =
\begin{pmatrix}
\fmU^{(t)} & 0 \\
0          & \fmV^{(t)}
\end{pmatrix},
\qquad
\widehat{\fmTheta}^{(t)} =
\begin{pmatrix}
\widehat{\fmU}^{(t)} & 0 \\
0                    & \widehat{\fmV}^{(t)}
\end{pmatrix},
\end{equation*}so that
\begin{equation*}
\widehat{\fmTheta}^{(t)} \widehat{\fmTheta}^{(t)H} - \fmTheta^{(t)} \fmTheta^{(t)H}
=
\begin{pmatrix}
\widehat{\fmU}^{(t)} \widehat{\fmU}^{(t)H} - \fmU^{(t)} \fmU^{(t)H} & 0 \\
0 & \widehat{\fmV}^{(t)} \widehat{\fmV}^{(t)H} - \fmV^{(t)} \fmV^{(t)H}
\end{pmatrix}.
\end{equation*}

Let $\fmU_{\perp}^{(t)} \in \mathbb{C}^{d_1 \times (d_1 - r)}$ and $\fmV_{\perp}^{(t)} \in \mathbb{C}^{d_2 \times (d_2 - r)}$ be orthogonal complements such that
$(\fmU^{(t)}, \fmU_{\perp}^{(t)})$ and $(\fmV^{(t)}, \fmV_{\perp}^{(t)})$ are unitary matrices.  
For any integer $s \geq 1$, following \cite{xia2021normal} we define
\begin{equation*}
(\boldsymbol{\mathfrak{P}}^{-s})^{(t)} =
\begin{cases}
\displaystyle
\begin{pmatrix}
\fmU^{(t)} \bigl(\fmS^{(t)}\bigr)^{-s} \fmU^{(t)H} & 0 \\
0 & \fmV^{(t)} \bigl(\fmS^{(t)}\bigr)^{-s} \fmV^{(t)H}
\end{pmatrix},
& \text{if } s \text{ is even}, \\[2ex]
\displaystyle
\begin{pmatrix}
0 & \fmU^{(t)} \bigl(\fmS^{(t)}\bigr)^{-s} \fmV^{(t)H} \\
\fmV^{(t)} \bigl(\fmS^{(t)}\bigr)^{-s} \fmU^{(t)H} & 0
\end{pmatrix},
& \text{if } s \text{ is odd},
\end{cases}
\end{equation*}
and
\begin{equation*}
(\boldsymbol{\mathfrak{P}}^{0})^{(t)}
= (\boldsymbol{\mathfrak{P}}^{\perp})^{(t)}
=
\begin{pmatrix}
\fmU_{\perp}^{(t)} \fmU_{\perp}^{(t)H} & 0 \\
0 & \fmV_{\perp}^{(t)} \fmV_{\perp}^{(t)H}
\end{pmatrix}.
\end{equation*}
 
As shown in \cite{xia2021normal}, if
$\|\fmP^{(t)}\| \leq \lambda_{\min}$ with $\lambda_{\min}$ being the smallest nonzero singular value of $\fmT^{(t)}$, then the spectral projector admits the convergent series expansion in $\fmP^{(t)}$.
We summarize this in the following lemma, stated in a form convenient for our tensor analysis.
\begin{lemma}[SVD representation]
\label{lem:svd_representation}
Fix $t \in \{1,\dots,d_3\}$, and suppose that the nonzero singular values of $\fmT^{(t)}$ are bounded below by $\lambda_{\min} > 0$ and that $\|\fmP^{(t)}\| \leq \lambda_{\min}$. Then \begin{multline}\label{eq:rep_svd}
    \widehat{\fmTheta}^{(t)} \widehat{\fmTheta}^{(t)H}
- \fmTheta^{(t)} \fmTheta^{(t)H}
\\=
\sum_{k=1}^{\infty}
\underbrace{
\sum_{\mathbf{s}:\,s_1+\cdots+s_{k+1}=k}
(-1)^{1+\tau(\mathbf{s})}
\,(\boldsymbol{\mathfrak{P}}^{-s_1})^{(t)} \fmP^{(t)}
(\boldsymbol{\mathfrak{P}}^{-s_2})^{(t)} \fmP^{(t)}
\cdots
(\boldsymbol{\mathfrak{P}}^{-s_k})^{(t)} \fmP^{(t)}
(\boldsymbol{\mathfrak{P}}^{-s_{k+1}})^{(t)}
}_{\mathcal{S}_{A,k}(\fmP^{(t)})},
\end{multline}
where the inner sum is taken over integer tuples
$\mathbf{s} = (s_1,\dots,s_{k+1})$ with $s_j \ge 0$ and
$s_1+\cdots+s_{k+1} = k$, and $\tau(\mathbf{s})=\sum_{j=1}^{k+1} \mathbf{1}\{s_j > 0\}$ denotes the number of positive indices in $\mathbf{s}$. Moreover, if the bound $\|\fmP^{(t)}\| \leq \lambda_{\min}$ holds for all $t = 1,\dots,d_3$, then the corresponding tensor-level relation
\begin{equation}
\label{eq:tensor_svd_rp}
\widehat{\tTheta}\,\widehat{\tTheta}^{\dagger} - \tTheta \tTheta^{\dagger}=\sum_{k=1}^{\infty}
\underbrace{\sum_{\mathbf{s}:\,s_1+\cdots+s_{k+1}=k}
(-1)^{1+\tau(\mathbf{s})}
\;\mathfrak{P}^{-s_1} \tP \mathfrak{P}^{-s_2} \cdots
\mathfrak{P}^{-s_k} \tP \mathfrak{P}^{-s_{k+1}}}_{\mathcal{S}_{A,k}(\tP)}
\end{equation}
holds, where all tensor-tensor products are t-products, and
$\mathfrak{P}^{-s}$ and $\tP$ are defined via their Fourier-domain
slices $(\boldsymbol{\mathfrak{P}}^{-s})^{(t)}$ and $\fmP^{(t)}$,
respectively.
\end{lemma}

The tensor-level identity \cref{eq:tensor_svd_rp} is obtained by the Fourier-domain characterization of the t-product, which is stated in \cref{prop:tprod} in the main paper. We give a review of \cref{prop:tprod} here: let $\mathcal{F}_3$ denote the discrete Fourier transform along the third mode, and let $\ftA = \mathcal{F}_3(\tA)$ denote the collection of frontal slices $\{\fmA^{(t)}\}_{t=1}^{d_3}$. For any pair of tensors $\tA$ and $\tB$ of compatible dimensions, the t-product satisfies\begin{equation*}
    \mathcal{F}_3(\tA * \tB)^{(t)} = \fmA^{(t)} \fmB^{(t)}, \qquad t=1,\dots,d_3.
\end{equation*}
Now, when the matrix relation \cref{eq:rep_svd} holds slice-wise for $t=1,\dots,d_3$, we can apply $\mathcal{F}_3^{-1}$ to both sides of each slice-wise identity. This establishes the asserted tensor-level SVD representation.

The expansion \cref{eq:tensor_svd_rp} serves as the basic SVD representation used in our subsequent analysis of the empirical singular subspaces. The verification that the assumed bound $\|\fmP^{(t)}\| \leq \lambda_{\min}$ holds with high probability under our model is deferred to later sections of the appendix.

\section{Proof of \texorpdfstring{\cref{thm:norm}}{Theorem~X}}

\label{proof:norm}

Denote $d^* = d_1 d_2 d_3$ and $n=2n_0$. Without loss of generality, we assume $n$ is even, and $n_0$ is an integer. For the two sample splits $a\in\{1,2\}$, denote the estimation error of the initial estimator $\tT_{\mathrm{init},a}$ as $\tZ_a = \tT - \tT_{\mathrm{init},a}.$
According to the observation model in \cref{eq:model} and the debiasing updates in \cref{alg:debias}, we obtain for $a=1,2$,\begin{multline*}
    \tT_{\mathrm{unbs},a}= \tT_{\mathrm{init},a}+ \frac{d^*}{n_0} \sum_{i \in I_a}\bigl( Y_i - \langle \tT_{\mathrm{init},a}, \tX_i \rangle \bigr) \tX_i = \tT + \frac{d^*}{n_0} \sum_{i \in I_a} \langle \tZ_a, \tX_i \rangle \tX_i - \tZ_a+ \frac{d^*}{n_0} \sum_{i \in I_a} \xi_i \tX_i ,
\end{multline*}
where $I_1 =\{n_0+1,\dots,n\} $ and $I_2 =\{1,\dots,n_0\}$ denotes how sample was split. 
It is convenient to isolate the two error components explicitly. We define
\begin{equation*}
    \tE_{\mathrm{init},a}
:= \frac{d^*}{n_0} \sum_{i\in I_a} \langle \tZ_a, \tX_i \rangle \tX_i - \tZ_a,
\qquad
\tE_{\mathrm{rn},a}
:= \frac{d^*}{n_0} \sum_{i\in I_a} \xi_i \tX_i,
\end{equation*}
With this notation, \begin{equation*}
    \tT_{\mathrm{unbs},a} = \tT + \tE_{\mathrm{init},a} + \tE_{\mathrm{rn},a},
\quad a = 1,2.
\end{equation*}
Now for the unbiased estimator $\tT_{\mathrm{unbs},a}$, the estimation error with respect of $\tT$ is decomposed into two parts: $\tE_{\mathrm{init},a}$ is the initialization error arising from using $\tT_{\mathrm{init},a}$ in place of $\tT$, and $\tE_{\mathrm{rn},a}$ is the accumulated random noise. We further write the estimation error of $\tT_{\mathrm{unbs},a}$, also called the \emph{debiasing error}, as\begin{equation*}
    \tE_a := \tE_{\mathrm{init},a} + \tE_{\mathrm{rn},a}, \quad a = 1,2.
\end{equation*}
By the projection step in \cref{alg:debias}, the final estimator can therefore be written as\begin{equation*}
   \widehat{\tT} = \frac{1}{2} \, \widehat{\tU}_1 \widehat{\tU}_1^{\dagger} (\tT + \tE_1) \widehat{\tV}_1 \widehat{\tV}_1^{\dagger}
+ \frac{1}{2} \, \widehat{\tU}_2 \widehat{\tU}_2^{\dagger} (\tT + \tE_2) \widehat{\tV}_2 \widehat{\tV}_2^{\dagger}.
\end{equation*}
Hence, for any fixed $\tM$,
\begin{multline}\label{aimofthm1}
\langle \tM, \widehat\tT \rangle - \langle \tM, {\tT} \rangle=\frac{1}{2} \langle \widehat{\tU}_1 \widehat{\tU}_1^{\dagger} \tE_1 \widehat{\tV}_1 \widehat{\tV}_1^{\dagger}, \tM \rangle
+ \frac{1}{2} \langle \widehat{\tU}_2 \widehat{\tU}_2^{\dagger} \tE_2 \widehat{\tV}_2 \widehat{\tV}_2^{\dagger}, \tM \rangle \\+ \frac{1}{2} \langle \widehat{\tU}_1 \widehat{\tU}_1^{\dagger} \tT \widehat{\tV}_1 \widehat{\tV}_1^{\dagger} - \tT, \tM \rangle
+ \frac{1}{2} \langle \widehat{\tU}_2 \widehat{\tU}_2^{\dagger} \tT \widehat{\tV}_2 \widehat{\tV}_2^{\dagger} - \tT, \tM \rangle.
\end{multline}

The proof proceeds in three steps.
First, we control the size of the debiasing errors $\tE_a$ ($a=1,2$) in spectral norm, so that their effect is small relative to the variance scale in \cref{thm:norm}.
Second, we bound the perturbation of the empirical singular subspaces $(\widehat{\tU}_a,\widehat{\tV}_a)$ ($a=1,2$) relative to $(\tU,\tV)$ and show that the resulting projected error terms (the first two terms in \cref{aimofthm1}) are of small orders.
Third, we expand the projection of $\tT$ (the last two terms in \cref{aimofthm1}) using the SVD representation introduced in \cref{Asec:SVDrepresentation}. In the expansion, we identify a leading linear term that satisfies a non-asymptotic Gaussian approximation, and we bound all higher–order terms so that their contribution also vanishes under the regime of \cref{thm:norm}.

\medskip

\noindent\paragraph{Step 1: Controlling the size of the debiasing error}
The following lemma provides a high-probability bound on the spectral norms of the two components of $\tE_k$. 
\begin{lemma}\label{lem:error_op_b}Under \cref{assump:sampling,assump:noise,assump:init_entry}, if $n> 6(d_1\wedge d_2)\log(d_1\wedge d_2)\log((d_1+d_2)d_3)$  , with probability at least $1-(d_1d_3+d_2d_3)^{-2}$, the following bounds hold for $a=1,2$: \begin{displaymath}
         \|\tE_{\mathrm{rn},a}\| \leq C_{\Bernstein}
    \cdot \sigma_{\xi} \sqrt{d_1d_2(d_1\vee d_2)}d_3 \sqrt{\frac{\log (d_1d_3+d_2d_3)}{n}},
    \end{displaymath}
 and\begin{displaymath}
         \|\tE_{\mathrm{init},a}\| \leq 
C_{\Bernstein} \cdot \|{\tZ}_a\|_{\max } \sqrt{d_1d_2(d_1\vee d_2)}d_3\sqrt{\frac{\log (d_1d_3+d_2d_3)}{n}} ,
    \end{displaymath}where $C_{\Bernstein>0}$ is an absolute constant.
    \end{lemma}
Intuitively, \cref{lem:error_op_b} states that, with high probability, both the random noise term and the initialization error are of order $\sigma_{\xi} \sqrt{d_1 d_2 (d_1 \vee d_2)} \, d_3 \sqrt{{\log(d_1d_3 + d_2d_3) }/{n}}.$ The proof of \cref{lem:error_op_b} is in \cref{proof:error_op_b}.
Recall that by \cref{assump:init_entry}, there exists an event $\Omega_{\mathrm{init}}$ with probability $\mathbb{P}(\Omega_{\mathrm{init}})\geq1-(d_1d_3 + d_2d_3)^{-2}$ such that, on $\Omega_{\mathrm{init}}$, we have:\begin{equation*}
    \|\tZ_a\|_{\max}\leq C_{\mathrm{init}}{\gamma_n}\cdot\sigma_{\xi},
\end{equation*}where $C_{\mathrm{init}}$ is an absolute constant and $\gamma_n\leq 1$. 
Let $\Omega_0$ be the intersection of $\Omega_{\mathrm{init}}$ and the event on which \cref{lem:error_op_b} holds. It follows that $\mathbb{P}(\Omega_1)\geq 1-2(d_1d_3 + d_2d_3)^{-2}$. Without loss of generality, we can enlarge $C_{\Bernstein}$ from \cref{lem:error_op_b} to absorb the constant $C_{\mathrm{init}}$ and the term $(1+\gamma_n)$. Then, for $a=1,2,$ the following bound holds on $\Omega_0$:\begin{equation}\label{eq:delta}
\|\tE_a\|
\leq
C_{\Bernstein} \sigma_{\xi}
\sqrt{d_1 d_2 (d_1 \vee d_2)} \, d_3
\sqrt{\frac{\log(d_1d_3 + d_2d_3) }{n}}
=: \delta.
\end{equation} 
The proof of \cref{lem:error_op_b} is deferred to \cref{proof:error_op_b}.

\medskip
\noindent\paragraph{Step 2: Perturbation of singular subspaces and projection of the error}
The projection in \cref{alg:debias} involves the empirical singular subspaces $(\widehat{\tU}_a, \widehat{\tV}_a)$, $a=1,2$. 
The next result provides a Davis–Kahan \cite{davis1970rotation} type bound on the distance between these empirical subspaces and the population subspaces $(\tU,\tV)$; see, for example, \cite[Chapter 2]{chen2021spectral} for related matrix-level perturbation bounds.
\begin{theorem} 
\label{thm:distance}
Under \cref{assump:sampling,assump:noise,assump:SNR,assump:init_entry,assump:incoh}, if \begin{displaymath}
    n\geq4\mu_{\max}^2(d_1\vee d_2)\log(\mu_{\max}(d_1\vee d_2))\log(d_1d_3+d_2d_3),
\end{displaymath}then with probability at least $1-\log(d_1\vee d_2)(d_1+d_2)^{-2}d_3^{-3}$, for $a=1,2$:\begin{displaymath}
\max_{j\in[d_1]}\|\te_j^{\dagger}\,
(\widehat{\tU}_a\widehat{\tU}_a^{\dagger}-\tU\tU^{\dagger})\| \leq
C_1\mu_{\max}\frac{\sigma_{\xi}}{\lambda_{\min}}d_3\sqrt{\frac{rd_2(d_1\vee d_2)}{n}}\sqrt{\log(d_1d_3+d_2d_3)},
\end{displaymath}
\begin{displaymath}
\max_{j\in[d_2]}\|\te_j^{\dagger}\,(\widehat{\tV}_a\widehat{\tV}_a^{\dagger}-\tV\tV^{\dagger})\|  \leq C_1\mu_{\max}\frac{\sigma_{\xi}}{\lambda_{\min}}d_3\sqrt{\frac{rd_1(d_1\vee d_2)}{n}}\sqrt{\log(d_1d_3+d_2d_3)} ,
\end{displaymath}where $C_1>0$ is an absolute constant.
\end{theorem}
The proof of \cref{thm:distance} is in \cref{proof:distance}.
Combining \cref{lem:error_op_b} and \cref{thm:distance}, the empirical subspaces $(\widehat{\tU}_a, \widehat{\tV}_a)$ are close to $(\tU, \tV)$, and the perturbation $\tE_a$ is small in spectral norm. The next lemma quantifies the resulting effect on the linear functional: it shows that the contribution of $\tE_a$ after projecting onto the estimated low-rank subspace is bounded.
\begin{lemma}
   Under \cref{assump:sampling,assump:noise,assump:SNR,assump:init_entry,assump:incoh}, and under the event of \cref{thm:distance}, when\begin{displaymath}
         n>8\mu_{\max}^4r^2d_3\log d_3\log(d_1d_3+d_2d_3),
   \end{displaymath}with probability at least $1 - 3(d_1 d_3 + d_2 d_3)^{-2}$ we have
    \begin{multline*}
            \max\{| \langle \widehat{\tU}_1\widehat{\tU}^{\dagger}_1 \tE_1\widehat{\tV}_1\widehat{\tV}^{\dagger}_1, \tM \rangle| ,| \langle \widehat{\tU}_2\widehat{\tU}^{\dagger}_2\tE_2\widehat{\tV}_2\widehat{\tV}^{\dagger}_2, \tM \rangle|\}
             \leq C_2\,\|\tM\|_{\ell_1}\mu_{\max}^2\sigma_{\xi}r\sqrt{\frac{d_3\log(d_1d_3+d_2d_3)}{n}} \\+ C_3\,\|\tM\|_{\ell_1}\mu_{\max}^2\frac{\sigma^2_{\xi}}{\lambda_{\min}}\frac{r\sqrt{d_1d_2}(d_1\vee d_2) d_3^2}{n}\log (d_1d_3+d_2 d_3),
    \end{multline*}
\label{lem:proj_error_term}
where $C_2>0$ and $C_3>0$ are absolute constants.
\end{lemma}
The proof of \cref{lem:proj_error_term} is in \cref{proof:proj_error_term}.
In view of \cref{aimofthm1}, \cref{lem:proj_error_term} gives a bound of the projected error terms. It therefore remains to analyze the terms where $\tT$ itself is projected onto the estimated singular subspaces.

\medskip
\noindent\paragraph{Step 3: Expansion of the projected signal and normal approximation}
We now analyze the contribution from projecting the true signal $\tT$ onto the estimated singular subspaces. Recall that the remaining terms in \cref{aimofthm1} are\begin{equation}\label{eq:last-two-terms}
    \frac{1}{2} \langle \widehat{\tU}_1\widehat{\tU}_1^{\dagger} \tT \widehat{\tV}_1\widehat{\tV}_1^{\dagger} -\tT, \tM\rangle
    +\frac{1}{2} \langle \widehat{\tU}_2\widehat{\tU}_2^{\dagger} \tT \widehat{\tV}_2\widehat{\tV}_2^{\dagger} -\tT, \tM\rangle.
\end{equation}
The idea is to expand these terms in powers of the perturbation of the singular subspaces (SVD representation introduced in \cref{Asec:SVDrepresentation}) and then isolate a leading, linear term that admits a Gaussian limit. All higher-order terms will be bounded.
According to \cref{lem:svd_representation}, the SVD representation holds if the perturbation is sufficiently small relative to the minimum non-zero singular value $\lambda_{\min}$. Under \cref{assump:SNR}, we have:\begin{equation*}
     \frac{\lambda_{\min}}{\sigma_\xi}\ \ge\ C\, \frac{d_3(d_1\vee d_2)^{3/2}}{\sqrt{n}}\,
    \sqrt{\log\!\big(d_1d_3+d_2d_3\big)}.
\end{equation*}Combining this SNR constraint with the bound on $\|\tE_a\|$ in \cref{eq:delta} satisfies the condition of \cref{lem:svd_representation}. Thus, we obtain the expansion: \begin{equation}\label{eq:svd_representation}
    \widehat{\tTheta}_a \widehat{\tTheta}_a^{\dagger} - \tTheta \tTheta^{\dagger}=\sum_{k=1}^{\infty} \mathcal{S}_{A,k}(\tP_a),\quad a=1,2.
\end{equation}
To apply the SVD representation \cref{eq:svd_representation} to the linear forms in \cref{eq:last-two-terms}, we define the augmented tensor $\widetilde{\tM} \in \mathbb{R}^{(d_1 + d_2) \times (d_1 + d_2) \times d_3}$ slice-wise in the frequency domain as:\begin{equation*}
    \widetilde{\fmM}^{(t)} =
\begin{pmatrix}
0 & \fmM^{(t)} \\
0 & 0
\end{pmatrix},
\qquad 1 \le t \le d_3,
\end{equation*}where $\fmM^{(t)}$ is the $t$-th slice of the query tensor $\tM$ after the Mode-3 DFT.
Then, by construction of $\tA$, $\widehat{\tTheta}_a$ and $\tTheta_a$ in \cref{Asec:SVDrepresentation}, we have\begin{equation}\label{eq:Theta-rewrite}
\langle \widehat{\tU}_a\widehat{\tU}_a^{\dagger} \tT \widehat{\tV}_a \widehat{\tV}_a^{\dagger} - \tT, \tM \rangle=\bigl\langle\widehat{\tTheta}_a\widehat{\tTheta}_a^{\dagger}\tA \widehat{\tTheta}_a \widehat{\tTheta}_a^{\dagger}- \tTheta \tTheta^{\dagger} \tA \tTheta \tTheta^{\dagger},\widetilde{\tM}\bigr\rangle.\end{equation}By SVD representation in \cref{eq:svd_representation}, for $a=1,2$,\begin{align*}
    \widehat{\tTheta}_a \widehat{\tTheta}_a^{\dagger} \tA \widehat{\tTheta}_a \widehat{\tTheta}_a^{\dagger}
- \tTheta \tTheta^{\dagger} \tA \tTheta \tTheta^{\dagger}&=
\mathcal{S}_{A,1}(\tP_a) \, \tA \tTheta \tTheta^{\dagger}
+ \tA \tTheta \tTheta^{\dagger} \tA \mathcal{S}_{A,1}(\tP_a) \\&\quad+ \sum_{k=2}^{\infty}\Bigl(\mathcal{S}_{A,k}(\tP_a) \, \tA \tTheta \tTheta^{\dagger}+ \tA \tTheta \tTheta^{\dagger} \tA \mathcal{S}_{A,k}(\tP_a)\Bigr) \\&\quad+ (\widehat{\tTheta}_a\widehat{\tTheta}_a^{\dagger} - \tTheta\tTheta^{\dagger})\, \tA \,(\widehat{\tTheta}_a \widehat{\tTheta}_a^{\dagger} - \tTheta \tTheta^{\dagger}).
\end{align*}
Combining this expansion with \cref{eq:Theta-rewrite}, we obtain the decomposition
\begin{equation}\label{eq:thm1dec}
\begin{aligned}&\frac{1}{2} \langle \widehat{\tU}_1 \widehat{\tU}_1^{\dagger} \tT \widehat{\tV}_1 \widehat{\tV}_1^{\dagger} - \tT, \tM \rangle
+ \frac{1}{2} \langle \widehat{\tU}_2 \widehat{\tU}_2^{\dagger} \tT \widehat{\tV}_2 \widehat{\tV}_2^{\dagger} - \tT, \tM \rangle \\&=
\frac{1}{2} \sum_{a\in\{1,2\}}
\bigl\langle
\mathcal{S}_{A,1}(\tP_a) \, \tA \tTheta \tTheta^{\dagger}
+ \tA \tTheta \tTheta^{\dagger} \tA \mathcal{S}_{A,1}(\tP_a),
\widetilde{\tM}
\bigr\rangle \\&+ \frac{1}{2} \sum_{a\in\{1,2\}} \sum_{k=2}^{\infty}
\bigl\langle
\mathcal{S}_{A,k}(\tP_a) \, \tA \tTheta \tTheta^{\dagger}+ \tA \tTheta \tTheta^{\dagger} \tA \mathcal{S}_{A,k}(\tP_a),
\widetilde{\tM}
\bigr\rangle \\&+ \frac{1}{2} \sum_{a\in\{1,2\}}\bigl\langle(\widehat{\tTheta}_a \widehat{\tTheta}_a^{\dagger} - \tTheta \tTheta^{\dagger})\, \tA \,(\widehat{\tTheta}_a \widehat{\tTheta}_a^{\dagger} - \tTheta \tTheta^{\dagger}),\widetilde{\tM}\bigr\rangle.\end{aligned}\end{equation}
The first line on the right-hand side of \cref{eq:thm1dec} will give the leading Gaussian approximation (\cref{lem:normal-t1}), while the remaining two lines will be shown to be of smaller order (\cref{lem:normal-t2,lem:normal-t3}).
For the leading term, the definition of $\mathcal{S}_{A,1}(\tP_a)$ implies\begin{multline*}
    \frac{1}{2}\sum_{a\in\{1,2\}} \langle\mathcal{S}_{A,1}(\tP_a) \tA{\tTheta}{\tTheta}^{\dagger} +\tA{\tTheta}{\tTheta}^{\dagger}\tA \mathcal{S}_{A,1}(\tP_a) ,\widetilde{\tM}\rangle\\=\langle \tU_{\perp} \tU_{\perp}^{\dagger}\frac{\tE_1+\tE_2}{2} \tV \tV^{\dagger}, \tM\rangle+\langle \tU \tU^{\top}\frac{\tE_1+\tE_2}{2} \tV_{\perp} \tV_{\perp}^{\dagger}, \tM\rangle,
\end{multline*}The normal approximation for this term is provided by the following lemma.
\begin{lemma}\label{lem:normal-t1}
     Under \cref{assump:sampling,assump:noise,assump:init_entry,assump:incoh,assump:M}, and if\begin{displaymath}
         n>6\mu_{\max}^2rd_3(d_1\vee d_2)\log^2(d_1d_3 +d_2d_3),
     \end{displaymath}we have\begin{multline*}
         \sup _{x \in \mathbb{R}}\left|\mathbb{P}\left(\frac{\langle \tU_{\perp} \tU_{\perp}^{\dagger}\frac{\tE_1+\tE_2}{2} \tV \tV^{\dagger}, \tM\rangle+\langle \tU \tU^{\dagger}\frac{\tE_1+\tE_2}{2} \tV_{\perp} \tV_{\perp}^{\dagger}, \tM\rangle}{\sigma_{\xi}(\|\tV^{\dagger} \tM^{\dagger}\|_{\mathrm{F}}^2+\|\tU^{\dagger} \tM\|_{\mathrm{F}}^2)^{1 / 2} \cdot \sqrt{d^* / n}} \leq x\right)-\Phi(x)\right|\\\leq C_{\BE}\,\mu_{\max}\sqrt{\frac{rd_3(d_1\vee d_2)}{n}} 
         +C_4\, \gamma_n\sqrt{\log(d_1d_3+d_2d_3)}
         \\+C_4\, \frac{\mu_{\max}^4\|\tM\|_{\ell_1}^2}{\alpha_{M}^2\|\tM\|_{\mathrm{F}}^2}\cdot{\frac{r\sqrt{\log(d_1d_3+d_2d_3)}}{d_1\wedge d_2}}+ \frac{4}{(d_1d_3+d_2d_3)^2},
     \end{multline*}where $C_{\BE}>0$ and $C_4>0$ are absolute constants.
\end{lemma}
\cref{lem:normal-t1} is proved in \cref{proof:normal-t1}. The remaining terms in \cref{eq:thm1dec} are controlled by the following bounds.

\begin{lemma}\label{lem:normal-t2}
Under \cref{assump:sampling,assump:noise,assump:SNR,assump:init_entry,assump:incoh,assump:M}, and under the event of \cref{thm:distance}, we have
\begin{multline*}
    |\sum_{a\in\{1,2\}} \sum_{k=2}^{\infty} \langle(\mathcal{S}_{A,k}(\tP_a) \tA{\tTheta}{\tTheta}^{\dagger} +\tA{\tTheta}{\tTheta}^{\dagger}\tA \mathcal{S}_{A,k}(\tP_a) ),\widetilde{\tM}\rangle| \\\quad
\leq 2C_{\Bernstein}^2\, \|\tM\|_{\ell_1} \mu_{\max}^2  \frac{\sigma_{\xi}^2}{\lambda_{\min}}\frac{r\sqrt{d_1d_2}(d_1\vee d_2)d_3^2}{n}\log(d_1d_3+d_2d_3)
\end{multline*}
where $C_{\Bernstein}>0$ is an absolute constant.
\end{lemma}

\begin{lemma}\label{lem:normal-t3}
Under \cref{assump:sampling,assump:noise,assump:SNR,assump:init_entry,assump:incoh,assump:M}, and under the event of \cref{thm:distance}, we have, for $a\in\{1,2\}$, \begin{multline*}
    \sum_{a\in\{1,2\}} |\langle(\widehat{\tTheta}_a \widehat{\tTheta}_a^{\dagger}-{\tTheta} \tTheta^{\dagger})\tA(\widehat{\tTheta}_a \widehat{\tTheta}_a^{\dagger}-{\tTheta} \tTheta^{\dagger}),\widetilde{\tM}\rangle| \\\quad
\leq C_1^2\,\|\tM\|_{\ell_1} \kappa_0\mu_{\max}^2 \frac{\sigma_{\xi}^2}{\lambda_{\min}}\frac{r\sqrt{d_1d_2}(d_1\vee d_2)d_3^2}{n}\log(d_1d_3+d_2d_3).
\end{multline*}where $C_1$ is an absolute constant.
\end{lemma}
The proof of \cref{lem:normal-t2,lem:normal-t3} is in \cref{proof:normal-t2,proof:normal-t3}, respectively.

\medskip

We now combine the decomposition of the debiased estimator with the bounds established above to derive the claimed normal approximation. By~\cref{aimofthm1} and~\cref{eq:thm1dec}, we have
\begin{multline}\label{eq:main-decomp-normalized}
\frac{\langle \widehat{\tT}, \tM\rangle-\langle \tT, \tM\rangle}
{\sigma_{\xi}\bigl(\|\tM\tV\|_{\mathrm{F}}^2+\|\tU^{\dagger}\tM\|_{\mathrm{F}}^2\bigr)^{1/2}\sqrt{d_1d_2d_3/n}}
\\
\begin{aligned}
&=
\frac{\sum_{a\in\{1,2\}}\frac{1}{2}\langle \widehat{\tU}_a\widehat{\tU}_a^{\dagger}\tE_a\widehat{\tV}_a\widehat{\tV}_a^{\dagger}, \tM\rangle}
{\sigma_{\xi}\bigl(\|\tM\tV\|_{\mathrm{F}}^2+\|\tU^{\dagger}\tM\|_{\mathrm{F}}^2\bigr)^{1/2}\sqrt{d_1d_2d_3/n}}
\\
&\quad+
\frac{\frac{1}{2}\sum_{a\in\{1,2\}}
\langle \mathcal{S}_{A,1}(\tP_a)\tA\tTheta\tTheta^{\dagger}
+\tA\tTheta\tTheta^{\dagger}\tA\mathcal{S}_{A,1}(\tP_a),\widetilde{\tM}\rangle}
{\sigma_{\xi}\bigl(\|\tM\tV\|_{\mathrm{F}}^2+\|\tU^{\dagger}\tM\|_{\mathrm{F}}^2\bigr)^{1/2}\sqrt{d_1d_2d_3/n}}
\\
&\quad+
\frac{\frac{1}{2}\sum_{a\in\{1,2\}}\sum_{k=2}^{\infty}
\langle \mathcal{S}_{A,k}(\tP_a)\tA\tTheta\tTheta^{\dagger}
+\tA\tTheta\tTheta^{\dagger}\tA\mathcal{S}_{A,k}(\tP_a),\widetilde{\tM}\rangle}
{\sigma_{\xi}\bigl(\|\tM\tV\|_{\mathrm{F}}^2+\|\tU^{\dagger}\tM\|_{\mathrm{F}}^2\bigr)^{1/2}\sqrt{d_1d_2d_3/n}}
\\
&\quad+
\frac{\frac{1}{2}\sum_{a\in\{1,2\}}
\langle(\widehat{\tTheta}_a\widehat{\tTheta}_a^{\dagger}-\tTheta\tTheta^{\dagger})
\tA
(\widehat{\tTheta}_a\widehat{\tTheta}_a^{\dagger}-\tTheta\tTheta^{\dagger}),\widetilde{\tM}\rangle}
{\sigma_{\xi}\bigl(\|\tM\tV\|_{\mathrm{F}}^2+\|\tU^{\dagger}\tM\|_{\mathrm{F}}^2\bigr)^{1/2}\sqrt{d_1d_2d_3/n}}.
\end{aligned}
\end{multline}
Under \cref{assump:M}, it holds that $\|\tM\tV\|_{\mathrm{F}}^2+\|\tU^{\dagger}\tM\|_{\mathrm{F}}^2\geq 2\alpha_{\tM}^2\|\tM\|_{\mathrm{F}}^2r/(d_1\vee d_2) $. Thus, the (approximate) variance in the denominator is bounded below as \begin{equation*}
    \sigma_{\xi}(\|\tM\tV\|_{\mathrm{F}}^2+\|\tU^{\dagger}\tM\|_{\mathrm{F}}^2)^{1/2}\, \sqrt{d_1d_2d_3/n}\geq \sqrt{2}\alpha_{\tM}\|\tM\|_{\mathrm{F}}\,\sigma_{\xi} \sqrt{r(d_1\wedge d_2)d_3/n}.
\end{equation*}
 Combining this lower bound with \cref{lem:proj_error_term,lem:normal-t2,lem:normal-t3} and enlarging the absolute constants $C_2$ and $C_3$ if necessary, we obtain that, with probability at least $1 - 4 (d_1 d_3 + d_2 d_3)^{-2}$,
 \begin{align*}
&\Biggl|
\frac{\langle \widehat{\tT}, \tM\rangle-\langle \tT, \tM\rangle}
{\sigma_{\xi}\bigl(\|\tM\tV\|_{\mathrm{F}}^2+\|\tU^{\dagger}\tM\|_{\mathrm{F}}^2\bigr)^{1/2}\sqrt{d_1d_2d_3/n}}
\\
&\qquad\qquad-
\frac{\frac{1}{2}\sum_{a\in\{1,2\}}
\langle \mathcal{S}_{A,1}(\tP_a)\tA\tTheta\tTheta^{\dagger}
+\tA\tTheta\tTheta^{\dagger}\tA\mathcal{S}_{A,1}(\tP_a),\widetilde{\tM}\rangle}
{\sigma_{\xi}\bigl(\|\tM\tV\|_{\mathrm{F}}^2+\|\tU^{\dagger}\tM\|_{\mathrm{F}}^2\bigr)^{1/2}\sqrt{d_1d_2d_3/n}}
\Biggr|
\\
&\le
C_2\,
\frac{\mu_{\max}^2\|\tM\|_{\ell_1}}
{\alpha_{\tM}\|\tM\|_{\mathrm{F}}}\,
\sqrt{\frac{r\log(d_1d_3+d_2d_3)}{d_1\wedge d_2}}
\\
&\quad+
C_3\,
\frac{\mu_{\max}^2\|\tM\|_{\ell_1}}
{\alpha_{\tM}\|\tM\|_{\mathrm{F}}} \,\kappa_0\,
\frac{\sigma_{\xi}}{\lambda_{\min}}
\frac{d_3^{3/2}(d_1\vee d_2)^{3/2}}{\sqrt{n}}
\log(d_1d_3+d_2d_3).
\end{align*}
Applying the normal approximation in \cref{lem:normal-t1} to the leading term in \cref{eq:main-decomp-normalized} and using the Lipschitz continuity of $\Phi(x)$, we obtain
\begin{multline*}
\sup_{x\in\mathbb{R}}
\left|
\mathbb{P}\!\left(
\frac{\langle\widehat{\tT},\tM\rangle-\langle\tT,\tM\rangle}
{\sigma_{\xi}\bigl(\|\tM\tV\|_{\mathrm{F}}^2+\|\tU^{\dagger}\tM\|_{\mathrm{F}}^2\bigr)^{1/2}\sqrt{d_1d_2d_3/n}}
\le x
\right)
-\Phi(x)
\right|
\\\le
C_{\BE}\,\mu_{\max}\sqrt{\frac{r\,d_3\,(d_1\vee d_2)}{n}}
+C_4\,\gamma_n\,\sqrt{\log(d_1d_3+d_2d_3)}
+
\frac{8}{(d_1d_3+d_2d_3)^2}
\\
+ C_4\,\frac{\mu_{\max}^4\|\tM\|_{\ell_1}^2}{\alpha_{\tM}^2\|\tM\|_{\mathrm{F}}^2}\,
\frac{r\sqrt{\log(d_1d_3+d_2d_3)}}{d_1\wedge d_2}
\\
+C_2\,\frac{\mu_{\max}^2\|\tM\|_{\ell_1}}{\alpha_{\tM}\|\tM\|_{\mathrm{F}}}\,
\sqrt{\frac{r\log(d_1d_3+d_2d_3)}{d_1\wedge d_2}}
\\+
C_3\,\frac{\mu_{\max}^2\|\tM\|_{\ell_1}}{\alpha_{\tM}\|\tM\|_{\mathrm{F}}}\,
\kappa_0\frac{\sigma_{\xi}}{\lambda_{\min}}\,
\frac{d_3^{3/2}(d_1\vee d_2)^{3/2}}{\sqrt{n}}\,
\log(d_1d_3+d_2d_3),
\end{multline*}where we use the fact that $\kappa_0$ is bounded, and we enlarge the absolute constants if necessary. Since $r\ge 1$, $\mu_{\max}\ge1$, $\alpha_{\tM}\le1$, and $\|\tM\|_{\ell_1}\ge \|\tM\|_{\mathrm{F}}$, the fourth term and fifth term can be combined as\begin{multline*}
    C_4\,\frac{\mu_{\max}^4\|\tM\|_{\ell_1}^2}{\alpha_{\tM}^2\|\tM\|_{\mathrm{F}}^2}\,
\frac{r\sqrt{\log(d_1d_3+d_2d_3)}}{d_1\wedge d_2}+
C_2\,\frac{\mu_{\max}^2\|\tM\|_{\ell_1}}{\alpha_{\tM}\|\tM\|_{\mathrm{F}}}\,
\sqrt{\frac{r\log(d_1d_3+d_2d_3)}{d_1\wedge d_2}}
\\\leq (C_4+ C_2)\, \frac{\mu_{\max}^4\|\tM\|_{\ell_1}^2}{\alpha_{\tM}^2\|\tM\|_{\mathrm{F}}^2}\,
r\sqrt{\frac{\log(d_1d_3+d_2d_3)}{d_1\wedge d_2}}.
\end{multline*} After renaming the absolute constants, we get the exact claimed bound:\begin{multline*}
\sup_{x\in\mathbb{R}}
\left|
\mathbb{P}\!\left(
\frac{\langle\widehat{\tT},\tM\rangle-\langle\tT,\tM\rangle}
{\sigma_{\xi}\bigl(\|\tM\tV\|_{\mathrm{F}}^2+\|\tU^{\dagger}\tM\|_{\mathrm{F}}^2\bigr)^{1/2}\sqrt{d_1d_2d_3/n}}
\le x
\right)
-\Phi(x)
\right|
\\\le \frac{8}{(d_1d_3+d_2d_3)^2}
+C_2\,\mu_{\max}\sqrt{\frac{r\,d_3\,(d_1\vee d_2)}{n}}
+C_3\,\gamma_n\,\sqrt{\log(d_1d_3+d_2d_3)}
\\
+C_4\,\frac{\mu_{\max}^4\|\tM\|_{\ell_1}^2}{\alpha_{\tM}^2\|\tM\|_{\mathrm{F}}^2}\,r
\sqrt{\frac{\log(d_1d_3+d_2d_3)}{d_1\wedge d_2}}
\\+
C_5\,\frac{\mu_{\max}^2\|\tM\|_{\ell_1}}{\alpha_{\tM}\|\tM\|_{\mathrm{F}}}\,
\kappa_0\,\frac{\sigma_{\xi}}{\lambda_{\min}}\,
\frac{d_3^{3/2}(d_1\vee d_2)^{3/2}}{\sqrt{n}}\,
\log(d_1d_3+d_2d_3),
\end{multline*}and hence completes the proof of \cref{thm:norm}.

\subsection{Proof of \texorpdfstring{\cref{lem:error_op_b}}{Lemma~X}}
\label{proof:error_op_b}

Without loss of generality, we bound $\|\tE_{\mathrm{rn},1}\|$ and $\|\tE_{\mathrm{init},1}\|$, the bounds for the $\|\tE_{\mathrm{rn},2}\|$ and $\|\tE_{\mathrm{init},2}\|$ follow by symmetry.

\paragraph{Step 1: Random-noise term $\tE_{\mathrm{rn},1}$}
Recall that
\begin{equation*}
    \tE_{\mathrm{rn},1}
    = \frac{d^*}{n_0} \sum_{i=n_0+1}^n \xi_i \tX_i,
\end{equation*}
where $\{(\xi_i,\tX_i)\}_{i=n_0+1}^n$ are i.i.d by the sampling mechanism in \cref{assump:sampling}. According to the definition of the tensor spectral norm and the t-SVD representation,
\begin{equation*}
    \|\tE_{\mathrm{rn},1}\|
    = \|\overbar{\fmE}_{\mathrm{rn},1}\|
    = \left\|\frac{d^*}{n_0} \sum_{i=n_0+1}^n \xi_i \overbar{\fmX}_i\right\|,
\end{equation*}
where $\overbar{\fmE}_{\mathrm{rn},1}:=\bdiag(\ftE_{\mathrm{rn},1})\in\mathbb{C}^{(d_1d_3)\times(d_2d_3)}$ and $ \overbar{\fmX}_i:=\bdiag(\ftX_i)\in\mathbb{C}^{(d_1d_3)\times(d_2d_3)}$ are block-diagonal matrices whose $s$-th diagonal block corresponds to the $s$-th ($s\in[d_3]$) frontal slice of the Fourier-transformed tensors $\ftE_{\mathrm{rn},1}$ and $\ftX_i$, respectively.

Thus, it suffices to bound the spectral norm of a sum of i.i.d. random matrices, to which we apply the matrix Bernstein inequality (see, e.g., \cite{koltchinskii2011neumann,minsker2017some,tropp2012user}).

Since $\xi_i$ is sub-Gaussian, we have
\begin{equation}\label{eq:Ernfirstmoment}
   \|\xi_i\|_{\psi_2}\lesssim \sigma_{\xi},
   \qquad
   \bigl\|\|\xi_i \overbar{\fmX}_i\|\bigr\|_{\psi_2}
   \leq \|\xi_i\|_{\psi_2}
   \lesssim \sigma_{\xi},
\end{equation}
where we use the fact that each $\overbar{\fmX}_i$ takes values in
\begin{equation*}
   \left\{
   \diag_{d_3}(e_{i_1})\,\fmW_k\,\diag_{d_3}(e_{i_2}^\top)
   : i_1\in[d_1],\, i_2\in[d_2],\, k\in[d_3]
   \right\},
\end{equation*}
with
\begin{equation*}
    \fmW_k
    = \operatorname{diag}\bigl(1, \omega^{k-1},\omega^{2(k-1)},\ldots,\omega^{(d_3-1)(k-1)}\bigr), \quad\omega=e^{-\frac{2\pi \sqrt{-1}}{d_3}}.
\end{equation*}

A direct calculation shows that\begin{align*}
\left\|\mathbb{E}\,\overbar{\fmX}_i\overbar{\fmX}_i^{H}\right\|
    &=
    \left\| \frac{1}{d_1 d_2 d_3} \sum_{i_1=1}^{d_1} \sum_{i_2=1}^{d_2}\sum_{k=1}^{d_3} 
    \diag_{d_3} (e_{i_1}) \fmW_k \diag_{d_3} (e_{i_2}^\top)
    \diag_{d_3} (e_{i_2}) \fmW_k^H \diag_{d_3} (e_{i_1}^\top)\right\|\\
    &=
    \left\| \frac{1}{d_1 d_2 d_3} \sum_{i_1=1}^{d_1} \sum_{i_2=1}^{d_2}\sum_{k=1}^{d_3} 
    \diag_{d_3} (e_{i_1}) \bigl(\fmW_k \fmW_k^H\bigr) \diag_{d_3} (e_{i_1}^\top)\right\|\\
    &=
    \left\| \frac{1}{d_1 d_2 d_3} \sum_{i_1=1}^{d_1} d_2
    \diag_{d_3} (e_{i_1}) \bigl(d_3I_{d_3}\bigr)\diag_{d_3} (e_{i_1}^\top)\right\|\\
    &=\frac{1}{d_1}\bigl\| I_{d_1d_3} \bigr\|
    =\frac{1}{d_1},\label{EmXH}
\end{align*}
where we used $\sum_{k=1}^{d_3} \fmW_k \fmW_k^H  = d_3I_{d_3}$. By symmetry, one similarly obtains
$\|\mathbb{E}\,\overbar{\fmX}_i^{H}\overbar{\fmX}_i\| = 1/d_2$.
Since $\xi_i$ and $\tX_i$ are independent, it follows that
\begin{equation}\label{eq:Ernsecmoment}
    \max\left\{ 
    \left\|\mathbb{E} \xi_i^2 \overbar{\fmX}_i\overbar{\fmX}_i^{H}\right\|,
    \left\|\mathbb{E} \xi_i^2 \overbar{\fmX}_i^{H}\overbar{\fmX}_i\right\|\right\}
    \leq \frac{\sigma_{\xi}^2}{d_1\wedge d_2}.
\end{equation}

With the first order bound \cref{eq:Ernfirstmoment} and the second moment bound \cref{eq:Ernsecmoment}, we can apply the matrix Bernstein inequality to the sum $\frac{1}{n_0} \sum_{i=n_0+1}^n \xi_i \overbar{\fmX}_i$.
We obtain that for any $t>0$, with probability at least $1-e^{-t}$,
\begin{multline*}
    \left\| \frac{1}{n_0}  \sum_{i=n_0+1}^n \xi_i \overbar{\fmX}_i\right\|
\leq
C_{\mathrm{Bernstein}}\Big(
\frac{\sigma_{\xi}}{\sqrt{d_1\wedge d_2}}
\sqrt{\frac{t+\log((d_1+d_2)d_3)}{n_0}}
\\\vee
\sigma_{\xi}\sqrt{\log(d_1\wedge d_2)}\,
\frac{t+\log((d_1+d_2)d_3)}{n_0}
\Big),
\end{multline*}
for some absolute constant $C_{\mathrm{Bernstein}}>0$.

Set $t=2\log(d_1d_3+d_2d_3)$ and assume
\begin{equation*}
    n_0 > 3(d_1\wedge d_2)\log(d_1\wedge d_2)\log(d_1d_3+d_2d_3),
\end{equation*}
so that the term in $n_0^{-1}$ is dominated by the square-root term. Thus, with probability at least $1-(d_1d_3+d_2d_3)^{-2}$,
\begin{equation*}
    \left\| \frac{1}{n_0}  \sum_{i=n_0+1}^n \xi_i \overbar{\fmX}_i\right\|
    \leq C_{\mathrm{Bernstein}} \sigma_{\xi} \sqrt{\frac{\log(d_1d_3+d_2d_3)}{n_0(d_1\wedge d_2)}}.
\end{equation*}
Scaling the sum by $d^*$ and using the relation between the matrix and tensor spectral norms, we obtain
\begin{equation}\label{eq:Ern}
    \|\tE_{\mathrm{rn},1}\|
   =\left\| \frac{d^*}{n_0}  \sum_{i=n_0+1}^n \xi_i \overbar{\fmX}_i\right\|
   \leq
    C_{\mathrm{Bernstein}}\sigma_{\xi}\sqrt{d_1 d_2(d_1\vee d_2)}\,d_3
    \sqrt{\frac{\log(d_1d_3+d_2d_3)}{n_0}},
\end{equation}
where in the last inequality we use $d^*=d_1d_2d_3$.
This yields the desired bound for $\tE_{\mathrm{rn},1}$.

\paragraph{Step 2: Initialization term $\tE_{\mathrm{init},1}$}
Recall that
\begin{equation}\label{eq:EinitDef}
    \tE_{\mathrm{init},1}
    = \frac{d^*}{n_0} \sum_{i=n_0+1}^n\left\langle{\tZ}_1, \tX_i\right\rangle \tX_i - {\tZ}_1.
\end{equation}By the design of \cref{alg:debias}, the initial estimator $\tT_{\mathrm{init},1}$ is computed by the sub-sample $\{(\tX_i,Y_i)\}_{i=1}^{n_0}$, thus its estimation error $\tZ_1=\tT_{\mathrm{init},1}-\tT$ is independent of the other half of the data including $\{\tX_i\}_{i=n_0+1}^n$. Hence, when quantifying the randomness of $\tX_i$ in \cref{eq:EinitDef}, we can treat $\tZ_1$ as fixed.

In the Fourier domain,
\begin{equation*}
    \left\|\tE_{\mathrm{init},1}\right\|
    =\|\overbar{\fmE}_{\mathrm{init},1}\|
    =\left\|\frac{\sum_{i=n_0+1}^n d^*\left\langle{\tZ}_1, \tX_i\right\rangle \overbar{\fmX}_i-\overbar{\fmZ}_1}{n_0}\right\|,
\end{equation*}
where $\overbar{\fmZ}_1:=\bdiag(\ftZ_1)$ is the block-diagonal Fourier representation of $\tZ_1$.

We first bound the (random) summands. We have
\begin{align*}
    \left\|d_1 d_2 d_3\left\langle{\tZ}_1, \tX_i\right\rangle \overbar{\fmX}_i-\overbar{\fmZ}_1\right\|
    &\leq d_1 d_2 d_3\|{\tZ}_1\|_{\max } \|\overbar{\fmX}_i\|
      +\|\overbar{\fmZ}_1\|\\
    &\leq d_1 d_2 d_3\|{\tZ}_1\|_{\max } 
      + \sqrt{(d_1\vee d_2)d_3}\,\|{\tZ}_1\|_{\max } \\
    &\leq 2d_1d_2d_3\|{\tZ}_1\|_{\max } .
\end{align*}

Next, we bound the variance proxies. A direct calculation gives
\begin{align*}
    & \left\|\mathbb{E}\Bigl(d_1 d_2d_3\langle{\tZ}_1, \tX_i\rangle \overbar{\fmX}_i-\overbar{\fmZ}_1\Bigr)
    \Bigl(d_1 d_2 d_3\langle{\tZ}_1, \tX_i\rangle \overbar{\fmX}_i-\overbar{\fmZ}_1\Bigr)^{H}\right\| \\
    &\quad  \leq\left\|d_1^2 d_2^2 d_3^2 \,\mathbb{E}\bigl[\langle{\tZ}_1, \tX_i\rangle^2 \overbar{\fmX}_i \overbar{\fmX}_i^{H}\bigr]\right\| \\
    &\quad  \leq d_1^2 d_2^2 d_3^2 \|\tZ_1\|_{\max}^2\cdot\left\|\mathbb{E} \overbar{\fmX}_i \overbar{\fmX}_i^{H}\right\| \\
    &\quad\leq d_1 d_2^2d_3^2 \|{\tZ}_1\|_{\max }^2,
\end{align*}
and similarly
\begin{equation*}
    \left\|\mathbb{E}\Bigl(d_1 d_2d_3\langle{\tZ}_1, \tX_i\rangle \overbar{\fmX}_i-\overbar{\fmZ}_1\Bigr)^{H}
    \Bigl(d_1 d_2 d_3\langle{\tZ}_1, \tX_i\rangle \overbar{\fmX}_i-\overbar{\fmZ}_1\Bigr)\right\|
    \leq d_1^2d_2d_3^2 \|{\tZ}_1\|_{\max }^2.
\end{equation*}

Applying the matrix Bernstein inequality to
$\frac{1}{n_0}\sum_{i=n_0+1}^n
    \bigl(d_1 d_2 d_3\langle\tZ_1, \tX_i\rangle \overbar{\fmX}_i-\overbar{\fmZ}_1\bigr)$,
we obtain that, with probability at least $1-e^{-t}$,
\begin{multline*}
      \left\|\frac{\sum_{i=n_0+1}^n d_1 d_2 d_3\langle\tZ_1, \tX_i\rangle \overbar{\fmX}_i-\overbar{\fmZ}_1}{n_0}\right\|
    \leq
    2\Biggl(
        \|{\tZ}_1\|_{\max } \sqrt{d_1d_2(d_1\vee d_2)}\,d_3
        \sqrt{\frac{t+\log (d_1d_3+d_2d_3)}{n_0}} \\
        \vee\; 2\|{\tZ}_1\|_{\max }d_1d_2d_3\,\frac{t+\log (d_1d_3+d_2d_3)}{n_0}
    \Biggr).
\end{multline*}
Choosing $t=2\log(d_1d_3+d_2d_3)$ and assuming
\begin{equation*}
    n_0 >12(d_1\wedge d_2)\log(d_1d_3+d_2d_3),
\end{equation*}
the linear term in $n_0^{-1}$ is dominated, and therefore, with probability at least $1-(d_1d_3+d_2d_3)^{-2}$,
\begin{equation}\label{eq:Einit}
     \left\|\tE_{\mathrm{init},1}\right\|
     \leq 
     2\|{\tZ}_1\|_{\max } \sqrt{d_1d_2(d_1\vee d_2)}\,d_3
     \sqrt{\frac{\log (d_1d_3+d_2d_3)}{n_0}}.
\end{equation}

Replacing $n_0$ with $n$ ($n=2n_0$) in \Cref{eq:Ern} and \cref{eq:Einit} affects only the absolute constants in the bounds. With this adjustment, the claim in \cref{lem:error_op_b} follows immediately.

\subsection{Proof of \texorpdfstring{\cref{thm:distance}}{Theorem~X}}
\label{proof:distance}

Recall that in \cref{eq:tensor_svd_rp}, $\tTheta$ and $\widehat{\tTheta}_a$ ($a=1,2$) are constructed as\begin{equation*}
    \tTheta:=\operatorname{diag}(\tU,\tV),
\qquad
\widehat{\tTheta}_a:=\operatorname{diag}(\widehat{\tU}_a,\widehat{\tV}_a).
\end{equation*}Then\begin{equation*}
    \tTheta\tTheta^\dagger
=\operatorname{diag}(\tU\tU^\dagger,\tV\tV^\dagger),
\qquad
\widehat{\tTheta}_a\widehat{\tTheta}_a^\dagger
=\operatorname{diag}(\widehat{\tU}_a\widehat{\tU}_a^\dagger,
\widehat{\tV}_a\widehat{\tV}_a^\dagger).
\end{equation*}
Hence, it suffices to control the row norms of
$\widehat{\tTheta}_a\widehat{\tTheta}_a^\dagger-\tTheta\tTheta^\dagger$, namely\begin{equation*}
    \max_{j\in[d_1]}\|\te_j^{\dagger}\,(\widehat{\tTheta}_a\widehat{\tTheta}_a^\dagger-\tTheta\tTheta^\dagger)\| \quad\text{and} \quad
    \max_{j\in[d_2]}\|\te_{d_1+j}^{\dagger}\,(\widehat{\tTheta}_a\widehat{\tTheta}_a^\dagger-\tTheta\tTheta^\dagger)\| .
\end{equation*}
Fix $a\in\{1,2\}$ and suppress the index $a$ throughout. 
By the representation formula in \cref{eq:tensor_svd_rp}, we have
\begin{equation}\label{eq:distance-expansion}
\widehat{\tTheta}\widehat{\tTheta}^{\dagger}-\tTheta\tTheta^{\dagger}
=
\sum_{k=1}^{\infty}
\sum_{\boldsymbol s:\,s_1+\cdots+s_{k+1}=k}
(-1)^{1+\tau(\boldsymbol s)}
\mathfrak P^{-s_1}\tP\mathfrak P^{-s_2}\cdots
\tP\mathfrak P^{-s_{k+1}}.
\end{equation}We next bound \begin{equation*}
    \max_{j\in[d_1]}\left\|\te_j^\dagger \, \mathfrak P^{-s_1}\tP\mathfrak P^{-s_2}\cdots
\tP\mathfrak P^{-s_{k+1}}\right\|
\quad\text{and}\quad
 \max_{j\in[d_2]}\left\|\te_{d_1+j}^\dagger \, \mathfrak P^{-s_1}\tP\mathfrak P^{-s_2}\cdots
\tP\mathfrak P^{-s_{k+1}}\right\|
\end{equation*}according to whether $s_1\ge 1$ or $s_1=0$.

\paragraph{Case 1: \(s_1\ge 1\)}
Fix $k\geq 1$ and $(s_1,\ldots,s_{k+1})$ satisfying
$s_1+\cdots+s_{k+1}=k$. Then
\begin{multline}\label{eq:thmAfactors}
     \left\| \te_j^{\dagger} \,
        \mathfrak{P}^{-s_1} \tP \mathfrak{P}^{-s_2} \cdots \mathfrak{P}^{-s_k} \tP \mathfrak{P}^{-s_{k+1}} \right\|\\
            \leq \bigl\|\te_j^{\dagger} \,\mathfrak{P}^{-s_1}\bigr\|\;
        \bigl\|\tP \mathfrak{P}^{-s_2} \cdots \mathfrak{P}^{-s_k} \tP \mathfrak{P}^{-s_{k+1}}\bigr\|
        \leq \bigl\|\te_j^{\dagger} \,\mathfrak{P}^{-s_1}\bigr\|\;
        \|\tP\|^k \|\tS^{-1}\|^{k-s_1}.
\end{multline}
Here,e the second inequality uses $s_2+\cdots+s_{k+1}=k-s_1$.

In this step, we consider the case $s_1\geq 1$. Recall from the definition of
$\mathfrak{P}^{-s}$ in \cref{eq:tensor_svd_rp} that, for $s\geq 1$, the first
$d_1$ rows of $\mathfrak P^{-s}$ are determined by $\tU$ and $\tS^{-s}$, while
the last $d_2$ rows are determined by $\tV$ and $\tS^{-s}$. Therefore, by the incoherence condition of $\tU$ and $\tV$,
\begin{equation}\label{eq:thmAincoU}
    \max_{j\in[d_1]} \bigl\|\te_j^{\dagger} \, \mathfrak{P}^{-s}\bigr\|
    \leq \max_{j\in[d_1]} \|\te_j^{\dagger} \, \tU\|\;\|\tS^{-s}\|
    \leq \mu_{\max}\sqrt{\frac{r}{d_1}}\;\|\tS^{-1}\|^{s},
\end{equation}and similarly,
\begin{equation}\label{eq:thmAincoV}
    \max_{j\in[d_2]} \bigl\|\te_{d_1+j}^{\dagger} \, \mathfrak{P}^{-s}\bigr\|
    \leq \max_{j\in[d_2]} \|\te_j^{\dagger} \, \tV\|\;\|\tS^{-s}\|
    \leq \mu_{\max}\sqrt{\frac{r}{d_2}}\;\|\tS^{-1}\|^{s}.
\end{equation}
Here, with a slight abuse of notation, $\te_j$ and $\te_{d_1+j}$ on the
left-hand sides denote canonical basis tensors in
$\mathbb{R}^{(d_1+d_2)\times 1\times d_3}$, while on the right-hand sides they
denote canonical basis tensors in $\mathbb{R}^{d_1\times 1\times d_3}$ and
$\mathbb{R}^{d_2\times 1\times d_3}$, respectively.

Next, recall from \cref{lem:error_op_b} that, with probability at least $1-2(d_1d_3+d_2d_3)^{-2}$,
\begin{equation*}
    \|\mathcal{E}\|
    \leq
    C_{\Bernstein}\,\sigma_{\xi}\,
    \sqrt{d_1d_2(d_1\vee d_2)}\,d_3
    \sqrt{\frac{\log(d_1d_3+d_2d_3)}{n}}
    \;=:\; \delta.
\end{equation*}Denote this event by $\Omega_0$. On $\Omega_0$, we have $\|\tP\|\leq \delta$
and $\|\tS^{-1}\|\leq 1/\lambda_{\min}$. Combining
\cref{eq:thmAfactors,eq:thmAincoU,eq:thmAincoV} yields, for $s_1\geq 1$,\begin{equation*}
    \max_{j\in[d_1]}
    \left\| \te_j^{\dagger} \,
    \mathfrak{P}^{-s_1} \tP \mathfrak{P}^{-s_2} \cdots \mathfrak{P}^{-s_k} \tP \mathfrak{P}^{-s_{k+1}} \right\|
    \leq
    \left(\frac{\delta}{\lambda_{\min}}\right)^k\,
    \mu_{\max}\sqrt{\frac{r}{d_1}},
\end{equation*}
and
\begin{equation*}
    \max_{j\in[d_2]}
    \left\| \te_{d_1+j}^{\dagger} \,
    \mathfrak{P}^{-s_1} \tP \mathfrak{P}^{-s_2} \cdots \mathfrak{P}^{-s_k} \tP \mathfrak{P}^{-s_{k+1}} \right\|
    \leq
    \left(\frac{\delta}{\lambda_{\min}}\right)^k\,
    \mu_{\max}\sqrt{\frac{r}{d_2}}.
\end{equation*}

\paragraph{Case 2: $s_1=0$}
When $s_1=0$, the direct incoherence argument in Step 1 \cref{eq:thmAincoU,eq:thmAincoV} is no longer available,
since the leading factors in $\mathfrak P^0$ are $\tU^{\perp}$ and $\tV^{\perp}$  rather than $\tU$ or $\tV$ with $s\ge 1$. Since $s_1+\cdots+s_{k+1}=k$, there exists an index
\begin{equation*}
   l_0 = \min\{l: s_{l+1}\geq 1\},\qquad 1\leq l_0\leq k.
\end{equation*}
Using $\mathfrak{P}^0 = \tI - \tTheta\tTheta^{\dagger} =: \mathfrak{P}^{\perp}$, we can write
\begin{align}
    &\left\| \te_j^{\dagger} \,
    (\mathfrak{P}^{\perp} \tP)^{l_0}
    \mathfrak{P}^{-s_{l_0+1}} \cdots \mathfrak{P}^{-s_k} \tP \mathfrak{P}^{-s_{k+1}} \right\| \nonumber\\
    &=\left\| \te_j^{\dagger} \,
    (\tI - \tTheta\tTheta^{\dagger})\tP
    (\mathfrak{P}^{\perp} \tP)^{l_0-1}
    \mathfrak{P}^{-s_{l_0+1}} \cdots \mathfrak{P}^{-s_k} \tP \mathfrak{P}^{-s_{k+1}} \right\| \nonumber\\
    &\leq
    \left\|\te_{j}^{\dagger}\, \tP(\mathfrak{P}^{\perp} \tP )^{l_0-1}
    \mathfrak{P}^{-s_{l_0+1}} \cdots \mathfrak{P}^{-s_k}  \tP\mathfrak{P}^{-s_{k+1}}\right\| \nonumber\\
    &\qquad\qquad
    +\left\|\te_j^{\dagger}\, \tTheta\tTheta^{\dagger}\tP(\mathfrak{P}^{\perp} \tP )^{l_0-1}
    \mathfrak{P}^{-s_{l_0+1}} \cdots \mathfrak{P}^{-s_k} \tP \mathfrak{P}^{-s_{k+1}}\right\| \nonumber\\
    &\leq \bigl\|\te_j^{\dagger}\,\tP(\mathfrak{P}^{\perp} \tP)^{l_0-1}\tTheta\bigr\|\,
    \|\tS^{-1}\|^{k}\|\tP\|^{k-l_0}
    + \bigl\|\te_j^{\dagger}\, \tTheta\bigr\|\,
    \|\tP\|^{k}\|\tS^{-1}\|^{k} \nonumber\\
    &\leq
    \bigl\|\te_j^{\dagger}\,\tP(\mathfrak{P}^{\perp} \tP)^{l_0-1}\tTheta\bigr\|
    \left(\frac{1}{\lambda_{\min}}\right)^{k}\delta^{k-l_0}
    + \bigl\|\te_j^{\dagger}\, \tTheta\bigr\|\,
    \left(\frac{\delta}{\lambda_{\min}}\right)^{k},
    \label{eq:thetadecom}
\end{align}where we again used $\|\tP\|\leq \delta$ and $\|\tS^{-1}\|\leq 1/\lambda_{\min}$
on $\Omega_0$.

The second term in \cref{eq:thetadecom} is controlled by the incoherence of $\tU$ and $\tV$ as factors in $\tTheta$:
\begin{equation*}   \max_{j\in[d_1]}\bigl\|\te_j^{\dagger}\, \tTheta\bigr\|
    \leq \mu_{\max}\sqrt{\frac{r}{d_1}},
    \qquad \max_{j\in[d_2]}\bigl\|\te_{d_1+j}^{\dagger}\, \tTheta\bigr\|
    \leq \mu_{\max}\sqrt{\frac{r}{d_2}}.
\end{equation*}
It remains to bound the first term $\|\te_j^{\dagger}\,\tP(\mathfrak{P}^{\perp} \tP)^{l_0-1}\tTheta\|$ in \cref{eq:thetadecom}.
This is summarized in the following lemma, whose proof is deferred to \cref{proof:s_1geq0}.

\begin{lemma}\label{lem:s_1geq0}
    On the event $\Omega_0$, for a fixed integer $l\geq1$ and for all integers $l_0$ with $1\leq l_0\leq l$, we have, with probability at least $1-l(d_1+d_2)^{-2}d_3^{-3}$,
    \begin{displaymath}
        \max_{j\in[d_1]}
        \bigl\|\te_j^{\dagger}\,\tP(\mathfrak{P}^{\perp} \tP)^{l_0-1}\tTheta\bigr\|
        \leq (C_1\delta)^{l_0}\,\mu_{\max}\sqrt{\frac{r}{d_1}},
    \end{displaymath}
        and
    \begin{displaymath}
        \max_{j\in[d_2]}
        \bigl\|\te_{d_1+j}^{\dagger}\,\tP(\mathfrak{P}^{\perp} \tP)^{l_0-1}\tTheta\bigr\|
        \leq (C_1\delta)^{l_0}\,\mu_{\max}\sqrt{\frac{r}{d_2}},
    \end{displaymath}
    where $\delta$ is defined in \cref{eq:delta} and $C_1$ is an absolute constant.
\end{lemma}

Combining \cref{lem:s_1geq0} with \cref{eq:thetadecom}, for any $0\leq l_0 \leq k$ we obtain
\begin{align*}
    &\max_{j\in[d_1]}
    \left\| \te_j^{\dagger}\,
    (\mathfrak{P}^{\perp} \tP)^{l_0}
    \mathfrak{P}^{-s_{l_0+1}} \cdots \mathfrak{P}^{-s_k} \tP \mathfrak{P}^{-s_{k+1}} \right\| \\
    &\leq
    (C_1\delta)^{l_0}\,\mu_{\max}\sqrt{\frac{r}{d_1}}
    \left(\frac{1}{\lambda_{\min}}\right)^k\delta^{k-l_0}
    + \mu_{\max}\sqrt{\frac{r}{d_1}}
    \left(\frac{\delta}{\lambda_{\min}}\right)^k \\
    &\leq
    \mu_{\max}\sqrt{\frac{r}{d_1}}
    \left(\frac{C_1\delta}{\lambda_{\min}}\right)^k.
\end{align*}
Thus, defined $k_{\max} = \lceil \log {(d_1\vee d_2)}\rceil$, for any $\boldsymbol{s}=(s_1,\ldots,s_{k+1})$ with $s_1+\cdots+s_{k+1}=k$ and $k\leq k_{\max}$,
\begin{equation*}
     \max_{j\in[d_1]}
     \left\| \te_j^{\dagger} \,
     \mathfrak{P}^{-s_1} \tP \mathfrak{P}^{-s_2} \cdots \mathfrak{P}^{-s_k} \tP \mathfrak{P}^{-s_{k+1}} \right\|
     \leq \mu_{\max}\sqrt{\frac{r}{d_1}}
     \left(\frac{C_1\delta}{\lambda_{\min}}\right)^k
\end{equation*}and
\begin{equation*}
     \max_{j\in[d_2]}
     \left\| \te_{d_1+j}^{\dagger}\,
     \mathfrak{P}^{-s_1} \tP \mathfrak{P}^{-s_2} \cdots \mathfrak{P}^{-s_k} \tP \mathfrak{P}^{-s_{k+1}} \right\|
     \leq \mu_{\max}\sqrt{\frac{r}{d_2}}
     \left(\frac{C_1\delta}{\lambda_{\min}}\right)^k
\end{equation*}
hold, on $\Omega_0$, with probability at least $1-k_{\max}(d_1+d_2)^{-2}d_3^{-3}$.

\medskip

We obtained bounds for every term in \cref{eq:distance-expansion}, covering both cases $s_1\geq 1$ and $s_1=0$.
We now sum over $k$.
\begin{multline*}
    \max_{j\in[d_1]}
        \|\te_j^{\dagger}\,
        (\widehat{\tTheta} \widehat{\tTheta}^{\dagger}-{\tTheta} \tTheta^{\dagger})\|\leq
        \sum_{k=1}^{k_{\max}}
        \sum_{\boldsymbol{s}:\, s_1+\cdots+s_{k+1}=k}
        \max_{j\in[d_1]}
        \left\| \te_j^{\dagger} \,
        \mathfrak{P}^{-s_1} \tP \cdots \mathfrak{P}^{-s_k} \tP \mathfrak{P}^{-s_{k+1}} \right\| \\
        + \sum_{k=k_{\max}+1}^{\infty}
        \sum_{\boldsymbol{s}:\, s_1+\cdots+s_{k+1}=k}
        \max_{j\in[d_1]}
        \left\| \te_j^{\dagger} \,
        \mathfrak{P}^{-s_1} \tP \cdots \mathfrak{P}^{-s_k} \tP \mathfrak{P}^{-s_{k+1}} \right\|.
\end{multline*}
For each fixed $k$,
\begin{equation*}
    \operatorname{Card}\bigl\{(s_1,\ldots,s_{k+1})\big|
    s_1+\cdots+s_{k+1}=k,\ s_1\geq0,\ldots,s_{k+1}\geq0\bigr\}
    \leq 4^k.
\end{equation*}
Therefore, on $\Omega_0$, with probability at least $1-k_{\max}(d_1+d_2)^{-2}d_3^{-3}$,\begin{align*}
        \max_{j\in[d_1]}
        \|\te_j^{\dagger}\,
        (\widehat{\tTheta} \widehat{\tTheta}^{\dagger}-{\tTheta} \tTheta^{\dagger})\|
        &\leq \sum_{k=1}^{k_{\max}} 4^k
        \left(\frac{C_1\delta}{\lambda_{\min}}\right)^k
        \mu_{\max}\sqrt{\frac{r}{d_1}}
        + \sum_{k=k_{\max}+1}^{\infty}
        4^k \left(\frac{\delta}{\lambda_{\min}}\right)^k \\
\end{align*}
Recall that the signal-to-noise condition in \cref{assump:SNR} ensures that
$\delta/\lambda_{\min}$ is sufficiently small; in particular, after enlarging the absolute constant if necessary, we may assume
$8C_1\delta/\lambda_{\min}\leq 1$. Assume also that $C_2\geq 2$. Then\begin{align*}
        \max_{j\in[d_1]}
        \|\te_j^{\dagger}\,
        (\widehat{\tTheta} \widehat{\tTheta}^{\dagger}-{\tTheta} \tTheta^{\dagger})\|
        &\leq
        \frac{4 C_1 \delta}{\lambda_{\min}}
        \sum_{k=1}^{k_{\max}}\left(\frac{1}{2}\right)^{k-1}
        \mu_{\max}\sqrt{\frac{r}{d_1}}
        +
        \left(\frac{4\delta}{\lambda_{\min}}\right)^{k_{\max}}
        \sum_{k=1}^{\infty}\left(\frac{1}{2}\right)^k \\
        &\leq
        \frac{8 C_1 \delta}{\lambda_{\min}}
        \mu_{\max}\sqrt{\frac{r}{d_1}}
        +  \left(\frac{4\delta}{\lambda_{\min}}\right)^{\lceil \log {(d_1\vee d_2)}\rceil} \\
        &\leq
        16C_1\,\frac{\delta}{\lambda_{\min}}\,
        \mu_{\max}\sqrt{\frac{r}{d_1}}.
    \end{align*}Here, the last inequality absorbs the tail term into the leading term.
Recalling the definition of $\delta$ in \cref{eq:delta},
\begin{equation*}
    \delta
    = C_{\Bernstein} \,\sigma_{\xi}\, d_3
    \sqrt{d_1d_2(d_1\vee d_2)}
    \sqrt{\frac{\log(d_1d_3+d_2d_3)}{n}},
\end{equation*}
we obtain
\begin{equation*}
        \max_{j\in[d_1]}
     \|\te_j^{\dagger}\,
     (\widehat{\tTheta} \widehat{\tTheta}^{\dagger}-{\tTheta} \tTheta^{\dagger})\|
     \leq
     C_1\,
     \frac{\sigma_{\xi}}{\lambda_{\min}}\,d_3\,
     \sqrt{\frac{d_1d_2(d_1\vee d_2)}{n}}\,
     \sqrt{\log(d_1d_3+d_2d_3)}\,
     \mu_{\max}\sqrt{\frac{r}{d_1}},
\end{equation*}holds with probability at least $1-\lceil \log {(d_1\vee d_2)}\rceil(d_1+d_2)^{-2}d_3^{-3}$, where we enlarge $C_1$ if necessary to absorb other absolute constants.
By symmetry, under the same event,
\begin{equation*}
        \max_{j\in[d_2]}
    \|\te_{d_1+j}^{\dagger}\,
    (\widehat{\tTheta} \widehat{\tTheta}^{\dagger}-{\tTheta} \tTheta^{\dagger})\|
    \leq
    C_1\,
    \frac{\sigma_{\xi}}{\lambda_{\min}}\,d_3\,
    \sqrt{\frac{d_1d_2(d_1\vee d_2)}{n}}\,
    \sqrt{\log(d_1d_3+d_2d_3)}\,
    \mu_{\max}\sqrt{\frac{r}{d_2}}.
\end{equation*}
Recalling the block-diagonal forms of
$\tTheta\tTheta^\dagger$ and $\widehat{\tTheta}\widehat{\tTheta}^\dagger$, the
last two displays are exactly the desired bounds for
$\max_{j\in[d_1]}\|\te_j^{\dagger}\,(\widehat{\tU}_a\widehat{\tU}_a^\dagger-\tU\tU^\dagger)\|$ and
$\max_{j\in[d_2]}\|\te_j^{\dagger}\,(\widehat{\tV}_a\widehat{\tV}_a^\dagger-\tV\tV^\dagger)\|$.
Since the argument is identical for $a=1,2$, a final union bound over
$a\in\{1,2\}$ completes the proof.

\subsection{Proof of \texorpdfstring{\cref{lem:proj_error_term}}{Lemma~X}}
\label{proof:proj_error_term}
We begin by bounding the term $\bigl|\langle \widehat{\tU}_1\widehat{\tU}_1^{\dagger} \tE_1\widehat{\tV}_1\widehat{\tV}_1^{\dagger}, \tM\rangle\bigr|$. The bound for the analogous second term, $\bigl|\langle \widehat{\tU}_1\widehat{\tU}_2^{\dagger} \tE_2\widehat{\tV}_2\widehat{\tV}_2^{\dagger}, \tM\rangle\bigr|$, will follow by a similar argument.

By Hölder's inequality,\begin{equation*}
    \bigl|\langle \widehat{\tU}_1\widehat{\tU}_1^{\dagger} \tE_1\widehat{\tV}_1\widehat{\tV}_1^{\dagger}, \tM\rangle\bigr|
\le \|\tM\|_{\ell_1}\,
\bigl\|\widehat{\tU}_1\widehat{\tU}_1^{\dagger} \tE_1\widehat{\tV}_1\widehat{\tV}_1^{\dagger}\bigr\|_{\max}.
\end{equation*} Because $\tM$ is fixed, it suffices to bound the max norm of the projected error $\widehat{\tU}_1\widehat{\tU}_1^{\dagger} \tE_1\widehat{\tV}_1\widehat{\tV}_1^{\dagger}$. 

To do this, we decompose the error by introducing the singular subspace perturbations. 
Recall that $\tU\tU^{\dagger}$ is the projection onto the left singular subspace of the true tensor $\tT$, while $\tU_1\tU_1^{\dagger}$ is the projection onto the left singular subspace of the estimator $\tT_{\mathrm{proj},1}$. We define these perturbations of the singular subspaces as:\begin{equation*}
    \Delta_{\tU}
=\widehat{\tU}_1\widehat{\tU}_1^{\dagger}-\tU\tU^{\dagger},
\qquad
\Delta_{\tV}
=\widehat{\tV}_1\widehat{\tV}_1^{\dagger}-\tV\tV^{\dagger}.
\end{equation*}
Here we omit the subtitle $1$ of $\tU_1$. Using these definitions, the projected error can be explicitly expanded into:
\begin{equation}\label{eq:split}
\widehat{\tU}_1\widehat{\tU}_1^{\dagger} \tE_1\widehat{\tV}_1\widehat{\tV}_1^{\dagger}
= \underbrace{\tU\tU^{\dagger}\tE_1\tV\tV^{\dagger}}_{\text{Oracle term}}
  + \underbrace{\Delta_{\tU}\tE_1\tV\tV^{\dagger}
  + \tU\tU^{\dagger}\tE_1\Delta_{\tV}
  + \Delta_{\tU}\tE_1\Delta_{\tV}}_{\text{perturbation terms}}.  
\end{equation}Our goal is to bound the four terms in \cref{eq:split} entrywise.


\medskip\noindent
\paragraph{Oracle term}We first analyze the oracle term $\tU\tU^{\dagger}\tE_1\tV\tV^{\dagger}$ in \cref{eq:split}. Recall the error $\tE_1$ is made up by two parts, $\tE_1:=\tE_{\mathrm{init},1}+\tE_{\mathrm{rn},1}$, and\begin{equation*}
   \tE_{\mathrm{rn},1}
:= \frac{d^*}{n_0} \sum_{i=n_0+1} ^{n_0} \xi_i \tX_i ,
\qquad
\tE_{\mathrm{init},1}
:= \frac{d^*}{n_0} \sum_{i=n_0+1} ^{n_0}\langle \tZ_1, \tX_i \rangle \tX_i - \tZ_1.
\end{equation*}Here, each sampling tensor takes the form $\tX_i=\te_{i_1}\tte_k\te_{i_2}^{\dagger}$ for some
$i_1\in[d_1]$, $i_2\in[d_2]$, $k\in[d_3]$, where $\tte_k\in\mathbb{R}^{1\times 1\times d_3}$ is the tube basis with its ${(1,1,k)}$ entry being one and zeros elsewhere.
Thus, the projected error can be decomposed into
\begin{equation}\label{eq:UUEVV}
\tU\tU^{\dagger}\tE_1\tV\tV^{\dagger}\,=\,\tU\tU^{\dagger}\tE_{\mathrm{rn},1}\tV\tV^{\dagger}\,+\,\tU\tU^{\dagger}\tE_{\mathrm{init},1}\tV\tV^{\dagger}.
\end{equation} We will use the Bernstein inequality to bound the max norm of the two parts in \cref{eq:UUEVV}. For the projected random noise part,\begin{equation*}
\tU\tU^{\dagger}\tE_{\mathrm{rn},1}\tV\tV^{\dagger}\,=\,\frac{d^*}{n_0}\sum_{i=n_0+1} ^{n_0}\xi_i\tU\tU^{\dagger}\tX_i\tV\tV^{\dagger}.
\end{equation*} 
For any fixed entry $(j_1,j_2,t)$ of $\tU\tU^{\dagger}\tE_{\mathrm{rn},1}\tV\tV^{\dagger}$, it can be written as\begin{equation*}
    \bigl\langle\tU\tU^{\dagger} \tE_{\mathrm{rn},1}\tV\tV^{\dagger}, \te_{j_1}\tte_t\te_{j_2}^{\dagger}\bigr\rangle\,=\,\frac{d^*}{n_0}\sum_{i=n_0+1} ^{n_0}\xi_i\bigl\langle\tU\tU^{\dagger} \tE_{\mathrm{rn},1}\tV\tV^{\dagger}, \te_{j_1}\tte_t\te_{j_2}^{\dagger}\bigr\rangle.
\end{equation*}The second moment can be bounded by\begin{equation*}
    \mathbb{E}\Bigl[\xi_i^2
  \bigl\langle\tU\tU^{\dagger} \tX_i\tV\tV^{\dagger},
              \te_{j_1}\tte_t\te_{j_2}^{\dagger}\bigr\rangle^2\Bigr]
= \frac{\sigma_{\xi}^2}{d^*}\,
  \bigl\|\tU\tU^{\dagger}\te_{j_1}\tte_t\te_{j_2}^{\dagger}\tV\tV^{\dagger}\bigr\|_{\mathrm{F}}^2
\le \frac{\sigma_{\xi}^2}{d^*}\,\mu_{\max}^4\frac{r^2}{d_1d_2},
\end{equation*}where we used the incoherence condition stated in \cref{assump:incoh}. Similarly, the sub-Gaussian norm of the summand satisfies\begin{equation*}
    \bigl\|\xi_i
  \bigl\langle\tU\tU^{\dagger} \tX_i\tV\tV^{\dagger},
              \te_{j_1}\tte_t\te_{j_2}^{\dagger}\bigr\rangle\bigr\|_{\psi_2}
\le \sigma_{\xi}\mu_{\max}^4\frac{r^2}{d_1d_2}.
\end{equation*} Now by Bernstein inequality, for any $t>0$, when $n>2\mu_{\max}^4r^2d_3\log(d_3)\big(t+\log(d_1d_3+d_2d_3)\big)$, we have,\begin{equation*}
    \mathbb{P}\left( \Big| \bigl\langle\tU\tU^{\dagger} \tE_{\mathrm{rn},1}\tV\tV^{\dagger}, \te_{j_1}\tte_t\te_{j_2}^{\dagger}\bigr\rangle\Big|\,\leq\,
   C_{\mathrm{Bernstein}}\mu_{\max}^2\sigma_{\xi}r\sqrt{d_3}\sqrt{\frac{t+\log(d_1d_3+d_2d_3)}{n}}\right)\geq 1-e^{-t}.
\end{equation*}Let $t=\log d^*+2\log(d_1d_3+d_2d_3)$, when $n>8\mu_{\max}^4r^2d_3\log(d_3)\log(d_1d_3+d_2d_3)$, we have\begin{equation}\label{eq:lem2rn}
    \Big| \bigl\langle\tU\tU^{\dagger} \tE_{\mathrm{rn},1}\tV\tV^{\dagger}, \te_{j_1}\tte_t\te_{j_2}^{\dagger}\bigr\rangle\Big|\,\leq\,
   C_{\mathrm{Bernstein}}\mu_{\max}^2\sigma_{\xi}r\sqrt{\frac{d_3\log(d_1d_3+d_2d_3)}{n}},
\end{equation}holds with probability at least $1-(d^*)^{-1}(d_1d_3+d_2d_3)^{-2}.$

For the projected $\tE_{\mathrm{init,1}}$, \begin{equation*}
{\tU\tU^{\dagger}\tE_{\mathrm{rn},1}\tV\tV^{\dagger}}\,=\,\frac{d^*}{n_0}\sum_{i=n_0+1} ^{n_0}\langle \tZ_1, \tX_i \rangle \tU\tU^{\dagger}\tX_i\tV\tV^{\dagger}-\langle \tU\tU^{\dagger}\tZ_1\tV\tV^{\dagger}, \tX_i \rangle .
\end{equation*}
An analogous argument, with $\xi_i$ replaced by $\langle\tZ_1,\tX_i\rangle$,
yields the same type of bound in \cref{eq:lem2rn} with $\sigma_{\xi}$ replaced by $\|\tZ_1\|_{\max}$. Thus we have, for any fixed entry $(j_1,j_2,t)$,\begin{equation}\label{eq:lem2init}
    \Big| \bigl\langle\tU\tU^{\dagger} \tE_{\mathrm{init},1}\tV\tV^{\dagger}, \te_{j_1}\tte_t\te_{j_2}^{\dagger}\bigr\rangle\Big|\,\leq\, C_{\mathrm{Bernstein}}\mu_{\max}^2\|\tZ_1\|_{\max}r\sqrt{\frac{d_3\log(d_1d_3+d_2d_3)}{n}}.
\end{equation}holds when $n>8\mu_{\max}^4r^2d_3\log d_3\log(d_1d_3+d_2d_3)$ with probability at least $1-(d^*)^{-1}(d_1d_3+d_2d_3)^{-2}$.

Since \cref{eq:lem2rn,eq:lem2init} hold simultaneously for all
$j_1\in[d_1]$, $j_2\in[d_2]$, $t\in[d_3]$ after a union bound,\begin{equation*}
    \bigl\|\tU\tU^{\dagger}\tE_1\tV\tV^{\dagger}\bigr\|_{\max}
\le C_{\mathrm{Bernstein}}\mu_{\max}^2(\sigma_{\xi}+\|\tZ_1\|_{\max}) r
   \sqrt{\frac{d_3\log(d_1d_3+d_2d_3)}{n}}
\end{equation*}
with probability at least $1-(d_1d_3+d_2d_3)^{-2}$.
Under \cref{assump:init_entry},\begin{equation*}
    \mathbb{P}\bigl(\|\tZ_1\|_{\max}\le C_{\mathrm{init}}\,\gamma_n\sigma_{\xi}\bigr)
\ge 1-(d_1d_3+d_2d_3)^{-2},
\end{equation*}where $C_{\mathrm{init}}>0$ is an absolute constant.
Thus, on the intersection of these events, after enlarging the constant if necessary, we have
\begin{equation*}
\bigl\|\tU\tU^{\dagger}\tE_1\tV\tV^{\dagger}\bigr\|_{\max}
\le
C_{\mathrm{Bernstein}}\mu_{\max}^2(1+\gamma_n)\sigma_{\xi} r
\sqrt{\frac{d_3\log(d_1d_3+d_2d_3)}{n}}.
\end{equation*}Since $\gamma_n\le 1$, it can be absorbed by the absolute constant. We thus conclude for the oracle term in \cref{eq:split},\begin{equation}\label{eq:oracle-max}
    \bigl\|\tU\tU^{\dagger}\tE_1\tV\tV^{\dagger}\bigr\|_{\max}
\le
C\mu_{\max}^2\sigma_{\xi} r
\sqrt{\frac{d_3\log(d_1d_3+d_2d_3)}{n}},
\end{equation}holds with probability at least $1-2(d_1d_3+d_2d_3)^{-2}$.

\medskip\noindent
\paragraph{Perturbation terms}
For the perturbation terms, the entrywise max norm can be reduced to a maximum over tubes. For example,
\begin{align*} \bigl\|\Delta_{\tU}\tE_1\tV\tV^{\dagger}\bigr\|_{\max}&=\max_{j_1\in[d_1],\,j_2\in[d_2]}\bigl\|\te_{j_1}^{\dagger}\Delta_{\tU}\tE_1\tV\tV^{\dagger}\te_{j_2}
   \bigr\|_{\max} \\
&\le\max_{j_1\in[d_1],\,j_2\in[d_2]}\bigl\|\te_{j_1}^{\dagger}\Delta_{\tU}\tE_1\tV\tV^{\dagger}\te_{j_2}
   \bigr\|,
\end{align*}where, for each $j_1\in[d_1]$ and $j_2\in[d_2]$,\[
\te_{j_1}^{\dagger}\Delta_{\tU}\tE_1\tV\tV^{\dagger}\te_{j_2}
\]
is a $1\times 1\times d_3$ tensor (i.e., a tube), and for any tube, its max norm is bounded above by its spectral norm. The same argument applies to the other perturbation terms. Thus, it suffices to bound the spectral norm of each corresponding tube.

Conditional on the event of \cref{thm:distance},
the deviation projections $\Delta_{\tU}$ and $\Delta_{\tV}$ satisfy
rowwise bounds of order
\[\mu_{\max}\sigma_{\xi}\lambda_{\min}^{-1}
  \sqrt{r(d_1\vee d_2)d_3\log(d_1d_3+d_2d_3)/n}.\]
Combining these bounds with the spectral norm bound for $\tE_1$ from \cref{lem:error_op_b} and the incoherence bounds on $\tU$ and $\tV$ from \cref{assump:incoh}, we obtain
\begin{multline}
    \label{eq:split-1}\bigl\|\te_{j_1}^{\dagger}\Delta_{\tU}\tE_1\tV\tV^{\dagger}\te_{j_2}\bigr\|\leq \bigl\|\te_{j_1}^{\dagger}\Delta_{\tU}\bigr\|\cdot\|\tE_1\|\cdot \bigl\| \tV^{\dagger}\te_{j_2}\bigr\|\\\le
C\,\mu_{\max}^2
\frac{\sigma_{\xi}^2}{\lambda_{\min}}\,
\frac{r\sqrt{d_1d_2}(d_1\vee d_2)d_3^2}{n}\,
\log(d_1d_3+d_2d_3).
\end{multline}Similarly,
\begin{equation}\label{eq:split-2}
\bigl\|\te_{j_1}^{\dagger}\tU\tU^{\dagger}\tE_1\Delta_{\tV}\te_{j_2}\bigr\|
\le
C\,\mu_{\max}^2
\frac{\sigma_{\xi}^2}{\lambda_{\min}}\,
\frac{r\sqrt{d_1d_2}(d_1\vee d_2)d_3^2}{n}\,
\log(d_1d_3+d_2d_3),
\end{equation}
and
\begin{equation}\label{eq:split-3}
\bigl\|\te_{j_1}^{\dagger}\Delta_{\tU}\tE_1\Delta_{\tV}\te_{j_2}\bigr\|
\le
C\,\mu_{\max}^2
\frac{\sigma_{\xi}^3}{\lambda_{\min}^2}\,
\frac{r d_1d_2 (d_1\vee d_2)^{3/2} d_3^3}{n^{3/2}}\,
\bigl(\log(d_1d_3+d_2d_3)\bigr)^{3/2}.
\end{equation}
Under the SNR condition\begin{equation*}
    \frac{\lambda_{\min}}{\sigma_\xi}\ \ge\ C\, \frac{d_3(d_1\vee d_2)^{3/2}}{\sqrt{n}}\,
    \sqrt{\log\!\big(d_1d_3+d_2d_3\big)},
\end{equation*}
the bound in \cref{eq:split-3} is of smaller order and can be absorbed into
\cref{eq:split-1,eq:split-2}. Thus, combining
\cref{eq:split,eq:oracle-max,eq:split-1,eq:split-3},
we obtain\begin{multline}\label{eq:split-123}
\bigl\|\widehat{\tU}_1\widehat{\tU}_1^{\dagger} \tE_1\widehat{\tV}_1\widehat{\tV}_1^{\dagger}\bigr\|_{\max}
\le
C\mu_{\max}^2\sigma_{\xi} r
\sqrt{\frac{d_3\log(d_1d_3+d_2d_3)}{n}} \\
\quad
+ C\,\mu_{\max}^2
\frac{\sigma_{\xi}^2}{\lambda_{\min}}\,
\frac{r\sqrt{d_1d_2}(d_1\vee d_2)d_3^2}{n}\,
\log(d_1d_3+d_2d_3),
\end{multline}
with probability at least $1-3(d_1d_3+d_2d_3)^{-2}$, on the intersection of
the events of \cref{lem:error_op_b,thm:distance}.

Finally, plugging \cref{eq:split-123} into the initial Hölder bound yields exactly the bound claimed in \cref{lem:proj_error_term}.

\subsection{Proof of \texorpdfstring{\cref{lem:normal-t1}}{Lemma~X}}
\label{proof:normal-t1}
We aim to establish the normal approximation for the quantity:
\begin{align*}
     S :&=\left\langle \tU_{\perp} \tU_{\perp}^{\dagger}\frac{\tE_1+\tE_2}{2} \tV \tV^{\dagger}, \tM\right\rangle + \left\langle \tU \tU^{\dagger}\frac{\tE_1+\tE_2}{2} \tV_{\perp} \tV_{\perp}^{\dagger}, \tM\right\rangle \\
     &=\underbrace{\left\langle \tU_{\perp} \tU_{\perp}^{\dagger}\frac{\tE_{\mathrm{rn},1}+\tE_{\mathrm{rn},2}}{2} \tV \tV^{\dagger}, \tM\right\rangle + \left\langle \tU \tU^{\dagger}\frac{\tE_{\mathrm{rn},1}+\tE_{\mathrm{rn},2}}{2} \tV_{\perp} \tV_{\perp}^{\dagger}, \tM\right\rangle}_{S_{\mathrm{rn}}} \\
    &\quad+ \underbrace{\left\langle \tU_{\perp} \tU_{\perp}^{\dagger}\frac{\tE_{\mathrm{init},1}+\tE_{\mathrm{init},2}}{2} \tV \tV^{\dagger}, \tM\right\rangle + \left\langle \tU \tU^{\dagger}\frac{\tE_{\mathrm{init},1}+\tE_{\mathrm{init},2}}{2} \tV_{\perp} \tV_{\perp}^{\dagger}, \tM\right\rangle}_{S_{\mathrm{init}}},
     \end{align*}
where we utilized the decomposition of the error $\tE_k$ into the initialization error $\tE_{\mathrm{init},k}$ and the i.i.d. random noise $\tE_{\mathrm{rn},k}$.

\paragraph{Step 1: Analysis of Random Noise Term ($S_{\mathrm{rn}}$)}
Recall that the random noise components are defined as:
\begin{equation*}
    \tE_{\mathrm{rn},1} = \frac{d^*}{n_0}\sum_{i=n_0+1}^n \xi_i\tX_i \quad \text{and} \quad \tE_{\mathrm{rn},2} = \frac{d^*}{n_0}\sum_{i=1}^{n_0} \xi_i\tX_i.
\end{equation*}
Substituting these into $S_{\mathrm{rn}}$, we have:
\begin{equation*} 
    S_{\mathrm{rn}} = \frac{d^*}{n} \sum_{i=1}^n \underbrace{\xi_i \left(\langle \tU_{\perp} \tU_{\perp}^{\dagger}\tX_i \tV \tV^{\dagger}, \tM\rangle + \langle \tU \tU^{\dagger}\tX_i \tV_{\perp} \tV_{\perp}^{\dagger}, \tM\rangle\right)}_{S_{\mathrm{rn},i}}.
\end{equation*}
This expression is a sum of i.i.d. random variables \begin{equation*}
    S_{\mathrm{rn},i}:= \xi_i (\langle \tU_{\perp} \tU_{\perp}^{\dagger}\tX_i \tV \tV^{\dagger}, \tM\rangle + \langle \tU \tU^{\dagger}\tX_i \tV_{\perp} \tV_{\perp}^{\dagger}, \tM\rangle).
\end{equation*} To apply the Berry-Esseen theorem, we compute the second and third moments of these terms.

\textit{Second Moment:}\begin{multline} \label{eq:sec-m}
    \mathbb{E}\left[S_{\mathrm{rn},i}^2\right] 
        = \sigma_{\xi}^2 \left( \mathbb{E} \langle \tU_{\perp} \tU_{\perp}^{\dagger}\tX_i \tV \tV^{\dagger}, \tM\rangle^2 + \mathbb{E}\langle \tU \tU^{\dagger}\tX_i \tV_{\perp} \tV_{\perp}^{\dagger}, \tM\rangle^2 \right) \\
        \quad + 2\sigma_{\xi}^2 \cdot \mathbb{E}\left[\langle \tU_{\perp} \tU_{\perp}^{\dagger}\tX_i \tV \tV^{\dagger}, \tM\rangle\langle \tU \tU^{\dagger}\tX_i \tV_{\perp} \tV_{\perp}^{\dagger}, \tM\rangle\right].
\end{multline}
For the interaction term, we have:\begin{align*}
     \mathbb{E}\left[\langle \tU_{\perp} \tU_{\perp}^{\dagger}\tX_i \tV \tV^{\dagger}, \tM\rangle\langle \tU \tU^{\dagger}\tX_i \tV_{\perp} \tV_{\perp}^{\dagger}, \tM\rangle\right]&=  \mathbb{E}\left[\langle \tX_i , \tU_{\perp} \tU_{\perp}^{\dagger}\tM\tV \tV^{\dagger}\rangle\langle \tX_i , \tU \tU^{\dagger}\tM\tV_{\perp} \tV_{\perp}^{\dagger}\rangle\right]
    \\&= \frac{1}{d_1d_2d_3}\langle\tU_{\perp} \tU_{\perp}^{\dagger}\tM\tV \tV^{\dagger}, \tU \tU^{\dagger}\tM\tV_{\perp} \tV_{\perp}^{\dagger} \rangle \\
    &= \frac{1}{d_1d_2d_3} \langle\tU_{\perp}^{\dagger}\tM\tV \tV^{\dagger}, \underbrace{\tU_{\perp}^{\dagger}\tU}_{0} \tU^{\dagger}\tM\tV_{\perp} \tV_{\perp}^{\dagger} \rangle = 0.
\end{align*}
For the squared terms, we obtain:
\begin{equation*}
     \mathbb{E} \langle \tU_{\perp} \tU_{\perp}^{\dagger}\tX_i \tV \tV^{\dagger}, \tM\rangle^2
   = \mathbb{E} \langle \tX_i , \tU_{\perp} \tU_{\perp}^{\dagger}\tM\tV \tV^{\dagger}\rangle^2=\frac{1}{d_1d_2d_3} \| \tU_{\perp} \tU_{\perp}^{\dagger}\tM\tV \tV^{\dagger}\|_{\mathrm{F}}^2 =\frac{1}{d_1d_2d_3} \|\tU_{\perp}^{\dagger}\tM\tV \|_{\mathrm{F}}^2.
\end{equation*}
Substituting these back into \cref{eq:sec-m}, the second moment is:
\begin{equation*}
\mathbb{E}\left[S_{\mathrm{rn},i}^2\right]   =\frac{\sigma_{\xi}^2}{d^*} (\|\tU_{\perp}^{\dagger}\tM\tV \|_{\mathrm{F}}^2+\|\tU^{\dagger}\tM\tV_{\perp} \|_{\mathrm{F}}^2),
\end{equation*}where $d^*=d_1d_2d_3$.
 Let $\tilde{s}_{\tM}^2 = \|\tU_{\perp}^{\top}\tM\tV \|_{\mathrm{F}}^2+\|\tU^{\top}\tM\tV_{\perp} \|_{\mathrm{F}}^2$.

\textit{Third Moment:}
By the incoherence assumption, for any sample $\tX_i$:\begin{multline}\label{eq:first-m}
     |\langle \tU_{\perp} \tU_{\perp}^{\dagger}\tX_i \tV \tV^{\dagger}, \tM\rangle| = |\langle\tU_{\perp}^{\dagger} \tX_i \tV,  \tU_{\perp}^{\dagger}\tM\tV \rangle| \\\leq \|\tU_{\perp}^{\dagger} \tX_i \tV \|_{\mathrm{F}} \cdot\|\tU_{\perp}^{\dagger}\tM\tV\|_{\mathrm{F}}\leq \mu_{\max} \sqrt{\frac{r}{d_2}}\|\tU_{\perp}^{\dagger}\tM\tV \|_{\mathrm{F}}.
\end{multline}
A similar bound holds for the second term: $|\langle \tU \tU^{\dagger}\tX_i \tV_{\perp} \tV_{\perp}^{\dagger}, \tM\rangle| \leq \|\tU^{\dagger}\tM\tV_{\perp} \|_{\mathrm{F}} \cdot \mu_{\max} \sqrt{r/d_1}$.
Using the sub-Gaussian assumption on $\xi$ and \cref{eq:first-m}:
\begin{equation*}
    \begin{aligned}
        \mathbb{E}&|S_{\mathrm{rn},i}|^3 \leq C\sigma_{\xi}^3 \cdot \left( \mathbb{E}|\langle \tU_{\perp} \tU_{\perp}^{\dagger}\tX_i \tV \tV^{\dagger}, \tM\rangle|^3 + \mathbb{E}|\langle \tU \tU^{\dagger}\tX_i \tV_{\perp} \tV_{\perp}^{\dagger}, \tM\rangle|^3 \right)\\
         &\leq C\sigma_{\xi}^3\mu_{\max}\sqrt{\frac{r}{d_1\wedge d_2}}\Big(\mathbb{E}\langle \tU_{\perp} \tU_{\perp}^{\dagger}\tX_i \tV \tV^{\dagger}, \tM\rangle^2\|\tU_{\perp}^{\dagger}\tM\tV \|_{\mathrm{F}}  + \mathbb{E}\langle \tU \tU^{\dagger}\tX_i \tV_{\perp} \tV_{\perp}^{\dagger}, \tM\rangle^2\|\tU^{\dagger}\tM\tV_{\perp} \|_{\mathrm{F}} \Big)
         \\
         &\leq C\sigma_{\xi}^3\mu_{\max}\frac{\sqrt{r}}{d^*\sqrt{d_1 \wedge d_2}}\left(\|\tU_{\perp}^{\dagger}\tM\tV \|_{\mathrm{F}}^3 + \|\tU^{\dagger}\tM\tV_{\perp} \|_{\mathrm{F}}^3\right).
    \end{aligned}
\end{equation*}
Applying the Berry-Esseen Theorem:\begin{equation}
    \begin{aligned}
        & \sup_{x\in\mathbb{R}} \left|\mathbb{P} \left(\frac{ S_{\mathrm{rn}} }{\sigma_{\xi}\tilde{s}_{\tM}\sqrt{d^* /n}} \leq x\right) -\Phi(x)\right|
     \\& \leq C_{\BE}\, \mu_{\max}\sqrt{\frac{rd_3(d_1\vee d_2)}{n}}\frac{\|\tU_{\perp}^{\dagger}\tM\tV \|_{\mathrm{F}}^3 + \|\tU^{\dagger}\tM\tV_{\perp} \|_{\mathrm{F}}^3}{(\|\tU_{\perp}^{\dagger}\tM\tV \|_{\mathrm{F}}^2 + \|\tU^{\dagger}\tM\tV_{\perp} \|_{\mathrm{F}}^2)^{\frac{3}{2}}} \\
        &\leq C_{\BE}\, \mu_{\max}\sqrt{\frac{rd_3(d_1\vee d_2)}{n}}.\label{eq:BESrn}
    \end{aligned}
\end{equation}

\paragraph{Step 2: Variance Replacement}The variance term in the application of Berry-Essen Theorem \cref{eq:BESrn} is $\tilde{s}_{\tM}=
(\|\tU_{\perp}^{\top}\tM\tV \|_{\mathrm{F}}^2+\|\tU^{\top}\tM\tV_{\perp} \|_{\mathrm{F}}^2 )^{1/2}$, we now replace it with an estimable proxy ${s}_{\tM} = (\|\tM\tV \|_{\mathrm{F}}^2+\|\tU^{\dagger}\tM \|_{\mathrm{F}}^2)^{1/2}$. We decompose the normalized term as:\begin{equation*}
     \frac{ S_{\mathrm{rn}} }{\sigma_{\xi}\tilde{s}_{\tM}\sqrt{d^*/n}}
        = \frac{ S_{\mathrm{rn}} }{\sigma_{\xi}{s}_{\tM}\sqrt{d^*/n}}
        + \frac{ S_{\mathrm{rn}} }{\sigma_{\xi}\tilde{s}_{\tM}\sqrt{d^*/n}} \left(1- \frac{\tilde{s}_{\tM}}{{s}_{\tM}}\right).
\end{equation*}
We bound the variance difference. Note that:
\begin{equation*} 
\begin{aligned}
    \|\tU^{\dagger}\tM\tV \|_{\mathrm{F}} &= \left\|\tU^{\dagger}\left(\sum_{j_1\in[d_1],j_2\in[d_2]} \dot{e}_{j_1} * \tM(j_1,j_2,:) * \dot{e}_{j_2}^{\dagger}\right)\tV\right\|_{\mathrm{F}} \\
    &\leq \sum_{j_1,j_2} \| \tU^{\dagger}e_{j_1} \| \cdot \|\tM(j_1,j_2,:)\|_{\mathrm{F}} \cdot \|e_{j_2}^{\dagger}\tV\| \\
    &\leq \mu_{\max}^2\frac{r}{\sqrt{d_1d_2}}\|\tM\|_{\ell_1}.
\end{aligned}
\end{equation*}
Using \cref{assump:M}, we obtain:
\begin{equation*} 
  0\leq 1- \frac{\tilde{s}_{\tM}}{{s}_{\tM}} \leq 1- \frac{\tilde{s}_{\tM}^2}{{s}_{\tM}^2}
    = \frac{\|\tU^{\dagger}\tM\tV \|_{\mathrm{F}}^2+\|\tU^{\dagger}\tM\tV \|_{\mathrm{F}}^2}{\|\tM\tV \|_{\mathrm{F}}^2+\|\tU^{\dagger}\tM \|_{\mathrm{F}}^2}
    \leq \frac{\mu_{\max}^4\|\tM\|_{\ell_1}^2}{\alpha_{M}^2\|\tM\|_{\mathrm{F}}^2}\cdot{\frac{r}{d_1\wedge d_2}}.
\end{equation*}Using Bernstein's inequality, when $n>6\mu_{\max}^2rd_3(d_1\vee d_2)\log^2(d_1d_3 +d_2d_3)$, there exists an event $\Omega_2$ with probability at least $1-2(d_1d_3+d_2d_3)^{-2}$ such that \begin{equation*}
    |S_{\mathrm{rn}}| \leq 2C_{\mathrm{Berstein}}\sigma_{\xi}\tilde{s}_{\tM}\sqrt{\frac{d^*}{n}\log(d_1d_3+d_2d_3)}.
\end{equation*} Under $\Omega_2$, the perturbation due to variance replacement is bounded by:
\begin{equation*}
\left|\frac{ S_{\mathrm{rn}} }{\sigma_{\xi}\tilde{s}_{\tM}\sqrt{d^*/n}}
- \frac{ S_{\mathrm{rn}} }{\sigma_{\xi}{s}_{\tM}\sqrt{d^*/n}}\right| \leq 2C_{\mathrm{Berstein}}\frac{\mu_{\max}^4\|\tM\|_{\ell_1}^2}{\alpha_{M}^2\|\tM\|_{\mathrm{F}}^2}\cdot{\frac{r\sqrt{\log(d_1d_3+d_2d_3)}}{d_1\wedge d_2}}.
\end{equation*}

\paragraph{Step 3: Initialization Error ($S_{\mathrm{init}}$)}
Recall $\tE_{\mathrm{init},1} = \frac{d^*}{n_0} \sum_{i=n_0+1}^n (\langle{\tZ}_1, \tX_i\rangle \tX_i-{\tZ}_1)$. The projection is linear, so:\begin{multline}
    \label{eq:Sinit_sum}
   \langle \tU_{\perp} \tU_{\perp}^{\dagger}\frac{\tE_{\mathrm{init},1}}{2} \tV \tV^{\dagger}, \tM\rangle\\= \frac{d^*}{n} \sum_{i=n_0+1}^n \left( \langle{\tZ}_1, \tX_i\rangle \langle\tU_{\perp} \tU_{\perp}^{\dagger}\tX_i \tV \tV^{\dagger} ,\tM\rangle - \langle \tU_{\perp} \tU_{\perp}^{\dagger}\tZ_1\tV \tV^{\dagger}, \tM\rangle \right),
\end{multline}and we have \begin{equation*}
    S_{\mathrm{init}}=\left\langle \tU_{\perp} \tU_{\perp}^{\dagger}\frac{\tE_{\mathrm{init},1}+\tE_{\mathrm{init},2}}{2} \tV \tV^{\dagger}, \tM\right\rangle + \left\langle \tU \tU^{\dagger}\frac{\tE_{\mathrm{init},1}+\tE_{\mathrm{init},2}}{2} \tV_{\perp} \tV_{\perp}^{\dagger}, \tM\right\rangle.
\end{equation*}
Using result \cref{eq:first-m}, the terms in \cref{eq:Sinit_sum} are bounded:
\begin{equation*}
    |\langle{\tZ}_1, \tX_i\rangle \langle\tU_{\perp} \tU_{\perp}^{\dagger}\tX_i \tV \tV^{\dagger} ,\tM\rangle - \langle \tU_{\perp} \tU_{\perp}^{\dagger}\tZ_1\tV \tV^{\dagger}, \tM\rangle| \leq 2d^*\|\tZ_1\|_{\max} \|\tU_{\perp}^{\dagger}\tM \tV\|_{\mathrm{F}} \mu_{\max}\sqrt{\frac{r}{d_2}}.
\end{equation*}
The variance of the the terms in \cref{eq:Sinit_sum} is bounded by\begin{align*}
     &\mathbb{E}\left(d^*\langle{\tZ}_1, \tX_i\rangle \cdot\langle\tU_{\perp} \tU_{\perp}^{\dagger}\tX_i \tV \tV^{\dagger} ,\tM\rangle - \langle \tU_{\perp} \tU_{\perp}^{\dagger}\tZ_1\tV \tV^{\dagger}, \tM\rangle\right)^2\\
    &\leq (d^*)^2\|\tZ_1\|_{\max}^2\cdot\mathbb{E}\langle\tU_{\perp} \tU_{\perp}^{\dagger}\tX_i \tV \tV^{\dagger} ,\tM\rangle^2\\
    &= d^*\|\tZ_1\|_{\max}^2\cdot\|\tU_{\perp}^{\dagger}\tM \tV\|^2_{\mathrm{F}}.
\end{align*}  By {Bernstein's inequality, when $n>6\mu_{\max}^2rd_3(d_1\vee d_2)\log^2(d_1d_3 +d_2d_3)$, with probability at least $1-2(d_1d_3+d_2d_3)^{-2}$:
\begin{equation*}
    \left| \frac{ S_{\mathrm{init}} }{\sigma_{\xi}\tilde{s}_{\tM}\sqrt{d^* /n}} \right| \leq 4\frac{\|Z_1\|_{\max}+\|Z_2\|_{\max}}{2}\sqrt{\log(d_1d_3+d_2d_3)}.
\end{equation*}
} Recall from \cref{assump:init_entry} that\begin{equation*}
    \mathbb{P}(\|\tZ_1\|_{\max}+\|\tZ_2\|_{\max}\leq C_{\mathrm{init}}\,\gamma_n\sigma_{\xi})\geq 1-(d_1d_3+d_2d_3)^{-2}.
\end{equation*}Therefore, with probability at least $1-2(d_1d_3+d_2d_3)^{-2}$:
\begin{equation*}
    \left| \frac{ S_{\mathrm{init}} }{\sigma_{\xi}\tilde{s}_{\tM}\sqrt{d^* /n}} \right| \leq 2C_{\mathrm{init}}\,\gamma_n\sqrt{\log(d_1d_3+d_2d_3)}.
\end{equation*}

\paragraph{Conclusion}
By Lipschitz property of $\Phi(x)$, the distributional approximation of the replacement of the variance term is\begin{multline*}
     \sup_{x\in\mathbb{R}} \left|\mathbb{P} \left(\frac{ S }{\sigma_{\xi}s_{\tM}\sqrt{d^* /n}} \leq x\right) -\Phi(x)\right|\\\leq 
     \sup_{x\in\mathbb{R}} \left|\mathbb{P} \left(\frac{ S_{\mathrm{rn}} }{\sigma_{\xi}\tilde s_{\tM}\sqrt{d^* /n}} \leq x\right) -\Phi(x)\right| + C_4 \left|\frac{ S_{\mathrm{init}} }{\sigma_{\xi}s_{\tM}\sqrt{d^* /n}} \right|,
\end{multline*}where $C_4>0$ is an absolute constant. Thus, we can combine the Berry-Esseen bound, the variance replacement error, and the initialization error to conclude:\begin{multline*}
     \sup_{x\in\mathbb{R}} \left|\mathbb{P} \left(\frac{ S }{\sigma_{\xi}s_{\tM}\sqrt{d^* /n}} \leq x\right) -\Phi(x)\right|\leq C_{\BE}\,\mu_{\max}\sqrt{\frac{rd_3(d_1\vee d_2)}{n}} + C_4\gamma_n\sqrt{\log(d_1d_3+d_2d_3)} \\+ C_4 \frac{\mu_{\max}^4\|\tM\|_{\ell_1}^2}{\alpha_{M}^2\|\tM\|_{\mathrm{F}}^2}\cdot{\frac{r\sqrt{\log(d_1d_3+d_2d_3)}}{d_1\wedge d_2}} + \frac{4}{(d_1d_3+d_2d_3)^2},
\end{multline*}where we enlarge the absolute constant $C_4$ if necessary.
This completes the proof.

\subsection{Proof of \texorpdfstring{\cref{lem:normal-t2}}{Lemma~X}}

\label{proof:normal-t2}
We next bound the higher-order expansion terms. Since $\widetilde{\tM}$ is constructed so that its only nonzero block is $\tM$ at upper left, entrywise Hölder's inequality yields
\begin{multline*}
    \left|\sum_{a\in\{1,2\}} \sum_{k=2}^{\infty}
    \left\langle
    \mathcal{S}_{A,k}(\tP_a)\tA{\tTheta}_a{\tTheta}_a^{\dagger}
    + \tA{\tTheta}_a{\tTheta}_a^{\dagger}\tA\mathcal{S}_{A,k}(\tP_a),
    \widetilde{\tM}
    \right\rangle\right| \\
    \le
    \|\tM\|_{\ell_1}
    \sum_{a\in\{1,2\}} \sum_{k=2}^{\infty}
    \max_{j_1\in[d_1],\,j_2\in[d_2]}
    \left\|
    \te_{j_1}^{\dagger}
    \left(
    \mathcal{S}_{A,k}(\tP_a)\tA{\tTheta}_a{\tTheta}_a^{\dagger}
    + \tA{\tTheta}_a{\tTheta}_a^{\dagger}\tA\mathcal{S}_{A,k}(\tP_a)
    \right)
    \te_{d_1+j_2}
    \right\|_{\max}.
\end{multline*}
Moreover, for every $a\in\{1,2\}$, $k\ge 2$, $j_1\in[d_1]$, and $j_2\in[d_2]$,\begin{equation*}
        \te_{j_1}^{\dagger}
    \left(
    \mathcal{S}_{A,k}(\tP_a)\tA{\tTheta}_a{\tTheta}_a^{\dagger}
    + \tA{\tTheta}_a{\tTheta}_a^{\dagger}\tA\mathcal{S}_{A,k}(\tP_a)
    \right)
    \te_{d_1+j_2}
\end{equation*}is a $1\times 1\times d_3$ tensor, i.e., a tube. Therefore,
\begin{multline*}
       \left\| \te_{j_1}^{\dagger}\left(
 \mathcal{S}_{A,k}(\tP_a)\tA{\tTheta}_a{\tTheta}_a^{\dagger}+ \tA{\tTheta}_a{\tTheta}_a^{\dagger}\tA\mathcal{S}_{A,k}(\tP_a)
\right)\te_{d_1+j_2}\right\|_{\max}\\
    \le\left\|\te_{j_1}^{\dagger}
    \left(\mathcal{S}_{A,k}(\tP_a)\tA{\tTheta}_a{\tTheta}_a^{\dagger}
    + \tA{\tTheta}_a{\tTheta}_a^{\dagger}\tA\mathcal{S}_{A,k}(\tP_a)\right)\te_{d_1+j_2} \right\|.
\end{multline*}Consequently,
\begin{multline}\label{eq:lemmaSM1.5}
    \left|\sum_{a\in\{1,2\}} \sum_{k=2}^{\infty}
    \left\langle\mathcal{S}_{A,k}(\tP_a)\tA{\tTheta}_a{\tTheta}_a^{\dagger}
    + \tA{\tTheta}_a{\tTheta}_a^{\dagger}\tA\mathcal{S}_{A,k}(\tP_a),
    \widetilde{\tM}
    \right\rangle\right| \\
    \le
    \|\tM\|_{\ell_1} \sum_{a\in\{1,2\}} \sum_{k=2}^{\infty}\max_{j_1\in[d_1],\,j_2\in[d_2]}\left\|\te_{j_1}^{\dagger}
    \left(\mathcal{S}_{A,k}(\tP_a)\tA{\tTheta}_a{\tTheta}_a^{\dagger}
    + \tA{\tTheta}_a{\tTheta}_a^{\dagger}\tA\mathcal{S}_{A,k}(\tP_a)\right)\te_{d_1+j_2}
    \right\|.
\end{multline}
We analyze the terms inside the summation in \cref{eq:lemmaSM1.5}. 

\paragraph{Bound of the summands by expansion of $\mathcal{S}_{A,k}$} Using the incoherence assumption, we can decouple the projection terms in \cref{eq:lemmaSM1.5}:
\begin{equation*}
    \begin{aligned}
        \|\te_{j_1}^{\dagger}\mathcal{S}_{A,k}(\tP_a) \tA{\tTheta}_a{\tTheta}_a^{\dagger}\te_{d_1+j_2}\|
        &\leq \|\te_{j_1}^{\dagger}\mathcal{S}_{A,k}(\tP_a) \tA{\tTheta}_a\| \cdot \|{\tTheta}_a^{\dagger}\te_{d_1+j_2}\| \\
        &\leq \|\te_{j_1}^{\dagger}\mathcal{S}_{A,k}(\tP_a) \tA{\tTheta}_a\|\cdot\mu_{\max}\sqrt{\frac{r}{d_2}} .
    \end{aligned}
\end{equation*}
Similarly, for the second term:
\begin{equation*}
        \|\te_{j_1}^{\dagger}{\tTheta}_a{\tTheta}_a^{\dagger}\tA\mathcal{S}_{A,k}(\tP_a) \te_{d_1+j_2}\|
        \leq \mu_{\max}\sqrt{\frac{r}{d_1}}\cdot \|{\tTheta}_a^{\dagger}\tA\mathcal{S}_{A,k}(\tP_a) \te_{d_1+j_2} \|.
\end{equation*}
Thus it left to bound $\|\te_{j}^{\dagger}\mathcal{S}_{A,k}(\tP_a) \tA{\tTheta}_a\|$ for all $j\in[d_1+d_2]$.
Recall the definition of the operator expansion, for any $k\geq2$, we have:
\begin{equation*}
   \mathcal{S}_{A,k}(\tP_a) = \sum_{\mathbf{s}:\, s_1+\cdots s_{k+1}=k}(-1)^{1+\tau(\mathbf{s})} \cdot \mathfrak{P}^{-s_1} \tP_a\mathfrak{P}^{-s_2} \cdots \mathfrak{P}^{-s_k} \tP_a\mathfrak{P}^{-s_{k+1}}.
\end{equation*}Thus, the term of our goal can be controlled by expansion:\begin{equation}
    \label{eq:goal_expansion}
  \bigl\|  \te_{j}^{\dagger}\,\mathcal{S}_{A,k}(\tP_a) \tA{\tTheta}_a\bigr\|\leq 
   \sum_{\mathbf{s}:\, s_1+\cdots s_{k+1}=k}  \bigl\|\te_{j}^{\dagger}\,\mathfrak{P}^{-s_1} \tP_a\mathfrak{P}^{-s_2} \cdots \mathfrak{P}^{-s_k} \tP_a\mathfrak{P}^{-s_{k+1}} \tA\tTheta_a\bigr\|.
\end{equation}
Consider the structure of $\tA$ in the frequency domain at slice $t$:
\begin{equation*}
    \fmA^{(t)}=\left(\begin{array}{cc}
    0 & {\fmT}^{(t)}  \\
    {\fmT}^{(t)H} & 0
    \end{array}\right).
\end{equation*}
Observe that $\mathfrak{P}^{\perp} \tA = 0$. Consequently, for terms in \cref{eq:goal_expansion} with $s_{k+1}=0$, they vanish since $\mathfrak{P}^{-s_{k+1}} \tA = \mathfrak{P}^{\perp} \tA = 0$. We restrict our attention to cases where $s_{k+1} \geq 1$. Applying \cref{lem:s_1geq0}, we have:\begin{multline*}
     \max_{j\in[d_1]}\|\te_{j}^{\dagger} \, \mathfrak{P}^{-s_1} \tP_a \mathfrak{P}^{-s_2} \cdots \mathfrak{P}^{-s_k} \tP_a\mathfrak{P}^{-s_{k+1}} \tA{\tTheta}_a\|\\
     \begin{aligned}
         &\leq \max_{j_1\in[d_1]}\|\te_{j}^{\dagger} \, \mathfrak{P}^{-s_1} \tP_a \mathfrak{P}^{-s_2} \cdots \mathfrak{P}^{-s_k} \tP_a\Theta\|\cdot\|\mathfrak{P}^{-s_{k+1}} \tA\|\\
         &\leq \mu_{\max}\sqrt{\frac{r}{d_1}}\left(\frac{C_2\delta}{\lambda_{\min}}\right)^{k-1}\delta,
     \end{aligned}
\end{multline*}
where $\delta$ is the upper bound of $\|\tE_a\|$ ($a=1,2$) defined in \cref{eq:delta}. Since for each fixed $k$,\[
\operatorname{Card}\left\{\left(s_1, \ldots, s_{k+1}\right) \mid s_1+\cdots+s_{k+1}=k, s_1 \geq 0, \cdots, s_k\geq0, s_{k+1} \geq 1\right\} \leq 4^{k-1} ,
\]we have for any fixed $k$,\begin{equation*}
    \max_{j\in[d_1]}   \bigl\|  \te_{j}^{\dagger}\,\mathcal{S}_{A,k}(\tP_a) \tA{\tTheta}_a\bigr\| \leq \mu_{\max}\sqrt{\frac{r}{d_1}}\left(\frac{4C_2\delta}{\lambda_{\min}}\right)^{k-1}\delta,
\end{equation*}and symmetrically\begin{equation*}
 \max_{j\in[d_2]}   \bigl\|  \te_{d_1+j}^{\dagger}\,\mathcal{S}_{A,k}(\tP_a) \tA{\tTheta}_a\bigr\| \leq \mu_{\max}\sqrt{\frac{r}{d_2}}\left(\frac{4C_2\delta}{\lambda_{\min}}\right)^{k-1}\delta.
\end{equation*} Thus we can conclude,\begin{multline*}
    \max_{j_1\in[d_1],\,j_2\in[d_2]}\left\|\te_{j_1}^{\dagger}
    \left(\mathcal{S}_{A,k}(\tP_a)\tA{\tTheta}_a{\tTheta}_a^{\dagger}
    + \tA{\tTheta}_a{\tTheta}_a^{\dagger}\tA\mathcal{S}_{A,k}(\tP_a)\right)\te_{d_1+j_2}
    \right\|\\\leq 2\mu_{\max}^2{\frac{r}{\sqrt{d_1d_2}}}\left(\frac{4C_2\delta}{\lambda_{\min}}\right)^{k-1}\delta
\end{multline*}hold for any $k\leq \log(d_1\vee d_2)$ under the event of \cref{thm:distance}.

\paragraph{Convergence analysis of the summation} We now bound the expansion in \cref{eq:lemmaSM1.5}.
We split the summation over $k$ into two regimes: $k \leq k_{\max}$ and $k > k_{\max}$ and choose $k_{\max}$ to be $k_{\max}=\log(d_1\vee d_2)$.

For small $k$ Regime ($2 \leq k \leq k_{\max}$):
Conditioned on the event where $\delta$ is small enough such that $8 C_4 \delta / \lambda_{\min} < 1$, we sum the geometric series:\begin{align*}
& \sum_{k=2}^{k_{\max }} \,\max _{j_1 \in[d_1],\,j_2 \in[d_2]}\left\|\dot{e}_{j_1}^{\dagger} \,\left(\mathcal{S}_{A, k}\left(\mathcal{P}_i\right) \mathcal{A} \Theta_i \Theta_i^{\dagger}+\mathcal{A} \Theta_i \Theta_i^{\dagger} \mathcal{A} \mathcal{S}_{A, k}\left(\mathcal{P}_i\right)\right) * \dot{e}_{d_1+j_2}\right\|\\&\leq
2\mu_{\max}^2\frac{r\delta}{\sqrt{d_1d_2}}\,\sum_{k=2}^{k_{\max }} \left(\frac{4C_2\delta}{\lambda_{\min}}\right)^{k-1}\\
&\leq 2\mu_{\max}^2\frac{r\delta}{\sqrt{d_1d_2}}\,\frac{\delta}{\lambda_{\min}},
\end{align*}hold with probability at least $1-k_{\max}(d_1+d_2)^{-2}d_3^{-3}$.

Large $k$ Regime ($k > k_{\max}$):
For the tail of the series, we use the fact that $k_{\max}$ is chosen such that $(1/2)^{k_{\max}}$ is negligible (specifically $(1 / 2)^{\log (d_1\vee d_2)} \leq 1/\sqrt{d_1d_2}$). By simply applying incoherence \cref{assump:incoh} and the bound for $\|\tE_a\|$ $(a=1,2)$ in \cref{eq:delta}, under the same event, we have
\begin{align*}
& \sum_{k=k_{\max }+1}^{\infty} \,\max _{j_1 \in[d_1],\,j_2 \in[d_2]}\left\|\dot{e}_{j_1}^{\dagger} \,\left(\mathcal{S}_{A, k}\left(\mathcal{P}_i\right) \mathcal{A} \Theta_i \Theta_i^{\dagger}+\mathcal{A} \Theta_i \Theta_i^{\dagger} \mathcal{A} \mathcal{S}_{A, k}\left(\mathcal{P}_i\right)\right) * \dot{e}_{d_1+j_2}\right\|\\&\leq
\sum_{k=k_{\max }+1}^{\infty}\sum_{\mathbf{s:}\,s_1+\cdots+s_{k+1=k}}
\mu_{\max}\sqrt{\frac{r}{d_1\wedge d_2}}\delta\left(\frac{\delta}{\lambda_{\min}}\right)^{k-1}\\
&\leq
\mu_{\max}\sqrt{\frac{r}{d_1\wedge d_2}}\delta \sum_{k=k_{\max }+1}^{\infty}\left(\frac{4\delta}{\lambda_{\min}}\right)^{k-1}\\
&\leq \mu_{\max}\sqrt{\frac{r}{d_1\wedge d_2}}\delta
\left(\frac{4\delta}{\lambda_{\min}}\right)^{k_{\max}}\\&\leq\mu_{\max}\sqrt{\frac{r}{(d_1\wedge d_2)d_1d_2}}\delta
\left(\frac{4\delta}{\lambda_{\min}}\right)
\end{align*}

Combining both regimes under the event of \cref{thm:distance}, the large $k$ term is dominated by the small $k$ term. We conclude:
\begin{equation*}
    \left|\sum_{i=1}^2 \sum_{k=2}^{\infty} \langle\mathcal{S}_{A,k}(\tP_a) \tA{\tTheta}_a{\tTheta}_a^{\dagger} +\tA{\tTheta}_a{\tTheta}_a^{\dagger}\tA \mathcal{S}_{A,k}(\tP_a) ,\widetilde{\tM}\rangle\right|
    \leq 2\|\tM\|_{\ell_1} \mu_{\max }^2\left(\frac{\delta}{\lambda_{\min}}\right) \frac{r \delta}{\sqrt{d_1 d_2}}.
\end{equation*}
Substituting $\delta = C_{\Bernstein}\sigma_{\xi} \sqrt{d_1 d_2(d_1 \vee d_2)} d_3 \sqrt{\frac{\log (d_1d_3+d_2d_3)}{n}}$ yields the final result.

\subsection{Proof of \texorpdfstring{\cref{lem:normal-t3}}{Lemma~X}}
\label{proof:normal-t3}
 By the block structure of the embedding, the inner product over the augmented tensor $\tA$ relates to the signal tensor $\tT$ as follows:
\begin{equation*}
 \langle  (\widehat{\tTheta}_a \widehat{\tTheta}^{\dagger}_a-{\tTheta} \tTheta^{\dagger}) \tA ( \widehat{\tTheta}_a \widehat{\tTheta}_a^{\dagger}-{\tTheta} \tTheta^{\dagger}),\widetilde{\tM}\rangle = \langle(\widehat{\tU}_a\widehat{\tU}_a^{\dagger}-\tU\tU^{\dagger})\tT(\widehat{\tV}_a\widehat{\tV}_a^{\dagger}-\tV\tV^{\dagger}),{\tM}\rangle,
\end{equation*}for $a\in\{1,2\}.$
Hölder's inequality yields\begin{multline*}
    \left| \langle(\widehat{\tU}_a\widehat{\tU}_a^{\dagger}-\tU\tU^{\dagger})\tT(\widehat{\tV}_a\widehat{\tV}_a^{\dagger}-\tV\tV^{\dagger}),{\tM}\rangle\right|\\
    \begin{aligned}
        &\leq \|\tM\|_{\ell_1}\, \max_{j_1\in[d_1],j_2\in[d_2]}\|\te_{j_1}^{\dagger}\,  (\widehat{\tU}_a\widehat{\tU}_a^{\dagger}-\tU\tU^{\dagger}) \tT (\widehat{\tV}_a\widehat{\tV}_a^{\dagger}-\tV\tV^{\dagger}) \, \te_{j_2}\|_{\max} \\
       &\leq \|\tM\|_{\ell_1}\, \max_{j_1\in[d_1],j_2\in[d_2]}\|\te_{j_1}^{\dagger}\,  (\widehat{\tU}_a\widehat{\tU}_a^{\dagger}-\tU\tU^{\dagger}) \tT (\widehat{\tV}_a\widehat{\tV}_a^{\dagger}-\tV\tV^{\dagger}) \,  \te_{j_2}\|,
    \end{aligned}
\end{multline*}where in the last step we used the fact that for a $1\times 1\times d_3$ tube, its maximum entry is dominated by its spectral norm.

Under the event of \cref{thm:distance}, the row-wise perturbation error is bounded. Furthermore, the spectral norm of the signal is bounded by $\|\tT\| \leq \kappa_0 \lambda_{\min}$. Thus, we obtain:
    \begin{align*}
        &\max_{j_1\in[d_1],j_2\in[d_2]}\bigl\|\te_{j_1}^{\dagger}\,  (\widehat{\tU}_a\widehat{\tU}_a^{\dagger}-\tU\tU^{\dagger}) \tT (\widehat{\tV}_a\widehat{\tV}_a^{\dagger}-\tV\tV^{\dagger}) \,  \te_{j_2}\bigr\|\\
        &\leq 
        \max_{j_1\in[d_1]} \bigl\|\te_{j_1}^{\dagger}\,(\widehat{\tU}_a\widehat{\tU}_a^{\dagger}-\tU\tU^{\dagger})\bigr\|\,
        \|\tT\|\,
        \max_{j_2\in[d_2]} \bigl\|\te_{j_2}^{\dagger}\,(\widehat{\tV}_a\widehat{\tV}_a^{\dagger}-\tV\tV^{\dagger})\bigr\|\,\\
        &\leq (\kappa_0\lambda_{\min}) \cdot 
        C_1^2\mu_{\max}^2\frac{\sigma_{\xi}^2}{\lambda_{\min}^2}d_3^2\frac{r\sqrt{d_1d_2}(d_1\vee d_2)}{n}\log(d_1d_3+d_2d_3)
    \end{align*}
This leads to the final result:
\begin{multline*}
     \langle  (\widehat{\tTheta}_a \widehat{\tTheta}_a^{\dagger}-{\tTheta} \tTheta^{\dagger}) \tA ( \widehat{\tTheta}_a \widehat{\tTheta}_a^{\dagger}-{\tTheta} \tTheta^{\dagger}),\widetilde{\tM}\rangle\\
     \leq \|\tM\|_{\ell_1}\,
     \kappa_0 
     C_1^2\mu_{\max}^2\frac{\sigma_{\xi}^2}{\lambda_{\min}}d_3^2\frac{r\sqrt{d_1d_2}(d_1\vee d_2)}{n}\log(d_1d_3+d_2d_3)
\end{multline*}
which completes the proof of \cref{lem:normal-t3}.

\subsection{Proof of \texorpdfstring{\cref{lem:s_1geq0}}{Lemma~X}} 
\label{proof:s_1geq0}

\paragraph{Case 1: $l=1$} We aim to bound $\max_{j\in[d_1+d_2]}\|\te_j^{\dagger}\tP\tTheta\|$. 
By the block structure of $\tP$ and $\tTheta$,
\begin{equation*}
    \max_{j\in[d_1]}\|\te_j^{\dagger}\,\tP\tTheta\|=
    \max_{j\in[d_1]}\|\te_j^{\dagger}\,\tE\tV\|,
    \qquad
    \max_{j\in[d_2]}\|\te_{d_1+j}^{\dagger}\,\tP\tTheta \|=
    \max_{j\in[d_2]}\|\te_j^{\dagger}\,\tE^{\dagger}\tU\|.
\end{equation*}
Here, with a slight abuse of notation, $\te_j$ and $\te_{d_1+j}$ on the left-hand side denote the canonical basis tensors in
$\mathbb R^{(d_1+d_2)\times 1\times d_3}$, whereas on the right-hand side $\te_j$ denotes the canonical basis tensors in
$\mathbb R^{d_1\times1\times d_3}$ and $\mathbb R^{d_2\times1\times d_3}$,
respectively.

Recall the decomposition of the perturbation tensor
\begin{equation*}
    \tE
    =
    \frac{d^*}{n_0}\sum_{i=n_0+1}^{n}\xi_i\tX_i
    +
    \left(
    \frac{d^*}{n_0}\sum_{i=n_0+1}^{n}\langle\tZ_1,\tX_i\rangle\tX_i-\tZ_1
    \right)
    =:\tE_{\mathrm{rn}}+\tE_{\mathrm{init}}.
\end{equation*}
Hence
\begin{equation*}
    \te_j^{\dagger}\,\tE\tV = \te_j^{\dagger}\,\tE_{\mathrm{rn}}\tV+\te_j^{\dagger}\,\tE_{\mathrm{init}}\tV.
\end{equation*}

As in the proof of the spectral-norm bound for $\|\tE\|$, we pass to the frequency
domain. For the random-noise part, for any fixed $j\in[d_1]$, we have\begin{equation*}
   \bigl\| \te_j^{\dagger}\,\tE_{\mathrm{rn}}\tV\bigr\| = \bigl\|\frac{d^*}{n_0} \sum_{j=n_0+1}^n \xi_j \diag(e_j^T) \overbar{\fmX}_j \overbar{\fmV}\bigr\|.
\end{equation*}
The $i$-th summand satisfies
\begin{equation*}
   \left\|\left\|\xi_i \diag(e_j^T)\overbar{\fmX}_i\overbar{\fmV}\right\|
   \right\|_{\psi_2}
   \leq
   \|\xi_j\|_{\psi_2}\cdot
   \left\|\diag(e_j^T)\overbar{\fmX}_j\overbar{\fmV}\right\|
   \leq
   \sigma_{\xi}\mu_{\max}\sqrt{\frac{r}{d_2}},
\end{equation*}
where we used the incoherence bound in \cref{assump:incoh}, \begin{equation*}
   \bigl\| \diag(e_j^T)\overbar{\fmV}\bigr\|=\bigl\|\te_j^{\dagger}\tV\bigr\|\leq \mu_{\max}\sqrt{{r}/{d_2}}.
\end{equation*}
For the variance proxy, the same computation as in the proof of
\cref{lem:error_op_b} gives
\begin{equation*}
 \left\|
 \mathbb{E}\,
 \xi_j^2 \diag(e_j^T)\overbar{\fmX}_j\overbar{\fmV}
\overbar{\fmV}^{T}\overbar{\fmX}_j^{H}\diag(e_j)
 \right\|=
\frac{\sigma_{\xi}^2}{d_1d_2}
\left\|
\sum_{j_2\in[d_2]}
\operatorname{diag}_{d_3}(e_{j_2}^{T})
\overbar{\fmV}\overbar{\fmV}^{H}
\operatorname{diag}_{d_3}(e_{j_2})
\right\| \leq
\frac{\sigma_{\xi}^2}{d_1d_2}\,r.
\end{equation*}
Therefore, by the matrix Bernstein inequality, for every fixed $j\in[d_1]$,
\begin{equation*}
    \mathbb P\left(
    \|\te_j^{\dagger}\,\tE_{\mathrm{rn}}\tV\|
    > C_{\mathrm{Berstein}}\,\sigma_{\xi}d_3\sqrt{\frac{rd_1d_2\log(d_1d_3+d_2d_3)}{n}}
    \right)
    \leq
    \frac{1}{2}(d_1d_3+d_2d_3)^{-3},
\end{equation*} when $n>4\mu_{\max}^2d_1\log^2(d_1d_3+d_2d_3)$.

For the initialization part, $\tZ_1$ is independent of the second half sample
$\{\tX_i\}_{i=n_0+1}^{n}$ so it can be treated as fixed. The same Bernstein argument applies with
$\xi_i$ replaced by $\langle\tZ_1,\tX_i\rangle$. We have, when $n>4\mu_{\max}^2d_1\log^2(d_1d_3+d_2d_3)$\begin{equation*}
        \mathbb P\left(
    \|\te_j^{\dagger}\,\tE_{\mathrm{init}}\tV\|
    > C_{\mathrm{Berstein}}\,\|\tZ_1\|_{\max}d_3\sqrt{\frac{rd_1d_2\log(d_1d_3+d_2d_3)}{n}}
    \right)
    \leq
    \frac{1}{2}(d_1d_3+d_2d_3)^{-3}.
\end{equation*}Recall that by \cref{assump:init_entry}, there exist an event $\Omega_0$ with probability $\mathbb{P}(\Omega_0)\geq1-(d_1d_3 + d_2d_3)^{-2}$ such that, on $\Omega_0$, we have:\begin{equation*}
    \|\tZ_a\|_{\max}\leq C_{\mathrm{init}}{\gamma_n}\cdot\sigma_{\xi},
\end{equation*}where $C_{\mathrm{init}}$ is an absolute constant and $\gamma_n\leq 1$. Thus, after enlarging the
absolute constant if necessary, we obtain for every fixed $j\in[d_1]$, conditioned on $\Omega_0$,
\begin{equation*}
    \mathbb P\left(
    \|\te_j^{\dagger}\,\tE_{\mathrm{init}}\tV\|
    >
C_{\mathrm{Berstein}}\,\sigma_{\xi}d_3\sqrt{\frac{rd_1d_2\log(d_1d_3+d_2d_3)}{n}}
    \right)
    \leq
    \frac{1}{2}(d_1d_3+d_2d_3)^{-3}.
\end{equation*}
Since
\begin{equation*}
    \sigma_{\xi}d_3\sqrt{\frac{rd_1d_2\log(d_1d_3+d_2d_3)}{n}}
    \leq
    \,\delta\sqrt{\frac{r}{d_1\vee d_2}}
    \leq
    \,\delta\,\mu_{\max}\sqrt{\frac{r}{d_1}},
\end{equation*}we conclude that, for every fixed $j\in[d_1]$, conditioned on $\Omega_0$,\begin{equation*}
      \mathbb P\bigl(
    \|\te_j^{\dagger}\,\tE_{\mathrm{rn}}\tV\|
    >
 \,\delta\,\mu_{\max}\sqrt{\frac{r}{d_1}}
    \bigr) \leq
   \frac{1}{2}(d_1d_3+d_2d_3)^{-3}
\end{equation*}
and
\begin{equation*}
   \quad   \mathbb P\bigl(
    \|\te_j^{\dagger}\,\tE_{\mathrm{init}}\tV\|
    >
 \,\delta\,\mu_{\max}\sqrt{\frac{r}{d_1}}
    \bigr) \leq
   \frac{1}{2}(d_1d_3+d_2d_3)^{-3}.
\end{equation*}
Combining the last two inequalities yields, for every fixed $j\in[d_1]$,
\begin{equation*}
    \mathbb P\left(
    \|\te_j^{\dagger}\,\tP\tTheta\|
    >
    C\,\delta\,\mu_{\max}\sqrt{\frac{r}{d_1}}
    \right)
    \leq
(d_1d_3+d_2d_3)^{-3}.
\end{equation*}

By symmetry, the same argument gives for every fixed $j\in[d_2]$,
\begin{equation*}
    \mathbb P\left(
    \|\te_{d_1+j}^{\dagger}\,\tP\tTheta\|
    >
    C\,\delta\,\mu_{\max}\sqrt{\frac{r}{d_2}}
    \right)
    \leq
(d_1d_3+d_2d_3)^{-3}.
\end{equation*}
Taking the union bound over all $j\in[d_1+d_2]$, on $\Omega_0$, with probability at least $1-(d_1+d_2)^{-2}d_3^{-3}$,
\begin{equation*}
    \max_{i\in[d_1]}
    \|\te_i^{\dagger}*\tP\tTheta\|
    \leq
    \,\delta\,\mu_{\max}\sqrt{\frac{r}{d_1}},
    \qquad
    \max_{j\in[d_2]}
    \|\te_{d_1+j}^{\dagger}*\tP\tTheta\|
    \leq
    \,\delta\,\mu_{\max}\sqrt{\frac{r}{d_2}}.
\end{equation*}

\paragraph{Case 2: $l=2$} We aim to bound $\max_{j\in[d_1+d_2]}\|\te_j^{\dagger}\,\tP\mathfrak P^{\perp}\tP\tTheta\|$. 
By the definition of $\mathfrak P^{\perp}$,
\begin{equation*}
    \max_{j\in[d_1]}\bigl\|\te_j^{\dagger}\tP\mathfrak P^{\perp}\tP\tTheta\bigr\|  =\max_{j\in[d_1]}\bigl\|\te_j^{\dagger}\,\tE\tV_{\perp}\tV_{\perp}^{\dagger}\tE^{\dagger}\tU\bigr\|,
\end{equation*}
and
\begin{equation*}
    \max_{j\in[d_2]}\bigl\|\te_{d_1+j}^{\dagger}\tP\mathfrak P^{\perp}\tP\tTheta\bigr\|  =\max_{j\in[d_2]}\bigl\|\te_j^{\dagger}\,\tE\tV_{\perp}\tV_{\perp}^{\dagger}\tE^{\dagger}\tU\bigr\|.
\end{equation*}As before, we use the notation $\te_j$ and $\te_{d_1+j}$ on the left-hand sides to denote the canonical basis tensors of size $\left(d_1+d_2\right) \times 1 \times d_3$, while on the right-hand sides $\te_j$ to the canonical basis tensors of sizes $d_1 \times 1 \times d_3$ and $d_2 \times 1 \times d_3$, respectively. 

We first treat the $d_1$ block. For $j\in[d_1]$, we write
\begin{equation*}
\te_j^{\dagger}\,\tE\tV_{\perp}\tV_{\perp}^{\dagger}\tE^{\dagger}\tU
=
\te_j^{\dagger}\,\tE\tV_{\perp}\tV_{\perp}^{\dagger}
\tE^{\dagger} (\te_j\te_j^{\dagger}) \tU +
\te_j^{\dagger}\tE\tV_{\perp}\tV_{\perp}^{\dagger}
\tE^{\dagger}( \tI_{d_1}-\te_j\te_j^{\dagger})\tU,
\end{equation*}where $\tI_{d_1}\in\mathbb{R}^{d_1\times d_1\times d_3}$ denotes the identity tensor defined in \cref{subsec:tubal-rank}.
For the first term, on $\Omega_0$,
\begin{equation*}
\bigl\|\te_j^{\dagger}\,\tE\tV_{\perp}\tV_{\perp}^{\dagger}
\tE^{\dagger}(\te_j\te_j^{\dagger})\tU\bigr \|
\leq
\bigl\|\te_j^{\dagger}\,\tE\tV_{\perp}\tV_{\perp}^{\dagger}
\tE^{\dagger}\te_j \|\,
\bigl\|\te_j^{\dagger}\tU\bigr \| 
\leq
\|\tE\|^2\,\|\te_i^{\dagger}*\tU\|
\leq
\delta^2\mu_{\max}\sqrt{\frac{r}{d_1}}.
\end{equation*}
For the second term, first recall the decomposition of the perturbation tensor
\begin{equation*}
    \tE  := \frac{d^*}{n_0}\sum_{i=n_0+1}^{n}\xi_i\tX_i+
    \left(
    \frac{d^*}{n_0}\sum_{i=n_0+1}^{n}\langle\tZ_1,\tX_i\rangle\tX_i-\tZ_1
    \right).
\end{equation*}Define the index sets
\begin{equation*}
    N_j=\Bigl\{i\in\{n_0+1,\ldots,n\}:\te_j^{\dagger}\tX_i\neq 0\Bigr\}.
\end{equation*}By a standard Chernoff bound, the sample-size assumption in the lemma implies that,
after enlarging the absolute constant if necessary, the event
\begin{equation*}
    \Omega_{\mathrm{cnt}}
    :=
    \left\{
    \frac{n_0}{2d_1}\leq |N_i|\leq \frac{2n_0}{d_1},\quad \forall j\in[d_1]
    \right\}
\end{equation*}
satisfies that for a absolute constant $C_{\Chernoff}>0$,
\begin{equation*}
    \mathbb P(\Omega_{\mathrm{cnt}}^{c})
    \leq e^{-C_{\Chernoff}n/d_1}.
\end{equation*}
Now condition on $N_j$, the observations
$\{(\tX_i,\xi_i)\}_{i\in N_i}$ are independent of $(\mathcal I_{d_1}-\te_j\te_j^{\dagger} )\tE$. Thus, using the decomposition of $\tE$, we have,\begin{multline*}
    \te_j^{\dagger}\tE\,\tV_{\perp}\tV_{\perp}^{\dagger}
\tE^{\dagger}(\mathcal I_{d_1}-\te_j\te_j^{\dagger} )\tU =\underbrace{\frac{d^*}{n_0}\sum_{i\in N_j}
\xi_i\,\te_j^{\dagger}\,\tX_i\,\tV_{\perp}\tV_{\perp}^{\dagger}
\tE^{\dagger}(\mathcal \tI_{d_1}-\te_j\te_j^{\dagger} )\tU }_{\te_j^{\dagger}\tE_{\mathrm{rn}}\,\tV_{\perp}\tV_{\perp}^{\dagger}
\tE^{\dagger}(\mathcal I_{d_1}-\te_j\te_j^{\dagger} )\tU}\\+
\underbrace{\left(
\frac{d^*}{n_0}\sum_{i\in N_j}
\langle\tZ_1,\tX_i\rangle\,
\te_j^{\dagger}\,\tX_i\,\tV_{\perp}\tV_{\perp}^{\dagger}
\tE^{\dagger}(\mathcal \tI_{d_1}-\te_j\te_j^{\dagger} )\tU
-
\te_j^{\dagger}\,\tZ_1\,\tV_{\perp}\tV_{\perp}^{\dagger}
\tE^{\dagger}(\mathcal \tI_{d_1}-\te_j\te_j^{\dagger} )\tU
\right)}_{\te_j^{\dagger}\tE_{\mathrm{init}}\,\tV_{\perp}\tV_{\perp}^{\dagger}
\tE^{\dagger}(\mathcal I_{d_1}-\te_j\te_j^{\dagger} )\tU},
\end{multline*}where the factor $\tV_{\perp}\tV_{\perp}^{\dagger}
\tE^{\dagger}(\mathcal \tI_{d_1}-\te_j\te_j^{\dagger} )\tU$ can be treated as fixed, conditioned on $N_j$. We first analyze the summands of $\te_j^{\dagger}\tE_{\mathrm{rn}}\,\tV_{\perp}\tV_{\perp}^{\dagger}
\tE^{\dagger}(\tI_{d_1}-\te_j\te_j^{\dagger} )\tU$ in then Fourier domain. Conditioned on $N_j$ and $(\tI_{d_1}-\te_j\te_j^{\dagger} )\tE$, we have\begin{equation*}
    \left\| \xi_i\,\diag_{d_3}(e_j^T)\,\overbar\fmX_i\,\overbar\fmV_{\perp}\overbar\fmV_{\perp}^{H}
\overbar\fmE^{H}\diag_{d_3}(I_{d_1}-e_je_j^{T}) \overbar\fmU 
    \right\|_{\psi_2}\leq \sigma_{\xi} \,\max_{j\in[d_1]}\bigl\|\te_j^{\dagger}\tV_{\perp}\tV_{\perp}^{\dagger}
\tE^{\dagger}( \tI_{d_1}-\te_j\te_j^{\dagger} )\tU 
    \bigr\|,
\end{equation*}and\begin{multline*}
     \bigl\| \mathbb E \xi_i^2  \,
   \diag_{d_3}(e_j^T)\,\overbar\fmX_i\,\overbar\fmV_{\perp}\overbar\fmV_{\perp}^{H}
\overbar\fmE^{H}\diag_{d_3}(I_{d_1}-e_je_j^{T}) \overbar\fmU \\
\overbar\fmU^H\diag_{d_3}(\tI_{d_1}-\te_j\te_j^{T})\overbar\fmE\overbar\fmV_{\perp}\overbar\fmV_{\perp}^{H}\,\overbar\fmX_i^H \,\diag_{d_3}(e_j) \bigr\|
\leq\frac{\sigma_{\xi}^2r}{d_2}\,\bigl\|\tE^{\dagger}(\tI_{d_1}-\te_j\te_j^{\dagger})\tU\bigr\|^2.
\end{multline*}By matrix Bernstein inequality, 
we have,\begin{multline*}
     \mathbb{P}\Bigl( \bigl\|
    \sum_{i\in N_j} \xi_i\,\diag_{d_3}(e_j^T)\,\overbar\fmX_i\,\overbar\fmV_{\perp}\overbar\fmV_{\perp}^{H}
\overbar\fmE^{H}\diag_{d_3}(I_{d_1}-e_je_j^{T}) \overbar\fmU 
\bigr\| \\> C_{\mathrm{Bernstein}} \sigma_{\xi}\sqrt{\frac{r}{d_2}} \bigl\|\tE^{\dagger}(\tI_{d_1}-\te_j\te_j^{\dagger})\tU\bigr\||N_j|^{1/2}\sqrt{\log(d_1d_3+d_2d_3)} 
   \Big|\ N_j,\mathcal (\mathcal I_{d_1}-\te_j\te_j^{\dagger} )\tE
   \Bigr )\\\leq \frac12(d_1d_3+d_2d_3)^{-3},
\end{multline*}when $n>$.
On $\Omega_0$, \begin{equation*}
    \bigl\|\tE^{\dagger}(\tI_{d_1}-\te_j\te_j^{\dagger})\tU\bigr\|\leq \|\tE\|\leq\delta,
\end{equation*}so we have,
\begin{multline*}
     \mathbb{P}\Bigl( \bigl\|
   \frac{d^*}{n_0} \sum_{i\in N_j} \xi_i\,\diag_{d_3}(e_j^T)\,\overbar\fmX_i\,\overbar\fmV_{\perp}\overbar\fmV_{\perp}^{H}
\overbar\fmE^{H}\diag_{d_3}(I_{d_1}-e_je_j^{T}) \overbar\fmU 
\bigr\| \\>\sqrt{\frac{r}{d_1}} \delta^2\,
|N_j|^{1/2}\sqrt{\frac{2d_1^2}{n_0(d_1\vee d_2)}} 
   \Big|\ N_j,\mathcal (\mathcal I_{d_1}-\te_j\te_j^{\dagger} )\tE
   \Bigr )\\\leq \frac12(d_1d_3+d_2d_3)^{-3}.
\end{multline*}Now we use $\Omega_{\mathrm{cnt}}$ to replace $|N_j|$ with its deterministic order $n_0/d_1$, we have, on $\Omega_0\cap \Omega_{\mathrm{cnt}}$, if\begin{equation*}
    |N_j|>{\frac{n_0(d_1\vee d_2)}{2d_1^2}},
\end{equation*}then,
\begin{equation*}
    \mathbb P\left(
    \|\te_j^{\dagger}\tE_{\mathrm{rn}}\tV_{\perp}\tV_{\perp}^{\dagger}
\tE^{\dagger}(\mathcal I_{d_1}-\te_j\te_j^{\dagger} )\tU\|
    >\delta^2\sqrt{\frac{r}{d_1}} )\tE
    \right)
    \leq
    \frac{1}{2}(d_1+d_2)^{-3}d_3^{-3}.
\end{equation*}Similarly, since $\tZ_1$ is independent with $\{(\tX_i,\xi_i)\}_{i=n_0+1}^n$, conditioned on $\Omega_0\cap\Omega_{\mathrm{cnt}}$,\begin{equation*}
    \mathbb P\left(
\|\te_j^{\dagger}\tE_{\mathrm{init}}\tV_{\perp}\tV_{\perp}^{\dagger}
\tE^{\dagger}(\mathcal I_{d_1}-\te_j\te_j^{\dagger} )\tU\|
    >\delta^2\sqrt{\frac{r}{d_1}}
    \right)
    \leq
    \frac{1}{2}(d_1+d_2)^{-3}d_3^{-3},
\end{equation*}as long as $\|\tZ_1\|\leq \sigma_{\xi}$. Thus, we obtain,
\begin{equation*}
    \mathbb P\left(
    \|\te_i^{\dagger}\,\tP\mathfrak P^{\perp}\tP\tTheta\|
    >
    C\,\delta^2\mu_{\max}\sqrt{\frac{r}{d_1}}
    \right)
    \leq
(d_1+d_2)^{-3}d_3^{-3}.
\end{equation*} The argument for the $d_2$ block is identical after replacing
$\tI_{d_1}$ by $\tI_{d_2}$, $\tU$ by $\tV$, and
$\tV_{\perp}$ by $\tU_{\perp}$. Hence, for every fixed $j\in[d_2]$,
\begin{equation*}
    \mathbb P\left(
    \|\te_{d_1+j}^{\dagger}\,\tP\mathfrak P^{\perp}\tP\tTheta\|
    >
    C\,\delta^2\mu_{\max}\sqrt{\frac{r}{d_2}}
    \right)
    \leq
(d_1+d_2)^{-3}d_3^{-3}.
\end{equation*}Taking the union bound over all rows, on $\Omega_0\cap \Omega_{\mathrm{cnt}}$, with probability at least $1-(d_1+d_2)^{-2}d_3^{-3},$
\begin{equation*}
    \max_{j\in[d_1]}
    \|\te_j^{\dagger}\,\tP\mathfrak P^{\perp}\tP\tTheta\|
    \leq
    \,\delta^2\mu_{\max}\sqrt{\frac{r}{d_1}},
    \quad
    \max_{j\in[d_2]}
    \|\te_{d_1+j}^{\dagger}\,\tP\mathfrak P^{\perp}\tP\tTheta\|
    \leq
    \,\delta^2\mu_{\max}\sqrt{\frac{r}{d_2}}.
\end{equation*}

\paragraph{General case: $l\ge 3$}
We prove by induction that there exist absolute constants $C_1>0$ such that, conditioned on $\Omega_0\cap\Omega_{\mathrm{cnt}}$, there exists an event $\Omega^{(l)}$ with probability\begin{equation*}
    \mathbb P (\Omega^{(l)})\geq 1-l(d_1+d_2)^{-2}d_3^{-3},
\end{equation*}
such that on $\Omega^{(l)}$, for every integer $1\le l_0\le l$, 
\begin{equation*}
    \max_{j\in[d_1]}
    \|\te_j^{\dagger}\tP(\mathfrak P^\perp\tP)^{l_0-1}\tTheta\|
    \le (C_1\delta)^{l_0}\mu_{\max}\sqrt{\frac{r}{d_1}},
\end{equation*}
and
\begin{equation*}
    \max_{j\in[d_2]}
    \|\te_{d_1+j}^{\dagger}\tP(\mathfrak P^\perp\tP)^{l_0-1}\tTheta\|
    \le (C_1\delta)^{l_0}\mu_{\max}\sqrt{\frac{r}{d_2}}.
\end{equation*}
The cases $l_0=1$ and $l_0=2$ were proved above.

Assume now that the claim holds for all $1\le l_0\le l$, where $l\ge 2$. We prove it for $l+1$.
It suffices to treat the $d_1$ block; the $d_2$ block follows by symmetry after exchanging $(\tU,d_1)$ and $(\tV,d_2)$.

Fix $j\in[d_1]$ and define
\begin{equation*}
    \tP_j
    :=
    \begin{pmatrix}
        0 & \te_j\te_j^\dagger\tE\\
        \tE^\dagger\te_j\te_j^\dagger & 0
    \end{pmatrix},
    \qquad
    \tP_j^\perp
    :=
    \tP-\tP_j
    =
    \begin{pmatrix}
        0 & (\tI_{d_1}-\te_j\te_j^\dagger)\tE\\
        \tE^\dagger(\tI_{d_1}-\te_j\te_j^\dagger) & 0
    \end{pmatrix}.
\end{equation*}
Thus $\tP=\tP_j+\tP_j^\perp $ and
\begin{align*}
    \te_j^\dagger\,\tP(\mathfrak P^\perp\tP)^l\tTheta
    &=
    \te_j^\dagger\,\tP\mathfrak P^\perp\tP(\mathfrak P^\perp\tP)^{l-1}\tTheta \\
    &=
    \te_j^\dagger\,\tP\mathfrak P^\perp\tP_j(\mathfrak P^\perp\tP)^{l-1}\tTheta
    +
     \te_j^\dagger\,\tP\mathfrak P^\perp\tP_j^{\perp}(\mathfrak P^\perp\tP)^{l-1}\tTheta.\\
     &:=\mathrm{I} + \mathrm{II}
\end{align*}

For the first term $\mathrm{I}$, on $\Omega_0$, since $\|\tP\|=\|\tE\|\le \delta$ and $\|\mathfrak P^\perp\|=1$, we have for any $j\in[d_1]$
\begin{align*}
    \|\mathrm{I}\| =\| \te_j^\dagger\,\tP\mathfrak P^\perp\tP_j(\mathfrak P^\perp\tP)^{l-1}\tTheta\|
    &\le
    \|\te_j^\dagger\,\tP\mathfrak P^\perp\te_j\|\,\max_{j\in[d_1]}\|\te_j^{\dagger}\tP(\mathfrak P^\perp\tP)^{l-1}\tTheta\| \\
    &\le
    \delta\,\max_{j\in[d_1]}\|\te_j^{\dagger}\tP(\mathfrak P^\perp\tP)^{l-1}\tTheta\| \\
    &\le
  (C_1\delta)^l\delta\,\mu_{\max}\sqrt{\frac{r}{d_1}},
\end{align*}
where the last step holds on $\Omega^{(l)}$ by the induction hypothesis.

For the second term $\mathrm{II}$, again expanding the remaining factors $\tP=\tP_j^\perp + \tP_j$ and notice that\begin{align*}
    (\mathfrak P^\perp\tP)^{l-1} &= (\mathfrak P^\perp \tP_j^\perp +\mathfrak P^\perp  \tP_j)^{l-1}
    \\&= (\mathfrak P^\perp \tP_j^\perp)^{l-1} + 
    \sum_{t=1}^{l-1} (\mathfrak P^\perp \tP_j^\perp)^t
    \,(\mathfrak P^\perp  \tP_j)
    \,(\mathfrak P^\perp \tP_j^\perp + \mathfrak P^\perp  \tP_j)^{l-1-t}\\
    &= (\mathfrak P^\perp \tP_j^\perp)^{l-1} + 
    \sum_{t=1}^{l-1} (\mathfrak P^\perp \tP_j^\perp)^t
    \,(\mathfrak P^\perp  \tP_j)
    \,(\mathfrak P^\perp \tP)^{l-1-t},
\end{align*}thus we may write
\begin{equation*}
   \mathrm{II}= \te_j^\dagger\,\tP\mathfrak P^\perp\tP_j^{\perp}(\mathfrak P^\perp\tP)^{l-1}\tTheta
 = \sum_{t=1}^{l-1} 
    \te_j^\dagger\,\tP
    \,(\mathfrak P^\perp \tP_j^\perp)^{t+1}
    \,\mathfrak P^\perp  \tP_j
    \,(\mathfrak P^\perp  \tP)^{l-1-t}\,\tTheta
     \,+\, 
     \te_j^\dagger\,\tP
    (\mathfrak P^\perp \tP_j^\perp)^{l}\,\tTheta
\end{equation*}
For each $1\le t\le l-1$, using again $\|\tP\|\le\delta$ and $\|\tP_j^\perp\|\le\|\tP\|\le\delta$, we get, for any $j\in[d_1]$,
\begin{align*}
   \|\mathrm{II}\|= \bigl\|
     \te_j^\dagger\,\tP
    \,(\mathfrak P^\perp \tP_j^\perp)^{t+1}
    \,\mathfrak P^\perp  \tP_j
    \,(\mathfrak P^\perp  \tP)^{l-1-t}\,\tTheta
    \bigr\|
    &\le
    \|\tP\|^{t+1}\,
    \max_{j\in[d_1]}\|\te_j^{\dagger}\tP(\mathfrak P^\perp\tP)^{l-1-t}\tTheta\|  \\
    &\le
    (C_1\delta)^{l-t}\delta^{t+1}\mu_{\max}\sqrt{\frac{r}{d_1}},
\end{align*}
where the last step again follows from the induction hypothesis. Hence, on $\Omega^{(l)}$, we have\begin{align}
\max_{j\in[d_1]}
\bigl\|\te_j^\dagger\,\tP(\mathfrak P^\perp\tP)^l&\tTheta\bigr\|
\le \|\mathrm{I}\|\,+\, \|\mathrm{II}\|\nonumber
\\ &\leq
    (C_1\delta)^l\delta\,\mu_{\max}\sqrt{\frac{r}{d_1}} 
    +
    \sum_{t=1}^{l-1} (C_1\delta)^{l-t}\delta^{t+1}\mu_{\max}\sqrt{\frac{r}{d_1}}
    +
    \max_{j\in[d_1]}\bigl\|\te_j^\dagger\,\tP(\mathfrak P^\perp\tP_j^{\perp})^l\tTheta\bigr\|\nonumber\\
    &=\frac{1}{2}(C_1\delta)^{l+1}\sqrt{\frac{r}{d_1}}
    +
    \max_{j\in[d_1]}\bigl\|\te_j^\dagger\,\tP(\mathfrak P^\perp\tP_j^{\perp})^l\tTheta\bigr\|\label{eq:induction_before}
\end{align}
provided that $C_1>4$.

It remains to control the term $\bigl\|\te_j^\dagger\,\tP(\mathfrak P^\perp\tP_j^{\perp})^l\tTheta\bigr\|$.
Conditioned on $N_j$, the tensor factor $(\mathfrak P^\perp\tP_j^\perp)^l\tTheta$ is independent of $\te_j^{\dagger}\tP$ thus can be treated as fixed.
Therefore, repeating the same matrix Bernstein argument as in Case~2 for the leftmost factor $\te_j^\dagger\tP$, we obtain,
\begin{multline*}
    \mathbb P\Bigl(
    \bigl\|\te_j^\dagger\,\tP
    (\mathfrak P^\perp \tP_j^\perp)^{l}\,\tTheta \bigr\|
    > C_{\Bernstein}
    \sigma_{\xi}\sqrt{\frac{r}{d_1\wedge d_2}}\|\mathfrak P^\perp\tP\tTheta\|^l 
    \\|N_j|^{1/2}\sqrt{\log (d_1d_3+d_2d_3)}
    \,\Big |\,
    N_j,\,P_j^{\perp}
    \Bigr)\leq (d_1+d_2)^{-3}d_3^{-3}.
\end{multline*}Now we use $\Omega_{\mathrm{cnt}}$ to replace $|N_j|$ with its deterministic order $O(n_/d_1)$ and use the fact that $\|\tP\|\leq\delta$ on $\Omega_{0}$. We have, on $\Omega_0\cap \Omega_{\mathrm{cnt}}$
\begin{equation*}
    \mathbb P\left(
        \bigl\|\te_j^\dagger\,\tP
    (\mathfrak P^\perp \tP_j^\perp)^{l}\,\tTheta \bigr\|>
        \frac{1}{2}(C_1\delta)^{l+1}\mu_{\max}\sqrt{\frac{r}{d_1}}
        \,\middle|\,
        N_j,\,P_j^{\perp}
    \right)
    \le
    (d_1+d_2)^{-3}d_3^{-3}.
\end{equation*}
The $d_2$-block bound is proved in the same way after exchanging
$(\tU,d_1)$ and $(\tV,d_2)$. Hence, Taking a union bound over all $j\in[d_1+d_2]$ yields, on $\Omega_0\cap\Omega_{\mathrm{cnt}}$, with probability at least
$1-(d_1+d_2)^{-2}d_3^{-3}$,\begin{equation*}
     \max_j\in[d_1]\bigl\|\te_j^\dagger\,\tP
    (\mathfrak P^\perp \tP_j^\perp)^{l}\,\tTheta \bigr\|\leq
        \frac{1}{2}(C_1\delta)^{l+1}\mu_{\max}\sqrt{\frac{r}{d_1}}
\end{equation*}and\begin{equation*}
        \max_j\in[d_2]\bigl\|\te_{d_1+j}^\dagger\,\tP
    (\mathfrak P^\perp \tP_j^\perp)^{l}\,\tTheta \bigr\|\leq
        \frac{1}{2}(C_1\delta)^{l+1}\mu_{\max}\sqrt{\frac{r}{d_2}}.
\end{equation*}

Combining the last display with the bound in \cref{eq:induction_before}, we get, on $\Omega_0\cap\Omega_{\mathrm{cnt}}$, with probability at least
$1-2(l+1)(d_1+d_2)^{-2}d_3^{-3}$,
\begin{equation*}
    \max_{j\in[d_1]}
    \|\te_j^\dagger\tP(\mathfrak P^\perp\tP)^l\tTheta\|
    \le
    (C_1\delta)^{l+1}\mu_{\max}\sqrt{\frac{r}{d_1}},
\end{equation*}
and
\begin{equation*}
    \max_{j\in[d_2]}
    \|\te_{d_1+j}^\dagger\tP(\mathfrak P^\perp\tP)^l\tTheta\|
    \le
    (C_1\delta)^{l+1}\mu_{\max}\sqrt{\frac{r}{d_2}}.
\end{equation*}
This closes the induction.

\section{Proof of \texorpdfstring{\cref{thm:infer}}{Theorem~X}}
\label{Asec:infer}

We aim to prove the normal approximation
\begin{equation*}
    \frac{\langle\widehat{\tT},\tM\rangle-\langle\tT,\tM\rangle}
    {\widehat{\sigma}_{\xi}\,\widehat{s}_{\tM}\,\sqrt{d_1 d_2 d_3/ n}}
    \;\toD\;\mathcal N(0,1),
\end{equation*}
where $\widehat{\sigma}_{\xi}$ and $\widehat{s}_{\tM}$ are the empirical estimators defined in \cref{subsec:Inferences about Linear Forms}. The argument proceeds by a decomposition and Slutsky’s theorem.

Denote\begin{equation*}
    s_{\tM} :=(\|\tM\tV\|_{\mathrm{F}}^2 +\|\tU^{\dagger}\tM\|_{\mathrm{F}}^2 \bigr)^{1 / 2},
\end{equation*}
and write\begin{multline}\label{eq:diff3.2}
     \frac{\langle\widehat{\tT},\tM\rangle-\langle\tT,\tM\rangle}
        {\widehat{\sigma}_{\xi} \widehat{s}_{\tM}\,\sqrt{d_1 d_2 d_3/ n}}
        = \frac{\langle\widehat{\tT},\tM\rangle-\langle\tT,\tM\rangle}
        {\sigma_{\xi}s_{\tM}\sqrt{d_1 d_2 d_3 / n}}
        +\frac{\langle\widehat{\tT},\tM\rangle-\langle\tT,\tM\rangle}
        {\widehat{\sigma}_{\xi} \widehat{s}_{\tM}\,\sqrt{d_1 d_2 d_3 / n}}
        \left(1-\frac{\widehat{\sigma}_{\xi}}{\sigma_{\xi}}\right) 
        \\+\frac{\langle\widehat{\tT},\tM\rangle-\langle\tT,\tM\rangle}
        {\sigma_{\xi}s_{\tM}\sqrt{d_1 d_2 d_3 / n}} 
        \cdot \left(\frac{s_{\tM}}{\widehat{s}_{\tM}}-1\right).
\end{multline}
By \cref{cor}, the first term on the right-hand side of \cref{eq:diff3.2} converges in distribution to $\mathcal N(0,1)$ under conditions \cref{eq:rate1,eq:rate2}. It therefore suffices to show that the second and third terms are $o_{\mathbb{P}}(1)$.

\paragraph{Step 1: Control of $\widehat{\sigma}_{\xi}$}
Recall that
\begin{equation}\label{eq:sigmahat-expansion}
    \begin{aligned}
        \widehat{\sigma}_{\xi}^2
        &=\frac{1}{n} \sum_{i=n_0+1}^n \bigl(Y_i-\langle\tT_{\mathrm{init},1},\tX_i\rangle\bigr)^2
          +\frac{1}{n} \sum_{i=1}^{n_0}\bigl(Y_i-\langle\tT_{\mathrm{init},2},\tX_i\rangle\bigr)^2 \\
        &= \frac{1}{n} \sum_{i=n_0+1}^n \bigl(Y_i-\langle\tT,\tX_i\rangle+\langle\tZ_1,\tX_i\rangle\bigr)^2
          +\frac{1}{n} \sum_{i=1}^{n_0} \bigl(Y_i-\langle\tT,\tX_i\rangle+\langle\tZ_2,\tX_i\rangle\bigr)^2\\
        &=\frac{1}{n} \sum_{i=1}^n \xi_i^2
          +\frac{1}{n} \sum_{i=n_0+1}^n\langle\tZ_1,\tX_i\rangle^2
          +\frac{1}{n} \sum_{i=1}^{n_0}\langle\tZ_2,\tX_i\rangle^2 \\
        &\quad +\frac{1}{n_0} \sum_{i=n_0+1}^n\xi_i\langle\tZ_1,\tX_i\rangle
             +\frac{1}{n_0} \sum_{i=1}^{n_0}\xi_i\langle\tZ_2,\tX_i\rangle,
    \end{aligned}
\end{equation}
where $\tZ_1=\tT_{\mathrm{init},1}-\tT$ and $\tZ_2=\tT_{\mathrm{init},2}-\tT$.

By construction, $\tZ_1$ is independent of $\{\tX_i,\xi_i\}_{i=n_0+1}^n$, and $\tZ_2$ is independent of $\{\tX_i,\xi_i\}_{i=1}^{n_0}$. Applying Bernstein’s inequality to each term in \cref{eq:sigmahat-expansion} yields, with probability at least $1-(d_1d_3+d_2d_3)^{-2}$,
\begin{equation*}
    \left|\widehat{\sigma}_{\xi}^2-\sigma_{\xi}^2\right|
    \leq \frac{2\bigl(\|\tZ_1\|_{\mathrm{F}}^2+\|\tZ_2\|_{\mathrm{F}}^2\bigr)}{d_1 d_2}
          +\frac{C_1 \sigma_{\xi}^2 \log (d_1d_3+d_2d_3)}{\sqrt{n}}.
\end{equation*}
Under \cref{assump:init_entry}, this further implies, with probability at least $1-2(d_1d_3+d_2d_3)^{-2}$,
\begin{equation}\label{eq:sigmahat2-bound}
    \left|\widehat{\sigma}_{\xi}^2-\sigma_{\xi}^2\right|
    \leq \frac{C_1 \sigma_{\xi}^2 \log (d_1d_3+d_2d_3)}{\sqrt{n}}
          + C_2\gamma_n^2\sigma_{\xi}^2.
\end{equation}
If $C_2 \gamma_n^2 \leq 1 / 3$, then \cref{eq:sigmahat2-bound} implies
$\left|\widehat{\sigma}_{\xi}^2-\sigma_{\xi}^2\right| \leq \sigma_{\xi}^2 / 2$, and hence
$\sigma_{\xi}^2/2 \leq \widehat{\sigma}_{\xi}^2 \leq 3\sigma_{\xi}^2/2$. In particular,
\begin{equation*}
    \left|1-\frac{\widehat{\sigma}_{\xi}}{\sigma_{\xi}}\right|
    = \frac{|\widehat{\sigma}_{\xi}-\sigma_{\xi}|}{\sigma_{\xi}}
    = \frac{|\widehat{\sigma}_{\xi}^2-\sigma_{\xi}^2|}{\sigma_{\xi}(\widehat{\sigma}_{\xi}+\sigma_{\xi})}
    \leq 2\,\frac{\left|\widehat{\sigma}_{\xi}^2-\sigma_{\xi}^2\right|}{\sigma_{\xi}^2}.
\end{equation*}
Combining this with \cref{eq:sigmahat2-bound} and absorbing the factor $2$ into the constants gives
\begin{equation}\label{eq:sigma-ratio-bound}
    \left|1-\frac{\widehat{\sigma}_{\xi}}{\sigma_{\xi}}\right|
    \leq \frac{C_1 \log (d_1d_3+d_2d_3)}{\sqrt{n}} + C_2\gamma_n^2.
\end{equation}
Therefore the factor $\left(1-\widehat{\sigma}_{\xi}/\sigma_{\xi}\right)$ in the second term of \cref{eq:diff3.2} is $o_{\mathbb{P}}(1)$ under condition \cref{eq:rate2}.

\paragraph{Step 2: Control of $\widehat{s}_{\tM}$}
We next bound the third term in \cref{eq:diff3.2}, which involves $\widehat{s}_{\tM}$. We start by controlling
\begin{equation*}
    \big|\|\tM\tV\|_{\mathrm{F}}^2-\|\tM\widehat{\tV}_1\|_{\mathrm{F}}^2\big|.
\end{equation*}
Since $\tV$ and $\widehat{\tV}_1$ are orthogonal, we can write
\begin{equation}\label{eq:MV-diff-decomp}
    \begin{aligned}
        \big|\|\tM\tV\|_{\mathrm{F}}^2-\|\tM\widehat{\tV}_1\|_{\mathrm{F}}^2\big|
        &=\big|\|\tM\tV\tV^{\dagger}\|_{\mathrm{F}}^2-\|\tM\widehat{\tV}_1\widehat{\tV}_1^{\dagger}\|_{\mathrm{F}}^2\big| \\
        &=\big|2\langle \tM(\tV\tV^{\dagger}-\widehat{\tV}_1\widehat{\tV}_1^{\dagger}) , \tM\tV\tV^{\dagger}\rangle\,-\,
        \|\tM(\tV\tV^{\dagger}-\widehat{\tV}_1\widehat{\tV}_1^{\dagger})\|_{\mathrm{F}}^2\big| \\
        &\leq \big\|\tM(\tV\tV^{\dagger}-\widehat{\tV}_1\widehat{\tV}_1^{\dagger})\big\|_{\mathrm{F}}^2
             \,+\,2\big|\langle\tM(\tV\tV^{\dagger}-\widehat{\tV}_1\widehat{\tV}_1^{\dagger}),\, \tM\tV\tV^{\dagger}\rangle\big|.
    \end{aligned}
\end{equation}
By the spectral subspace perturbation bound in \cref{thm:distance}, we have
\begin{equation}\label{eq:tMtVtV}
    \begin{aligned}
        \big\|\tM(\tV\tV^{\dagger}-\widehat{\tV}_1\widehat{\tV}_1^{\dagger})\big\|_{\mathrm{F}}^2
        &\leq \|\tM\|_{\ell_1}^2\, \max_{j\in[d_2]}
        \bigl\|\te_j^{\dagger}\,(\widehat{\tV}_1\widehat{\tV}^{\dagger}_1-\tV\tV^{\dagger})\bigr\|_{\mathrm{F}}^2 \\
        &\leq \|\tM\|_{\ell_1}^2\, \max_{j\in[d_2]}
        \bigl\|\te_j^{\dagger}\,(\widehat{\tV}_1\widehat{\tV}^{\dagger}_1-\tV\tV^{\dagger})\bigr\|^2\\
        &\leq \|\tM\|_{\ell_1}^2\,C_1^2\mu_{\max}^2
        \frac{\sigma_{\xi}^2}{\lambda_{\min}^2}\,d_3^2
        \frac{rd_1(d_1\vee d_2)}{n}\,
        \log(d_1d_3+d_2d_3),
    \end{aligned}
\end{equation}
and
\begin{equation}\label{eq:tMtVtV2}
    \begin{aligned}
        &\big|\langle\tM(\tV\tV^{\dagger}-\widehat{\tV}_1\widehat{\tV}_1^{\dagger}),\, \tM\tV\tV^{\dagger}\rangle\big|\\
        &\leq  \|\tM\tV\|_{\mathrm{F}} \|\tM\|_{\ell_1}\,
        \max_{j\in[d_2]}\bigl\|\te_j^{\dagger}\,(\widehat{\tV}_1\widehat{\tV}^{\dagger}_1-\tV\tV^{\dagger})\bigr\| \\
        &\leq \|\tM\tV\|_{\mathrm{F}} \|\tM\|_{\ell_1}\,
        C_1\mu_{\max}\frac{\sigma_{\xi}}{\lambda_{\min}}\,
        d_3\sqrt{\frac{rd_1(d_1\vee d_2)}{n}}\,
        \sqrt{\log(d_1d_3+d_2d_3)}.
    \end{aligned}
\end{equation}
Combining \cref{eq:MV-diff-decomp,eq:tMtVtV,eq:tMtVtV2}, we obtain
\begin{multline*}
        \big|\|\tM\tV\|_{\mathrm{F}}^2-\|\tM\widehat{\tV}_1\|_{\mathrm{F}}^2\big|
        \leq \|\tM\|_{\ell_1}^2\,C_1^2\mu_{\max}^2
        \frac{\sigma_{\xi}^2}{\lambda_{\min}^2}\,d_3^2
        \frac{rd_1(d_1\vee d_2)}{n}\,
        \log(d_1d_3+d_2d_3) \\
         + \|\tM\tV\|_{\mathrm{F}} \|\tM\|_{\ell_1}\,
        C_1\mu_{\max}\frac{\sigma_{\xi}}{\lambda_{\min}}\,
        d_3\sqrt{\frac{rd_1(d_1\vee d_2)}{n}}\,
        \sqrt{\log(d_1d_3+d_2d_3)}. 
\end{multline*}
A completely analogous argument yields a bound for
$\big|\|\tU^{\dagger}\tM\|_{\mathrm{F}}^2-\|\widehat{\tU}_1^{\dagger}\tM\|_{\mathrm{F}}^2\big|$.
Hence, on the high-probability event of \cref{thm:distance},
\begin{multline}\label{eq:sM-sq-diff}
        \big| \widehat{s}_{\tM}^2 - s_{\tM}^2\big|
        \leq 2C_1^2\mu_{\max}^2\|\tM\|_{\ell_1}^2\,
        \frac{\sigma_{\xi}^2}{\lambda_{\min}^2}\,d_3^2
        \frac{r(d_1\vee d_2)^2}{n}\,
        \log(d_1d_3+d_2d_3) \\
        \quad + \bigl(\|\tM\tV\|_{\mathrm{F}} +\|\tU^{\dagger}\tM\|_{\mathrm{F}}\bigr)\|\tM\|_{\ell_1}
         C_1\mu_{\max}\frac{\sigma_{\xi}}{\lambda_{\min}}\,
        d_3\sqrt{\frac{r(d_1\vee d_2)^2}{n}}\,
        \sqrt{\log(d_1d_3+d_2d_3)}.
\end{multline}

By \cref{assump:M},
\begin{equation*}
    \|\tM\tV\|_{\mathrm{F}}^2 +\|\tU^{\dagger}\tM\|_{\mathrm{F}}^2
    \;\geq\; 2\alpha_{\tM}^2\|\tM\|_{\mathrm{F}}^2\,\frac{r}{d_1\vee d_2}.
\end{equation*}
Dividing both sides of \cref{eq:sM-sq-diff} by
$\|\tM\tV\|_{\mathrm{F}}^2 +\|\tU^{\dagger}\tM\|_{\mathrm{F}}^2$ and using the fact that $(\|\tM\tV\|_{\mathrm{F}} +\|\tU^{\dagger}\tM\|_{\mathrm{F}})\leq \sqrt{2(\|\tM\tV\|_{\mathrm{F}}^2 +\|\tU^{\dagger}\tM\|_{\mathrm{F}}^2)}$, we obtain
\begin{multline}\label{eq:sM-ratio-bound}
        \bigg| \frac{\widehat{s}_{\tM}^2}{s_{\tM}^2} -1 \bigg|
        \leq C_1^2\frac{\mu_{\max}^2\|\tM\|_{\ell_1}^2}{\alpha_{\tM}^2\|\tM\|_{\mathrm{F}}^2}\,
        \frac{\sigma_{\xi}^2}{\lambda_{\min}^2}\,d_3^2
        \frac{(d_1\vee d_2)^3}{n}\,
        \log(d_1d_3+d_2d_3) \\
        \quad + C_1\,\frac{\mu_{\max}\|\tM\|_{\ell_1}}{\alpha_{\tM}\|\tM\|_{\mathrm{F}}}
         \frac{\sigma_{\xi}}{\lambda_{\min}}\,
        d_3\sqrt{\frac{(d_1\vee d_2)^3}{n}}\,
        \sqrt{\log(d_1d_3+d_2d_3)}.
\end{multline}Given the SNR threshold in \cref{assump:SNR},\begin{equation*}
    \frac{\lambda_{\min}}{\sigma_\xi}\ \ge\ C\, \frac{d_3(d_1\vee d_2)^{3/2}}{\sqrt{n}}\,
    \sqrt{\log\!\big(d_1d_3+d_2d_3\big)},
\end{equation*}and the fact that $\|\tM\|_{\ell_1}/\|\tM\|_{\mathrm{F}}\ge1$, we can simplify the bound in \cref{eq:sM-ratio-bound} as:
\begin{equation}\label{eq:sM-ratio-bound_simple}
    \bigg| \frac{\widehat{s}_{\tM}^2}{s_{\tM}^2} - 1\bigg|
    \;\leq\; C_1\, \frac{\mu_{\max}^2\|\tM\|_{\ell_1}^2 }{\alpha_{\tM}^2\|\tM\|_{\mathrm{F}}^2}\,
    \frac{\sigma_{\xi}}{\lambda_{\min}}\,
    \frac{d_3(d_1\vee d_2)^{3/2}}{\sqrt n}\,
    \sqrt{{\log(d_1d_3+d_2d_3)}}.
\end{equation}
Under conditions~\cref{eq:rate1}, the right-hand side of \cref{eq:sM-ratio-bound_simple} tends to zero.

\paragraph{Step 3: Conclusion}
By \cref{cor}, the first term on the right-hand side of \cref{eq:diff3.2} converges in distribution to $\mathcal N(0,1)$ as $d_1,d_2,n\to\infty$ under conditions~\cref{eq:rate1,eq:rate2}. The bounds \cref{eq:sigma-ratio-bound} and \cref{eq:sM-ratio-bound_simple}, together with conditions~\cref{eq:rate1,eq:rate2}, imply that the second and third terms in \cref{eq:diff3.2} converge to zero. An application of Slutsky’s theorem then yields
\begin{equation*}
    \frac{\langle\widehat{\tT},\tM\rangle-\langle\tT,\tM\rangle}
    {\widehat{\sigma}_{\xi} \widehat{s}_{\tM}\,\sqrt{d_1 d_2 d_3/ n}}
    \;\toD\; \mathcal N(0,1),
\end{equation*}
which completes the proof of \cref{thm:infer}.

\section{Additional simulation results}
To further assess the robustness of our uncertainty quantification procedure, we report additional simulation results under three more challenging variants of the main setting in \cref{sub:simu}: a longer tensor with $d_3=1000$, a higher missingness level with $n=0.2\,d_1d_2d_3$, and a higher noise level with $\sigma_\xi=0.8$. Across all these settings, the confidence intervals constructed for both $\langle \tT,\tM^{(k)}\rangle$ and $Y_{\tM^{(k)}}$ maintain empirical coverage close to the nominal 95\% level, with interval widths adapting to the increased difficulty of the problem. These results provide additional evidence that our UQ framework remains reliable under \textbf{larger dimensionality}, \textbf{sparser observations}, and \textbf{noisier measurements}.

\begin{table}[H]
\centering
\small
\begin{tabular}{lccccc}
\toprule
Estimator & CI width($\langle\tT,\tM\rangle$) & Coverage($\langle\tT,\tM\rangle$) & CI width($Y_{\tM}$) & Coverage($Y_{\tM}$) \\
\midrule
$\langle \widehat{\tT},\tM^{(1)}\rangle$ &0.4948$\pm$ 0.0005  &0.951& 2.5159$\pm$ 0.0002 & 0.96 \\
$\langle \widehat{\tT},\tM^{(2)}\rangle$ &0.4908$\pm$0.0005 & 0.952& 2.5151 $\pm$ 0.0002& 0.965 \\
$\langle \widehat{\tT},\tM^{(3)}\rangle$ & 0.6969$\pm$ 0.0005 & 0.947&3.5575$\pm$ 0.0003& 0.96\\
$\langle \widehat{\tT},\tM^{(4)}\rangle$& 0.8649$\pm$ 0.0012& 0.95&4.3593$\pm$ 0.0004 & 0.968 \\
\bottomrule
\end{tabular}
\caption{\small Empirical 95\% confidence interval (CI) performance over 1000 Monte Carlo replications. The columns “CI width($\langle\tT,\tM\rangle$)” and “Coverage($\langle\tT,\tM\rangle$)” report, respectively, the average interval width and empirical coverage of the confidence intervals in~\cref{CI1} for the target $\langle\tT,\tM^{(k)}\rangle$. The columns “CI width($Y_{\tM}$)” and “Coverage($Y_{\tM}$)” report the corresponding quantities for the noisy linear functional $Y_{\tM^{(k)}}=\langle\tT,\tM^{(k)}\rangle+\xi$ based on the intervals in~\cref{CI2}. The simulation parameters are $d_1=d_2=500$, $\boldsymbol{d_3=1000}$ \textbf{(larger dimensionality)}, $r=4$, $\sigma_\xi=0.6$, and $n=0.4\,d_1d_2d_3$.
}
\end{table}

\begin{table}[H]
\centering
\small
\begin{tabular}{lccccc}
\toprule
Estimator & CI width($\langle\tT,\tM\rangle$) & Coverage($\langle\tT,\tM\rangle$) & CI width($Y_{\tM}$) & Coverage($Y_{\tM}$) \\
\midrule
$\langle \widehat{\tT},\tM^{(1)}\rangle$ &
0.7464 $\pm$ 0.0016  &  0.956
& 2.7130 $\pm$ 0.0007  &  0.977\\
$\langle \widehat{\tT},\tM^{(2)}\rangle$ &
0.7263 $\pm$ 0.0016  &  0.944
& 2.7075 $\pm$ 0.0007  &  0.972 \\
$\langle \widehat{\tT},\tM^{(3)}\rangle$ & 
1.0415 $\pm$ 0.0016  &  0.951
& 3.8328 $\pm$ 0.0009  &  0.968\\
$\langle \widehat{\tT},\tM^{(4)}\rangle$&
1.3034 $\pm$ 0.0039  &  0.956
& 4.7019 $\pm$ 0.0014  &  0.978\\
\bottomrule
\end{tabular}
\caption{\small Empirical 95\% confidence interval (CI) performance over 1000 Monte Carlo replications. The columns “CI width($\langle\tT,\tM\rangle$)” and “Coverage($\langle\tT,\tM\rangle$)” report, respectively, the average interval width and empirical coverage of the confidence intervals in~\cref{CI1} for the target $\langle\tT,\tM^{(k)}\rangle$. The columns “CI width($Y_{\tM}$)” and “Coverage($Y_{\tM}$)” report the corresponding quantities for the noisy linear functional $Y_{\tM^{(k)}}=\langle\tT,\tM^{(k)}\rangle+\xi$ based on the intervals in~\cref{CI2}. The simulation parameters are $d_1=d_2=d_3=500$, $r=4$, $\sigma_\xi=0.6$, and $\boldsymbol{n=0.2\,d_1d_2d_3}$ \textbf{(sparser observations)}.
}
\end{table}

\begin{table}[H]
\centering
\small
\begin{tabular}{lccccc}
\toprule
Estimator & CI width($\langle\tT,\tM\rangle$) & Coverage($\langle\tT,\tM\rangle$) & CI width($Y_{\tM}$) & Coverage($Y_{\tM}$) \\
\midrule
$\langle \widehat{\tT},\tM^{(1)}\rangle$ &
0.6658 $\pm$ 0.0013  &  0.956
&  3.3568 $\pm$ 0.0005  &  0.966\\
$\langle \widehat{\tT},\tM^{(2)}\rangle$ &
0.6478 $\pm$ 0.0012  &  0.93
&  3.3533 $\pm$ 0.0005  &  0.964\\
$\langle \widehat{\tT},\tM^{(3)}\rangle$ & 
0.9290 $\pm$ 0.0013  &  0.946
& 4.7447 $\pm$ 0.0006  &  0.954\\
$\langle \widehat{\tT},\tM^{(4)}\rangle$& 
1.1626 $\pm$ 0.0032  &  0.959
&  5.8160 $\pm$ 0.0009  &  0.969\\
\bottomrule
\end{tabular}
\caption{\small Empirical 95\% confidence interval (CI) performance over 1000 Monte Carlo replications. The columns “CI width($\langle\tT,\tM\rangle$)” and “Coverage($\langle\tT,\tM\rangle$)” report, respectively, the average interval width and empirical coverage of the confidence intervals in~\cref{CI1} for the target $\langle\tT,\tM^{(k)}\rangle$. The columns “CI width($Y_{\tM}$)” and “Coverage($Y_{\tM}$)” report the corresponding quantities for the noisy linear functional $Y_{\tM^{(k)}}=\langle\tT,\tM^{(k)}\rangle+\xi$ based on the intervals in~\cref{CI2}. The simulation parameters are $d_1=d_2=d_3=500$, $r=4$, $\boldsymbol{\sigma_\xi=0.8}$ \textbf{(noisier measurements)}, and $n=0.4\,d_1d_2d_3$.
}
\end{table}

\section*{Acknowledgments}
YC acknowledges funding from NASA. The authors thank Professor Shasha Zou for helpful discussions on TEC data.

\bibliographystyle{siam} 
\bibliography{references} 
\end{document}